\documentclass[a4paper, 12pt]{article}
\usepackage{authblk}
\usepackage[top=28truemm,bottom=28truemm,left=27truemm,right=27truemm]{geometry}
\usepackage{setspace}
\usepackage{amsmath,amssymb,mathrsfs,amsbsy,latexsym,amsfonts,amsthm}
\usepackage{fancyhdr}
\usepackage[Symbol]{upgreek}

\pdfoutput=1                        
\usepackage{graphicx}           %   usepackage without driver option
\usepackage{color}
\usepackage{slashed}
\usepackage{tensor}
\usepackage{comment}
\usepackage{braket}

\usepackage{ascmac}
\usepackage{here}
\usepackage{bm}
\usepackage{hyperref}
\newtheorem{theorem}{Theorem}
\newtheorem{definition}{Definition}
\newtheorem{lemma}[theorem]{Lemma}
\newtheorem{corollary}[theorem]{Corollary}
\newtheorem{example}{Example}
\usepackage[utf8]{inputenc}
\usepackage{mathtools}
\newcommand{\epn}{\epsilon}
\newcommand{\be}{\begin{equation}}
\newcommand{\ee}{\end{equation}}
\newcommand{\ba}[1]{\begin{align}#1\end{align}}
\makeatletter

\@addtoreset{equation}{section}
\makeatother

\begin{document}
	\setlength{\abovedisplayskip}{10pt}
	\setlength{\belowdisplayskip}{10pt}

%%%%%%%%%%%%%%%%%%%% TITLE %%%%%%%%%%%%%%%%%%%%
\title{
	\bf \Large
Shrinking of Operators in\\ Quantum Error Correction and AdS/CFT
	\vskip 0.5cm
}

%\noindent\rule{\columnwidth}{1pt}

\author{
	 Hayato~Hirai\thanks{\tt hirai@het.phys.sci.osaka-u.ac.jp}
}
\affil{\it\normalsize Department of Physics, Osaka University, Toyonaka, Osaka, 560-0043, Japan}
\setcounter{Maxaffil}{0}

\date{}

\maketitle
\thispagestyle{fancy}
\renewcommand{\headrulewidth}{0pt}

\begin{abstract}
%We show that for bipartite pure states with non-vanishing Schmidt coefficients on $\mathcal{H}_{A}\otimes\mathcal{H}_{B}$ with $\text{dim}(\mathcal{H}_{A})\leq\text{dim}(\mathcal{H}_{B})$, nonlocal operators on $\mathcal{H}_{A}\otimes\mathcal{H}_{B}$ can shrink their supports only to $\mathcal{H}_{B}$. Using this fact, we show how to systematically construct the decoder of quantum error corrections against erasure errors.
We first show that a class of operators acting on a given bipartite pure state on $\mathcal{H}_{A}\otimes\mathcal{H}_{B}$ can shrink its supports on $\mathcal{H}_{A}\otimes\mathcal{H}_{B}$ to only  $\mathcal{H}_{A}$ or $\mathcal{H}_{B}$ while keeping its mappings. 
Using this result, we show how to systematically construct the decoders of the quantum error-correcting codes against erasure errors. 
The implications of the results for the operator dictionary in the AdS/CFT correspondence are also discussed.
The ``subalgebra code with complementary recovery" introduced  in the recent work of Harlow is a quantum error-correcting code that shares many common features with the AdS/CFT correspondence. 
We consider it under the restriction of  the bulk (logical) Hilbert space to a subspace that generally has no tensor factorization into subsystems.
In this code, the central operators of the reconstructed algebra on the boundary subregion can emerge as a consequence of the restriction of   the bulk Hilbert space.
Finally, we show a theorem in this code which implies the validity of not only the entanglement wedge reconstruction but also its converse statement with the central operators. 
%We also show the analogous results to the Ryu-Takayanagi formula and its quantum correction in the code.
%with non-vanishing Schmidt coefficients on $\mathcal{H}_{A}\otimes\mathcal{H}_{B}$ with $\text{dim}(\mathcal{H}_{A})\leq\text{dim}(\mathcal{H}_{B})$,
\end{abstract}

\newpage
\setcounter{tocdepth}{2}
\tableofcontents

\newpage

\setlength{\parskip}{5pt} 
%%%%%%%%%%%%%%%%%%%%%%%%%%%%%%%%%%%%%%%%%%%%%%%%%%%%%%%%%%%%%%
%%%%%%%%%%%%%%%%%%%%%%%%%%%%%%%%%%%%%%%%%%%%%%%%%%%%%%%%%%%%%%
\section{Introduction}
In this paper, we first address a question of ``what kind of operators can \textit{shrink} the supports on a given state (or on a given space) while keeping their mapping".
More precisely, given a bipartite pure state $|\psi\rangle_{AB}\in \mathcal{H}_{A}\otimes\mathcal{H}_{B}$, for what kind of operators $\mathcal{O}_{AB}$, does there exist  an operator $\mathcal{O}_{A}$ such that
\ba{
\mathcal{O}_{AB}|\psi\rangle_{AB}=\mathcal{O}_{A}\otimes I_{B}|\psi\rangle_{AB},
}
and how can we construct such $\mathcal{O}_{A}$\hspace{0.5mm}?
We show a result  as an answer to the question  and also show the conditions for the Hermiticity and the unitarity of such $\mathcal{O}_{A}$ in section \ref{Shrinkingteleporting}.
We then apply it to
%in order to increase our understanding of   quantum error-correcting codes and the  AdS/CFT correspondence.
the construction of decoders in quantum error-correcting codes against erasure errors,
and also to the study of  the qualitative properties of the operator dictionary in the AdS/CFT correspondence.

\begin{center} \textit{Quantum error-correcting codes}
\end{center}\vspace{-4mm}

%A quantum computer is a desired device which is expected to be able to solve interesting problems that cannot be solved with classical computers in a reasonable amount of time. 
%However, it is extremely difficult to built a quantum computer.  
The major obstacle to building a quantum computer is originated from the fact that quantum information is extremely fragile against the disturbances caused by its environment. 
Since it is impossible to perfectly isolate a quantum device from its environment, the quantum information on the quantum device is rapidly transferred to nonlocal correlations between the device and its environment via the interactions between the two. 
The schemes to protect quantum information against such information outflow are called \emph{quantum error-correcting codes} \cite{PhysRevA.52.R2493}.
Thus both theoretical and experimental constructions of quantum error-correcting codes are crucial to realizing quantum computers.

An essential idea of the quantum error-correcting codes is that quantum information encoded in nonlocal correlations, $i.e$ quantum entanglement, is invulnerable to local errors.
Quantum error-correcting codes map a quantum state, which we want to protect against errors, into a largely entangled state spanned by the \emph{logical} basis so that local errors cannot destroy the quantum information about the encoded state.
Finding out and correcting  the errors during  quantum computations is called the \emph{decoding} process.
A central problem of the theoretical construction of quantum error-correcting codes is how to find the pairs of a logical basis and a decoder.

In this paper, we  focus on quantum error-correcting codes that are able to protect against erasure errors where the position of the erroneous degrees of freedom is known. This error model is called the \emph{quantum erasure channel} \cite{Grassl:1996eh} and the quantum error-correcting codes are called the \emph{quantum secret sharing scheme} \cite{Hillery:1998yq, Cleve:1999qg}.
The decoder against the quantum erasure channel is a unitary operator that is able to recover an encoded state without touching the erased degrees of freedom. 
We briefly review  quantum error-correcting codes against the erasure errors, especially focusing on the encoding and decoding process, in subsection \ref{introQTS}.

One of the main results of this paper is the systematic constructions of decoders of the quantum secret sharing schemes from the given logical bases.
Our approach to the problem starts with noticing that it is very easy to construct a decoder if it is allowed to act on all the degrees of freedom on a code subspace. We call such a kind of decoders \emph{trivial} decoders in the sense that it cannot decode even a single erasure error.
Then we reduce the problem of how to construct decoders of the quantum error-correcting codes to the problem of how to shrink the support of trivial decoders while keeping their action  on the code subspaces. %and its unitarity. 
%As an answer to the problem, we will show a general result of how to shrink the supports of nonlocal operators on bipartite systems, in section \ref{Shrinkingteleporting}.
%Then we will apply it to the quantum error-correcting codes and show how to construct the decoders systematically in section \ref{Systematicconstruction}.
Finally, by using the results of how to shrink the supports of operators in section \ref{Shrinkingteleporting}, we show a formula (in Theorem \ref{makedecoder}) about how to construct the decoders systematically in section \ref{Systematicconstruction}.

%\noindent\underline{AdS/CFT correspondence}
\begin{center} \textit{AdS/CFT correspondence}
\end{center}\vspace{-4mm}

The AdS/CFT correspondence \cite{Maldacena:1997re} is a duality between quantum gravity on a $(d+1)$-dimensional asymptotically anti-de Sitter space(AdS) and a $d$-dimensional conformal field theory (CFT) defined on the boundary of the AdS spacetime. 
In this correspondence, it is believed that there is an exact map between the observables of the two theories, which is sometimes called the operator dictionary.
The dictionary is still under investigation, 
but it has been known that the dictionary has the following two important qualitative features in the low-energy sector of the theories at least 
when the bulk theory is the weakly interacting semiclassical field theory on AdS background,
; (i) the local operators acting deep inside the bulk correspond to highly nonlocal operators in the CFT,  (ii) the map from a single bulk operator to a boundary operator is not unique \cite{Hamilton:2006az, Morrison:2014jha}. 
These two features are encapsulated by the  \textit{causal wedge reconstruction} \cite{Hamilton:2006az, Morrison:2014jha}
: given a subregion $A$ in the boundary on a time-slice $\Sigma$,  then a bulk operator $\phi(x)$ acting inside the \textit{causal wedge} of $A$ \cite{Hubeny:2012wa} can be reconstructed as a boundary operator with nontrivial support only on $A$. 
Moreover it has been proposed that the bulk operators that can be reconstructed only on $A$  are not just the ones acting inside the causal wedge of $A$ but also inside the larger region, the \textit{entanglement wedge} of $A$ \cite{Czech:2012bh, Wall:2012uf, Headrick:2014cta}. This proposal is called the \textit{entanglement wedge reconstruction hypothesis}. 
The validity of the entanglement wedge reconstruction was argued based on the equivalence of the bulk and the boundary relative entropy in \cite{Jafferis:2015del}

In \cite{Almheiri:2014lwa},  the non-uniqueness of the operator dictionary has been reinterpreted in the language of quantum error-correcting codes against erasure errors.
In this interpretation,  the bulk low-energy Hilbert space on  semiclassical AdS background is regarded as the logical Hilbert space, and it is encoded into the dual boundary Hilbert space.
It opened the possibility of the intimate relation between the  AdS/CFT and quantum error-correcting codes, and this relation has been further studied from many different perspectives \cite{ Mintun:2015qda, Pastawski:2015qua, Hayden:2016cfa, Freivogel:2016zsb, Harlow:2016vwg, Donnelly:2016qqt, Pastawski:2016qrs}. 
In particular, the quantum error-correcting features of AdS/CFT argued in \cite{Almheiri:2014lwa}  was first explicitly realized in toy models \cite{Pastawski:2015qua}.
This is a quantum error-correcting code model based on the tensor network and it realize the many important relations between the bulk geometry and the entanglement structures of boundary CFT on a time-slice $e.g.$ the Ryu-Takayanagi formula \cite{Ryu:2006bv}, the holographic formula for the EoP \cite{Takayanagi:2017knl, Nguyen:2017yqw}
\footnote{It has been proposed that the holographic dual of the entanglement of purification (EoP) is  given by the minimal cross-section of entanglement wedge in \cite{Takayanagi:2017knl, Nguyen:2017yqw}.  This conjecture actually holds in the holographic code model \cite{Hirai:2018jwy}.}
, and in particular the entanglement wedge reconstruction.
Furthermore, a theorem in quantum information theory which implies the entanglement wedge reconstruction was proven in \cite{Dong:2016eik} based on the arguments in \cite{Jafferis:2015del}, and this theorem was further generalized in \cite{Harlow:2016vwg}.
Based on the \textit{operator-algebra quantum error correction} of \cite{PhysRevLett.98.100502, PhysRevA.76.042303}, a class of error-correcting codes called  the \textit{subsystem code with complementarity recovery}  was introduced in \cite{Harlow:2016vwg}.
This code qualitatively realizes the subregion duality in the bulk reconstruction, the Ryu-Takayanagi formula including quantum corrections and also  the equivalence of bulk and boundary relative entropies. 
The mathematical equivalence of the above three features was also revealed.
It was  also shown that in general there exist the nontrivial central operators in the reconstructed algebra,
which implies the existence of the bulk operators  reconstructable on a boundary subregion and also on the complementary of the subregion.
A concrete example of such subalgebra code with complementary recovery with such nontrivial centers was constructed in \cite{Donnelly:2016qqt}.

In order to further study  the qualitative features of the operator dictionary in AdS/CFT from the viewpoint of quantum error-correcting codes, we consider the subsystem code with complementarity recovery in which the logical system is restricted to a subspace that generally does not have the tensor factorization into subsystems.
In this code, we point out that the nontrivial central operators of the reconstructed algebra can emerge from the restriction of  the  logical Hilbert space.
We also show a theorem in this code, which implies the validity of the entanglement wedge reconstruction and also its converse statement.

\vspace{6mm}

This paper is organized as follows.
In section \ref{Shrinkingteleporting}, we  show how to shrink the supports of nonlocal operators in bipartite systems while keeping their mappings and also their unitarity or Hermiticity. 
In section \ref{Systematicconstruction}, we start with reviewing briefly the ``quantum secret sharing schemes" as a typical example of the quantum error-correcting codes that are able to correct erasure errors, especially by focusing on the encoding and the decoding procedures.
Next, we show a useful expression of a logical basis in the quantum error-correcting codes against erasure errors.
We then show how to systematically construct a decoder against erasure errors.
In section \ref{adscft}, we study the subalgebra code with complementary recovery under the restrictions of the bulk (logical) Hilbert space to a subspace.
We first explain our model.
Secondly, we review the basic operator dictionary and also how the Ryu-Takayanagi-like formula holds in a concrete way.
We then discuss the emergence of the central operators from the restriction of bulk Hilbert space.
Finally, we prove a theorem for the code, which implies the validity of the entanglement wedge reconstruction and also its converse statement with nontrivial centers.
Appendix \ref{sec:Teleporting} shows the formula for how to teleport the support of operators supported on $B$ to only $A$ on a state in $\mathcal{H}_{A}\otimes\mathcal{H}_{B}$.
Section \ref{SandD} contains a summary and some open questions.
In Appendix \ref{sec:TFD}, we calculate the teleported operator on the thermofield double state based on the formula in Appendix \ref{sec:Teleporting}  and see that the answer reproduces the known result.

%%%%%%%%%%%%%%%%%
\subsection{Notation}
We will use $|A|$ to denote the dimensionality of $\mathcal{H}_{A}$. We will also label the physical degrees of freedom ($e.g.$ qudits) associated with the Hilbert space $\mathcal{H}_{A}$ by $A$. 
We will write operators supported on $\mathcal{H}_{A}$ and states in $\mathcal{H}_{A}$ with a subscript $A$, for example $\mathcal{O}_{A}$ is a linear operator on $\mathcal{H}_{A}$ and $|\psi\rangle_{A}$ is a state in $\mathcal{H}_{A}$.
We do not distinguish between $\mathcal{O}_{A}\hspace{0.5mm}\otimes\hspace{0.5mm} I_{\overline{A}}$ and $\mathcal{O}_{A}$. In most cases we will omit the identity operator, but sometimes write it explicitly to emphasize the support of the operator. 
In order to lighten our notation we will often write $\sum_{i=1}^{|A|}\sum_{j=1}^{|A|}\cdots$ as $\sum_{i,j}^{|A|}\cdots$, and also write the statement 
``$\  ^{\forall}|\psi\rangle \in \mathcal{H}\ ,\ \mathcal{O}|\psi\rangle \in \mathcal{H}$\ "
as
``\ $\mathcal{O}\mathcal{H}\subseteq\mathcal{H}$\ ".

We will write $\mathcal{L}(\mathcal{H})$ as the set of linear operators acting on $\mathcal{H}$, and also write $\mathcal{A}(\mathcal{H})$ as the algebra of the Hermitian linear operators supported on  $\mathcal{H}$ and acting within $\mathcal{H}$.

%%%%%%%%%%%%%%%%%%%%%%%%%%%%%%%%%%%%%%%%%%%%
\section{Shrinking the supports of operators in bipartite systems}\label{Shrinkingteleporting}
%Simple proof of Reeh-Schlieder theorem
In this section, we  show when and how to shrink the supports of nonlocal operators in bipartite systems while keeping their mappings and also their unitarity or Hermiticity. 
\subsection{Shrinking the supports of nonlocal operators }
Consider a bipartite finite-dimensional Hilbert space $\mathcal{H}_{AB}=\mathcal{H}_{A}\otimes\mathcal{H}_{B}$. 
Then any state in $\mathcal{H}_{AB}$ can be expressed in the Schmidt basis as,  
\be\label{state}
|\psi\rangle_{AB}=\sum_{i=1}^{N}\psi_{i}|i\rangle_{A}|i\rangle_{B}
\ee
where $N$ is a positive integer equal to or smaller than $min\{|A|,|B|\}$,  $\psi_i$'s are positive real numbers with $\sum_{i=1}^{N}\psi_i^{2}=1$, and  $\{|i\rangle_{A/B} \}_{i=1,\cdots, N}$  is a set of orthonormal states in $\mathcal{H}_{A/B}$. It is important that the choice of these Schmidt basis generally depends on the state $|\psi\rangle_{AB}$.

Then we can always prepare  bases $\{|i\rangle_{A} \}_{i=1,\cdots, |A|}$ on  $\mathcal{H}_{A}$ and $\{|i\rangle_{B} \}_{i=1,\cdots, |B|}$ on $\mathcal{H}_{B}$ by adding  $|A|-N$ number of new orthonormal states in $\mathcal{H}_{A}$ to $\{|i\rangle_{A} \}_{i=1,\cdots, |N|}$ and $|B|-N$ number of new orthonormal states in $\mathcal{H}_{B}$ to $\{|i\rangle_{B} \}_{i=1,\cdots, |N|}$, respectively. 
Defining $\mathcal{L}(\mathcal{H})$ as the set of the linear operators on $\mathcal{H}$,
we can write any operator $\mathcal{O}_{AB}$ in $\mathcal{L}(\mathcal{H}_{AB})$  as 
\ba{\label{operatorO_AB}
\mathcal{O}_{AB}=\sum_{i,k=1}^{|A|}\sum_{j,l=1}^{|B|}\mathcal{O}^{ij,kl}\ |i\rangle_{A}|j\rangle_{B\hspace{0.5mm}A}\langle k|_{B}\langle l|.
}
The density matrix for $B$ is given by
\ba{
\rho_{B}={\mathop{\mathrm{Tr}}_{A}}|\psi\rangle_{AB\hspace{0.5mm}AB}\langle \psi |=\sum_{i}^{N}\psi_{i}^{2}|i\rangle_{B\hspace{0.5mm} B}\langle i|\ .
}
The projection operator onto the kernel of $\rho_{B}$ in $\mathcal{H}_{B}$ is then given by
\ba{
P_{\text{ker}\rho_{B}} =\sum_{i>N}^{|B|} |i\rangle_{B\hspace{0.5mm} B}\langle i|\ .
}
We then define  two sets of operators, $\mathcal{V}(\hspace{0.5mm}|\psi\rangle_{AB}\hspace{0.5mm})$ and $\mathcal{S}_{\mathcal{L}}(A\hspace{0.5mm}; |\psi\rangle_{AB})$, as follows:

\newpage

%%%%%%%%% Definition%%%%%%%%%%%%%%
\noindent\rule{\columnwidth}{1pt}\vspace{-2mm}
%\begin{screen}
\begin{definition} %\ \\
\ Given a state $|\psi\rangle_{AB}$ in $\mathcal{H}_{AB}$,  two set of operators, $\mathcal{V}(\hspace{0.5mm}|\psi\rangle_{AB}\hspace{0.5mm})$ and $\mathcal{S}_{\mathcal{L}}(A\hspace{0.5mm}; |\psi\rangle_{AB})$, are defined as
\normalfont
\ba{
&\bullet\  \mathcal{V}(\hspace{0.5mm}|\psi\rangle_{AB}\hspace{0.5mm})
\equiv \{\ \mathcal{O}_{AB}\in \mathcal{L}(\mathcal{H}_{AB})\ |\ \mathcal{O}_{AB}|\psi\rangle_{AB}=0\ \}\ ,\ \\[5mm]
\label{defofSpsiwithKer}
&\bullet\  \mathcal{S}_{\mathcal{L}}(A\hspace{0.5mm}; |\psi\rangle_{AB})
 \equiv \{\ \mathcal{O}_{AB}\in \mathcal{L}(\mathcal{H}_{AB})\ |\  P_{\text{ker}\rho_{B}}\mathcal{O}_{AB}|\psi\rangle_{AB}=0\   \}.
}
\end{definition}
\vspace{-5mm}
\noindent\rule{\columnwidth}{1pt}
%%%%%%%%%%%%%%%%%%%%%%%%%%

 $\mathcal{V}(\hspace{0.5mm}|\psi\rangle_{AB}\hspace{0.5mm})$ is the set of the operators that vanish when they act on $|\psi\rangle_{AB}$.
Any element $\overline{\mathcal{O}}_{AB}(\psi)\in \mathcal{V}(\hspace{0.5mm}|\psi\rangle_{A\overline{A}}\hspace{0.5mm})$ can  generally be written as
\ba{
\overline{\mathcal{O}}_{AB}(\psi)
=\sum_{i,k=1}^{|A|}\sum_{j,l=1}^{|B|}\overline{\mathcal{O}}^{ij,kl}\ |i\rangle_{A}|j\rangle_{B\hspace{0.5mm}A}\langle k|_{B}\langle l|\ ,
}
with the condition,
\ba{
\sum_{k=1}^{N}\psi_{k}\overline{\mathcal{O}}^{ij,kk}=0\ , \ \ \text{for}\ \ 1\le i\le |A|\ ,\  1\le j\le |B|\ .
}
%for all possible $i$'s and $j$'s.
Any element  $\mathcal{O}_{AB} \in \mathcal{S}_{\mathcal{L}}(A\hspace{0.5mm}; |\psi\rangle_{AB})$ can also be written as
\ba{
\mathcal{O}_{AB} 
=\sum_{i=1}^{|A|}\sum_{j,k=1}^{N}\mathcal{O}^{ij,kk}\ |i\rangle_{A}|j\rangle_{B\hspace{0.5mm}A}\langle k|_{B}\langle k|
+\overline{\mathcal{O}}_{AB}(\psi)\ ,\ \label{operatorABpsiA}
}
where $\overline{\mathcal{O}}_{AB}(\psi) \in \mathcal{V}(\hspace{0.5mm}|\psi\rangle_{AB}\hspace{0.5mm})$.
The meaning of  $\mathcal{S}_{\mathcal{L}}(A\hspace{0.5mm}; |\psi\rangle_{AB})$ will become clear in the next lemma, which shows the condition for ``shrinking" the support of operators while keeping their mappings :

%%%%%%%%%  Lemma 1%%%%%%%%%%%%%%
\noindent\rule{\columnwidth}{1pt}\vspace{-2mm}
%\begin{screen}
\begin{lemma} \label{generalgatetel}%\ \\
\ For an given arbitrary state $|\psi\rangle_{AB} \in \mathcal{H}_{AB}$ with the Schmidt decomposition \eqref{state}, the following two statements are equivalent:
%with the Schmidt rank $N$ that is written in the Schmidt basis as \eqref{state}, the following two statements are equivalent,
\ba{
(A)\  \ &\text{For}\  \mathcal{O}_{AB}\in\mathcal{L}(\mathcal{H}_{AB}), \  ^{\exists}\mathcal{O}_{AB\rightarrow A}(\psi)\in\mathcal{L}(\mathcal{H}_{A}) \ s.t. \nonumber\\[5pt]
\label{GeneGateTel}
&\mathcal{O}_{AB}|\psi\rangle_{AB}=\mathcal{O}_{AB\rightarrow A}(\psi) \otimes I_{B} |\psi\rangle_{AB}\ . \\[0.3cm]
(B)\ \ &\mathcal{O}_{AB}\in \mathcal{S}_{\mathcal{L}}(A\hspace{0.5mm}; |\psi\rangle_{AB})\ \  \text{and}\ \  
\mathcal{O}_{AB\rightarrow A}(\psi)=\sum_{i=1}^{|A|}\sum_{j,k=1}^{N}\psi^{-1}_{j}\mathcal{O}^{ij,kk}\psi_{k}| i\rangle_{A\hspace{0.5mm}A}\langle j|
+
\overline{\mathcal{O}}_{A}(\psi),\label{shrunkopABtoA}\ \ \\[-0.5cm]
&\text{where}\ \mathcal{O}^{ij,kk}\  \text{is the matrix element of $\mathcal{O}_{AB}$ defined in \eqref{operatorABpsiA}, and $\overline{\mathcal{O}}_{A}(\psi)$ }\nonumber \\ 
&\text{is an arbitrary operator such that}\ \  \overline{\mathcal{O}}_{A}(\psi)\otimes I_{B} \in \mathcal{V}(|\psi\rangle_{AB}). 
\nonumber
}
\end{lemma}
\vspace{-5mm}
\noindent\rule{\columnwidth}{1pt}
%%%%%%%%%%%%%%%%%%%%%%%%%%

\begin{proof}\ \\
\noindent$(A)\Rightarrow(B)$: 
We can generally write an operator $\mathcal{O}_{AB}$ in $\mathcal{L}(\mathcal{H}_{AB})$ as
\ba{
\mathcal{O}_{AB}=\sum_{i,k}^{|A|}\sum_{j,l}^{|B|}\mathcal{O}^{ij,kl}\ |i\rangle_{A}|j\rangle_{B\hspace{0.5mm}A}\langle k|_{B}\langle l|\ ,
}
and also write an operator $\mathcal{O}_{AB\rightarrow A}(\psi)$ in $\mathcal{L}(\mathcal{H}_{A})$ as
\ba{
\mathcal{O}_{AB\rightarrow A}(\psi)=\sum_{i,j}^{|A|}\overline{\mathcal{O}}^{i,j} |i\rangle_{A\hspace{0.5mm}A}\langle j|\ ,
}
with $\mathcal{O}^{ij,kl}, \overline{\mathcal{O}}^{i,j}\in \mathbb{C}$.
Then we put these operators into eq.\eqref{GeneGateTel}.
The lefthand side of \eqref{GeneGateTel} is given by
\ba{
\mathcal{O}_{AB}|\psi\rangle_{AB}
=\sum_{i}^{|A|}\sum_{j}^{|B|}\sum_{k}^{N}\psi_{k}\mathcal{O}^{ij,kk} |i\rangle_{A}|j\rangle_{B}\ ,
}
and the righthand side of \eqref{GeneGateTel} is given by
\ba{
\mathcal{O}_{AB\rightarrow A}(\psi) \otimes I_{B} |\psi\rangle_{AB}
=\sum_{i}^{|A|}\sum_{j}^{N}\psi_{j}\overline{\mathcal{O}}^{i,j}|i\rangle_{A}|j\rangle_{B}\ .
}
By comparing the both sides, we obtain the following conditions
\ba{\label{shrinkingcd1}
&\sum_{k}^{N}\psi_{k}\mathcal{O}^{ij,kk} =\psi_{j}\overline{\mathcal{O}}^{i,j}\ \ \text{for}\ 1\le j\le N\ , \\\label{shrinkingcd2}
&\sum_{k}^{N}\psi_{k}\mathcal{O}^{ij,kk}=0\ \ \ \ \ \ \  \  \text{for}\ N< j\le |B|\ ,
}
for all possible $i$'s.
Since the condition \eqref{shrinkingcd1} gives $\overline{\mathcal{O}}^{i,j}=\sum_{k}^{N}\psi_{j}^{-1}\mathcal{O}^{ij,kk}\psi_{k}$ for $1\le j\le N$, $\mathcal{O}_{AB\rightarrow A}(\psi)$ can be written as
\ba{\label{expansionOfShrunkOp}
\mathcal{O}_{AB\rightarrow A}(\psi)=\sum_{i=1}^{|A|}\sum_{j,k=1}^{N}\psi^{-1}_{j}\mathcal{O}^{ij,kk}\psi_{k}| i\rangle_{A\hspace{0.5mm}A}\langle j|
%+\sum_{i=1}^{|A|}\sum_{j>N}^{|B|}\psi^{-1}_{j}\mathcal{O}^{ij,kk}\psi_{k}| i\rangle_{A\hspace{0.5mm}A}\langle j|\ ,
+\sum_{i=1}^{|A|}\sum_{j>N}^{|A|}\overline{\mathcal{O}}^{i,j}| i\rangle_{A\hspace{0.5mm}A}\langle j|\ ,
}
and the second term is an operator in $\mathcal{V}(|\psi\rangle_{AB})$.
The condition \eqref{shrinkingcd2} is exactly the condition for $\mathcal{O}_{AB}$  to be included in $\mathcal{S}_{\mathcal{L}}(A\hspace{0.5mm}; |\psi\rangle_{AB})$.
Thus we conclude $(A)\Rightarrow(B)$.
\vspace{3mm}

\noindent$(B)\Rightarrow(A)$: 
Since $\overline{\mathcal{O}}_{AB}(\psi) $ in \eqref{operatorABpsiA} vanishes when it acts on $|\psi\rangle_{AB}$ defined in \eqref{state}, $\mathcal{O}_{AB}\in\mathcal{S}_{\mathcal{L}}(A\hspace{0.5mm}; |\psi\rangle_{AB})$ acts on $|\psi\rangle_{AB}$ as
\ba{\label{opst}
\mathcal{O}_{AB}|\psi\rangle_{AB}
&=\sum_{i}^{|A|}\sum_{j,k}^{N}\mathcal{O}^{ij,kk}\psi_{k}|i\rangle_{A}|j\rangle_{B}
\nonumber\\
&=\sum_{j}^{N}\psi_{j}\sum_{i,k}^{N}\psi_{j}^{-1}\mathcal{O}^{ij,kk}\psi_{k}|i\rangle_{A}|j\rangle_{B}
}
where we have inserted $\psi_{j}\psi_{j}^{-1}(=1)$ in the second equality.
It then follows from the definition \eqref{shrunkopABtoA} and eq.\eqref{opst} that
\ba{
\mathcal{O}_{AB}\hspace{0.5mm}|\psi\rangle_{AB}=\mathcal{O}_{AB\rightarrow A}(\psi) \otimes I_{B} \hspace{0.5mm}|\psi\rangle_{AB}\ ,
}
since $\overline{\mathcal{O}}_{A}(\psi)$ in \eqref{shrunkopABtoA} vanishes when it acts on $|\psi\rangle_{AB}$.
\end{proof}
%%%%%%%%%%%%%%%%%%%%%%%%%%%%%%%%%%%%
In this way, any operator $\mathcal{O}_{AB}$ in $\mathcal{S}_{\mathcal{L}}(A\hspace{0.5mm}; |\psi\rangle_{AB})$ can shrink its support on $|\psi\rangle_{AB}$ such that it acts only on $A$ nontrivially while keeping its mapping.
In this sense, we can redefine $\mathcal{S}_{\mathcal{L}}(A\hspace{0.5mm}; |\psi\rangle_{AB})$ as
\ba{
 \mathcal{S}_{\mathcal{L}}&(A; |\psi\rangle_{AB})\\
\equiv &\{\ \mathcal{O}_{AB}\in \mathcal{L}(\mathcal{H}_{AB})\ |\  
  ^{\exists}\mathcal{O}_{A B\rightarrow A}(\psi)\in \mathcal{L}(\mathcal{H}_{A})\  s.t. \ \mathcal{O}_{A B}\hspace{0.5mm}|\psi\rangle_{A B}=\mathcal{O}_{A B\rightarrow A}(\psi)\otimes I_{ B}\hspace{0.5mm}|\psi\rangle_{A B}\ \}\nonumber 
}
We will call $\mathcal{O}_{AB\rightarrow A}(\psi)$ ``shrunk operator'' for $\mathcal{O}_{AB}$ on $|\psi\rangle_{AB}$.
One may regard Lemma \ref{generalgatetel} as the finite dimensional version of the Reeh-Schlieder theorem\cite{article} (see also \cite{Witten:2018lha} for a review)\footnote{The relation to the Reeh-Schlieder theorem will be more clearer in Corollary \ref{maximalgatetel}.}.

It should be emphasized that the shrunk operator $\mathcal{O}_{AB\rightarrow A}(\psi)$ is ``state-dependent" in the sense that the operator depends on the state $|\psi\rangle$ through the matrix elements of the operator and also through the choice of the basis in \eqref{GeneGateTel}. 
Note that since the system $A$ and the $B$ are equally treated at the level of the Hilbert space $\mathcal{H}_{AB}$ and the state $|\psi\rangle_{AB}$ in this Lemma, this Lemma can apply to the case where operators shrink their supports from $AB$ to $B$ just by exchanging the indices associated with $A$ for those associated with $B$. %in the statement in Lemma \ref{generalgatetel}. 
Therefore, the set of operators that can shrink their supports to $A$ and also to $B$ can be defined as follows.

%%%%%%%%% Definition%%%%%%%%%%%%%%
\noindent\rule{\columnwidth}{1pt}\vspace{-2mm}
%\begin{screen}
\begin{corollary} \label{shrinktoAandB}
Any operator $\mathcal{O}_{AB} \in \mathcal{S}_{\mathcal{L}}(A\hspace{0.5mm}; |\psi\rangle_{AB}) \cap  \mathcal{S}_{\mathcal{L}}(B\hspace{0.5mm}; |\psi\rangle_{AB})$ can be written as
\ba{
\mathcal{O}_{AB} 
=\sum_{i,j,k=1}^{N}\mathcal{O}^{ij,kk}\ |i\rangle_{A}|j\rangle_{B\hspace{0.5mm}A}\langle k|_{B}\langle k|
+\overline{\mathcal{O}}_{AB}(\psi)
}
with $ \mathcal{O}^{ij,kk} \in \mathbb{C}$ and $\overline{\mathcal{O}}_{AB}(\psi) \in \mathcal{V}(\hspace{0.5mm}|\psi\rangle_{AB}\hspace{0.5mm})$.
Then the operators $\mathcal{O}_{AB\rightarrow A}\in \mathcal{L}(\mathcal{H}_{A})$ and $\mathcal{O}_{AB\rightarrow B}\in \mathcal{L}(\mathcal{H}_{B})$ that satisfy
\ba{
\mathcal{O}_{AB} |\psi\rangle_{AB}=\mathcal{O}_{AB\rightarrow A}\otimes I_{B} |\psi\rangle_{AB} = I_{A}\otimes\mathcal{O}_{AB\rightarrow B} |\psi\rangle_{AB}\ ,
}
can be written as
\ba{
&\mathcal{O}_{AB\rightarrow A}(\psi)=\sum_{i,j,k=1}^{N}\psi^{-1}_{j}\mathcal{O}^{ij,kk}\psi_{k}| i\rangle_{A\hspace{0.5mm}A}\langle j|
+\sum_{i=1}^{|A|}\sum_{j>N}^{|A|}\overline{\mathcal{O}}_{A}^{i,j}| i\rangle_{A\hspace{0.5mm}A}\langle j|\ , \\
&\mathcal{O}_{AB\rightarrow B}(\psi)=\sum_{i,j,k=1}^{N}\psi^{-1}_{j}\mathcal{O}^{ji,kk}\psi_{k}| i\rangle_{B\hspace{0.5mm}B}\langle j|
+\sum_{i=1}^{|B|}\sum_{j>N}^{|B|}\overline{\mathcal{O}}_{B}^{i,j}| i\rangle_{A\hspace{0.5mm}A}\langle j|\ ,
}
with arbitrary complex numbers, $\overline{\mathcal{O}}_{A}^{i,j}\in\mathbb{C}$ and $\overline{\mathcal{O}}_{B}^{i,j}\in\mathbb{C}$.
\end{corollary}
\vspace{-5mm}
\noindent\rule{\columnwidth}{1pt}
%%%%%%%%%%%%%%%%%%%%%%%%%%

Moreover, considering an extreme case where $N=min\{|A|, |B|\}$, $i.e.$, $|\psi\rangle_{AB}$ has the maximal Schmidt rank, in Lemma \ref{generalgatetel}, we obtain a corollary below.
\ \\
\rule{\columnwidth}{1pt}\vspace{-2mm}
%%%%%%% corollary %%%%%%%
%\begin{screen}
\begin{corollary}\label{maximalgatetel}%\ \\ 
\ Let $|\psi\rangle_{AB}$ be a state with Schmidt rank $|B|$ in $\mathcal{H}_{AB}$ with $|B|\le |A|$ that can be written in the Schmidt basis as
\ba{
 |\psi\rangle_{AB}=\sum_{i=1}^{|B|}\psi_{i}|i\rangle_{A}|i\rangle_{B},
 }
 with positive real numbers, $\psi_{i}$.
 Then for any operator $\mathcal{O}_{AB}$ in $\mathcal{L}(\mathcal{H}_{AB})$, there exists $\mathcal{O}_{AB\rightarrow A}(\psi)\in \mathcal{L}(\mathcal{H}_{A})$ such that
 \ba{
\mathcal{O}_{AB}\hspace{0.5mm}|\psi\rangle_{AB}=\mathcal{O}_{AB\rightarrow A}(\psi) \otimes I_{B} \hspace{0.5mm}|\psi\rangle_{AB}\ .
}
Such $\mathcal{O}_{AB\rightarrow A}(\psi)$ is calculated by eq.\eqref{shrunkopABtoA} with $N=|B|$.
\end{corollary}\vspace{-5mm}
%\end{screen}
\noindent\rule{\columnwidth}{1pt}

\vspace{5mm}
This corollary tells us a fact that even though the state is not a maximally entangled state\footnote{This condition means that $|\psi\rangle_{AB}$ is separating for the algebra of linear operators acting on $\mathcal{H}_{B}$. %See  for the relation between this condition and Reeh-Schlieder theorem\cite{article}. 
This condition is known as one of the crucial conditions for Reeh-Schlieder theorem\cite{article, Witten:2018lha}.
}, 
any operator acting on the state can shrink the support to the larger subsystem as long as the state has small entanglement spread throughout all the Schmidt basis on the smaller system
\footnote{If we consider an infinite-dimensional bipartite systems in which both $|A|$ and $|B|$ are infinite, it would be hard to compare $|A|$ and $|B|$. If we naively assume that $|A|$ is equal to $|B|$ in this case, then it follows from Corollary \ref{maximalgatetel} that any operator can shrink its support to $A$ if the state is separating for the algebra of linear operators acting on $B$. In this sense, we expect that the counterpart of Corollary \ref{maximalgatetel} in such infinite-dimensional systems is the Reeh-Schlieder theorem. 
} .

However the essence of Lemma \ref{generalgatetel} and its corollaries is the well-known fact that operators can shrink their supports to either subsystem on a maximally entanglement state. 
For example, defining a maximally entanglement state as
\ba{
|\Psi\rangle_{AB}\equiv\sum_{i}^{|B|}|i\rangle_{A}|i\rangle_{B}\ , %\frac{1}{\sqrt{|B|}}
}
we can rewrite $|\psi\rangle_{AB}$ in \eqref{state} as
\ba{
|\psi\rangle_{AB}=(\rho_{B})^{1/2}|\Psi\rangle_{AB}\ .
}
Moreover we define
\ba{
\widetilde{\mathcal{O}}_{AB}\equiv (\rho_{B})^{-1/2}\hspace{0.5mm}\mathcal{O}_{AB}\hspace{0.5mm}(\rho_{B})^{1/2}
}
with
\ba{
(\rho_{B})^{-1/2}\equiv\sum_{i}^{N}\psi^{-2}|i\rangle_{B\hspace{0.5mm} B}\langle i|\ .
}
Then $\mathcal{O}_{AB}\in \mathcal{S}_{\mathcal{L}}(A\hspace{0.5mm}; |\psi\rangle_{AB})$ acts on $|\psi\rangle_{AB}$ as
\ba{\label{ShrinkonMstate}
\mathcal{O}_{AB}|\psi\rangle_{AB}
=\mathcal{O}_{AB}\hspace{0.5mm}(\rho_{B})^{1/2}|\Psi\rangle_{AB}
%=(\rho_{B})^{1/2}(\rho_{B})^{-1/2}\hspace{0.5mm}\mathcal{O}_{AB}\hspace{0.5mm}(\rho_{B})^{1/2}|\Psi\rangle_{AB}
=(\rho_{B})^{1/2}\widetilde{\mathcal{O}}_{AB}|\Psi\rangle_{AB}
} %P_{\text{Im}\rho_{B}}
where we have used $I_{B}-P_{\text{ker}\rho_{B}}= \sum_{i}^{N}|i\rangle_{B\hspace{0.5mm} B}\langle i|=(\rho_{B})^{1/2}(\rho_{B})^{-1/2}$ acts on $\mathcal{O}_{AB}$ trivially, namely $(I_{B}-P_{\text{ker}\rho_{B}})\mathcal{O}_{AB}=\mathcal{O}_{AB}$, in the second equality.
Using the shrunk operator $\widetilde{\mathcal{O}}_{AB\rightarrow A}(\Psi)$ for $\widetilde{\mathcal{O}}_{AB}$ on $|\Psi\rangle_{AB}$, which satisfies
\ba{
\widetilde{\mathcal{O}}_{AB}|\Psi\rangle_{AB}=\widetilde{\mathcal{O}}_{AB\rightarrow A}(\Psi)\otimes I_{B}\hspace{0.5mm}|\Psi\rangle_{AB}\ ,
}
we can rewrite eq.\eqref{ShrinkonMstate} as
\ba{
\mathcal{O}_{AB}|\psi\rangle_{AB}=\widetilde{\mathcal{O}}_{AB\rightarrow A}(\Psi)\otimes I_{B}|\psi\rangle_{AB}\ .
}
One can easily check that $\widetilde{\mathcal{O}}_{AB\rightarrow A}(\Psi)$ is actually identical to $\mathcal{O}_{AB\rightarrow A}(\psi)$ in \eqref{shrunkopABtoA}.

Before proceeding, let us see how the shrinking works in a simple example.
%%%%%%%%   Example  %%%%%%%%%%%%%%%%%%%%
\begin{example}
\normalfont
Consider a state in $\mathcal{H}_{AB}=(\mathbb{C}^{2})^{\otimes 2}$ given by
\ba{
\ket{\psi}\equiv\sqrt{1-\epn}\ket{00}+\sqrt{\epn}\ket{11}\equiv \psi_{0}\ket{00}+\psi_1\ket{11}\ ,
}
and an operator,
\ba{
\mathcal{O}_{AB}&\equiv X_{A}Y_{B}=(\ket{1}\bra{0}+\ket{0}\bra{1})\otimes(-i\ket{1}\bra{0}+i\ket{0}\bra{1})\nonumber\\
&=-i\ket{11}\bra{00}+i\ket{10}\bra{01}-i\ket{01}\bra{10}+i\ket{00}\bra{11}\nonumber\\
&\equiv \sum_{i,j,k,l=0,1}\mathcal{O}^{ij,kl}\ket{ij}\bra{kl}
}
Then the operator $\mathcal{O}_{AB\rightarrow A}$ which satisfies
\ba{
\mathcal{O}_{AB}\ket{\psi}=\mathcal{O}_{AB\rightarrow A}\otimes I_{B} \ket{\psi}\label{shrinkeq}
}
can be computed by eq.(2.9) as
\ba{
\mathcal{O}_{AB\rightarrow A}\equiv \sum_{i,j=0,1}\mathcal{O}^{i,j}\ket{i}\bra{j}\ \ 
\text{,}\ \  \mathcal{O}^{i,j}\equiv \sum_{k=0,1}\psi_{j}^{-1}\mathcal{O}^{ij,kk}\psi^{k}.
}
Nonzero matrix components are given by
\ba{
\mathcal{O}^{0,0}=\sum_{k=0,1}\psi_{0}^{-1}\mathcal{O}^{00,kk}\psi^{k}
%=\psi_{0}^{-1}\mathcal{O}^{00,11}\psi^{1}
=i\sqrt{\frac{\epn}{1-\epn}}\ \  ,\ \ 
\mathcal{O}^{1,1}=\sum_{k=0,1}\psi_{1}^{-1}\mathcal{O}^{11,kk}\psi^{k}
%=\psi_{1}^{-1}\mathcal{O}^{11,00}\psi^{0}
=-i\sqrt{\frac{1-\epn}{\epn}}}
Then we obtain
\ba{
\mathcal{O}_{AB\rightarrow A}=i\sqrt{\frac{\epn}{1-\epn}}\ket{0}\bra{0}-i\sqrt{\frac{1-\epn}{\epn}}\ket{1}\bra{1}
}
This actually satisfies eq.\eqref{shrinkeq}.  \vspace{-6mm}

\rightline{$\square$}
%\rightline{$\square$}
%\vspace{-10mm}\begin{flushright}$\square$\end{flushright}
\end{example}
%%%%%%%%   End Example  %%%%%%%%%%%%%%%%%%%%
Note that even though $XY$ is a Hermitian operator and also a unitary operator, the shrunk operator $\mathcal{O}_{AB\rightarrow A}$ is neither of them except for when $\epn=1/2$ in this example
\footnote{
For $\mathcal{O}_{AB}=XX$, the shrunk operator is $\mathcal{O}_{AB\rightarrow A}=\sqrt{\frac{\epn}{1-\epn}}\ket{0}\bra{0}+\sqrt{\frac{1-\epn}{\epn}}\ket{1}\bra{1}$. This is a case where the Hermiticity is preserved but the unitarity is not. For $\mathcal{O}_{AB}=ZZ$, the shrunk operator is $\mathcal{O}_{AB\rightarrow A}=I_{A}$. This is the case where both properties are preserved.
}.
Thus, the unitarity and the Hermiticity of the operators are not generally preserved through the shrinking in Lemma \ref{generalgatetel}.
We will study the conditions for preserving the unitarity or the Hermiticity through the shrinking in the next subsection.

%%%%%%%%%%%%%%%%%%%%%%%%%%%%%%%%%%%%%%%%%%%%%%%%%%%%%%%%%%%
%%%%%%%%%%%%%%%%%%%%%%%%%%%%%%%%%%%%%%%%%%%%%%%%%%%%%%%%%%%
\subsection{The conditions for preserving unitarity or Hermiticity through the shrinking of operators }
The shrinking will be valid in physics only when either unitarity or Hermiticity is preserved.
One can show the conditions for preserving the unitarity or the Hermiticity with straightforward calculations by imposing unitarity or Hermiticity on $\mathcal{O}_{AB\rightarrow A}$ in Lemma \ref{generalgatetel}.
The results are summarized as the following Lemmas:

%%%%%%%%%  Lemma %%%%%%%%%%%%%%
\noindent\rule{\columnwidth}{1pt}\vspace{-2mm}
%\begin{screen}
\begin{lemma} \label{preservingUnitarity}%\ \\
\ %When $\mathcal{O}_{AB}\in \mathcal{S}_{\mathcal{L}}(A\hspace{0.5mm}; |\psi\rangle_{AB})$ is a unitary operator, 
For $\mathcal{O}_{AB}\in \mathcal{S}_{\mathcal{L}}(A\hspace{0.5mm}; |\psi\rangle_{AB})$, $\mathcal{O}_{AB\rightarrow A}$ defined in \eqref{expansionOfShrunkOp} %\eqref{shrunkopABtoA} 
also becomes a unitary operator on $\mathcal{H}_A$ if  $\mathcal{O}_{AB}$ and $ \overline{\mathcal{O}}_{A}(\psi)$  satisfy the following conditions:
%\begin{enumerate}
%\item 
%\normalfont
\ba{
&\text{{\rm (i)}}\ \ \sum_{j}^{|A|}\sum_{n,m}^{N}\psi_{i}^{-1}\psi_{k}^{-1}\mathcal{O}^{ji,nn\ast}\mathcal{O}^{jk,mm}\psi_{n}\hspace{0.5mm}\psi_{m}=\delta_{i,k} \ ,\hspace{3mm} 
%\text{for\ \ $i , k=1,\cdots,N$}\\
\text{for\ \  $1\le i \le N$\ , $1\le k \le N$} \\
&\text{{\rm (ii)}}\ \ \sum_{j}^{|A|}\sum_{n}^{N}\mathcal{O}^{ji,nn\ast}\overline{\mathcal{O}}^{j,k}\psi_{n}=0\ ,\hspace{3mm}  
%\text{for\ \ $i=1,\cdots,N$\ ,\ $k=N+1,\cdots,|A|$}\\
\text{for\ \  $1\le i \le N$\ , $N< k \le |A|$}\\
&\text{{\rm (iii)}}\ \ \sum_{j}^{|A|}\overline{\mathcal{O}}^{j,i\ast}\overline{\mathcal{O}}^{j,k}=\delta_{i,k}\ ,
\hspace{3mm} \text{for\ \  $N< i \le |A|$\ , $N< k \le |A|$}
}
%\end{enumerate}
\end{lemma}
\vspace{-5mm}
\noindent\rule{\columnwidth}{1pt}
%%%%%%%%%%%%%%%%%%%%%%%%%%

%%%%%%%%%  Lemma %%%%%%%%%%%%%%
\noindent\rule{\columnwidth}{1pt}\vspace{-2mm}
%\begin{screen}
\begin{lemma} \label{preservingHermiticity}%\ \\
\ For $\mathcal{O}_{AB}\in \mathcal{S}_{\mathcal{L}}(A\hspace{0.5mm}; |\psi\rangle_{AB})$, $\mathcal{O}_{AB\rightarrow A}$ defined in \eqref{expansionOfShrunkOp} %\eqref{shrunkopABtoA} 
also becomes a Hermitian operator on $\mathcal{H}_A$ if  $\mathcal{O}_{AB}$ and $ \overline{\mathcal{O}}_{A}(\psi)$ satisfy the following conditions,
%\begin{enumerate}
%\item 
%\normalfont
\ba{
&\text{{\rm (i)}}\ \ \sum_{j}^{N}\psi_{i}^{-1}\mathcal{O}^{ki,jj\ast}\psi_{j}
=\sum_{j}^{N}\psi_{k}^{-1}\mathcal{O}^{ik,jj}\psi_{j}
\ ,\hspace{3mm} \text{for\ \  $1\le i \le N$\ , $1\le k \le N$}, \label{Hermiticitycond1}\\
&\text{{\rm (ii)}}\ \ \sum_{j}^{N}\psi_{i}^{-1}\mathcal{O}^{ki,jj\ast}\psi_{j}=\overline{\mathcal{O}}^{i,k} ,\hspace{3mm} \text{for\ \  $1\le i \le N$\ , $N< k \le |A|$}, \\
&\text{{\rm (iii)}}\ \ \overline{\mathcal{O}}^{k,i\ast}=\overline{\mathcal{O}}^{i,k}\ ,
\hspace{3mm} \text{for\ \  $N< i \le |A|$\ , $N< k \le |A|$}.\label{Hermiticitycond3}
}
%\end{enumerate}
\end{lemma}
\vspace{-5mm}
\noindent\rule{\columnwidth}{1pt}
%%%%%%%%%%%%%%%%%%%%%%%%%%

Lemma \ref{preservingUnitarity} and Lemma \ref{preservingHermiticity} show the conditions for the unitarity and the Hermiticity of the shrunk operator $\mathcal{O}_{AB\rightarrow A}$, respectively. 
These constraints look strong, but we will see that the shrinking that preserves the unitarity is realized in the code subspaces of  quantum error-correcting codes against erasure errors in section \ref{Systematicconstruction} and the shrinking that preserves the Hermiticity is also realized on the code subspace of the AdS/CFT-like subalgbra code in section \ref{adscft}.
Here, let us give easy examples of the shrinking that preserves the Hermiticity.
%%%%%%%%   Example  %%%%%%%%%%%%%%%%%%%%
\begin{example} \normalfont
If the matrix elements of $\mathcal{O}_{AB}$ take the following form,
\ba{
\mathcal{O}^{ijkl}=a\delta^{i,j}\delta^{k,l}\hspace{0.5mm}\hspace{0.5mm}\psi_{i}\hspace{0.5mm}\psi_{k}\ \  \text{for $1\le i, j, k, l \le N$}\ \ ,\ \ \text{others}=0\ ,
}
with a real number $a$, one can check that these actually satisfy the conditions \eqref{Hermiticitycond1}-\eqref{Hermiticitycond3}.
In this case $\mathcal{O}_{AB}$ can be written as 
\ba{
\mathcal{O}_{AB}=\sum_{i,k}^{N}a\hspace{0.5mm}\psi_{i}\hspace{0.5mm}\psi_{k}\ 
|i\rangle_{A}|i\rangle_{B\hspace{0.5mm}A}\langle k|_{B}\langle k|\ ,
}
and $\mathcal{O}_{AB\rightarrow A}$ can be written as 
\ba{
\mathcal{O}_{AB\rightarrow A}=a\hspace{0.5mm} I_{A} + \overline{\mathcal{O}}_{A}(\psi)
}
where $\overline{\mathcal{O}}_{A}(\psi)\otimes I_{B} \in \mathcal{V}(|\psi\rangle_{AB})$.
\vspace{-5.3mm}

\rightline{$\square$}
\end{example}
%%%%%%%%   Example  %%%%%%%%%%%%%%%%%%%%

$\mathcal{O}_{AB}$ and $\mathcal{O}_{AB\rightarrow A}$ in the above example are a kind of trivial solution since they act on $\ket{\psi}_{AB}$ as just multiplying a real number $a$.
However we can make this example a bit more non-trivial by introducing other states as follows.
%%%%%%%%   Example  %%%%%%%%%%%%%%%%%%%%
\begin{example} \label{essenceofShrinkHermitian}
\normalfont
Consider $\mathcal{H}_{A}$ and  $\mathcal{H}_{B}$ spanned by the orthonormal bases $\{ |b\hspace{0.5mm};k\rangle_{A}\}^{b=1,\cdots, N_{A}}_{k=1,\cdots, N}$ and $\{ |\overline{b}\hspace{0.5mm};k\rangle_{B}\}^{\overline{b}=1,\cdots, N_{B}}_{k=1,\cdots, N}$, respectively.
Assume that these are orthonormal in the sense that
\ba{
& _{A}\langle a\hspace{0.5mm};i| b;j\rangle_{A}
=\delta_{ a,  b}\delta_{i,j}\ \ ,\ \  _{B}\langle \overline{a}\hspace{0.5mm};i| \overline{b}\hspace{0.2mm};j\rangle_{B}
=\delta_{ \overline{a},  \overline{b}\hspace{0.2mm}}\delta_{i,j}\ .
}
Then we consider the states in $\mathcal{H}_{A}\otimes \mathcal{H}_{B}$ whose Schmidt decomposition is written as
\ba{\label{ew2.eq2}
| b\hspace{0.3mm},\hspace{0.3mm}\overline{b}\rangle_{AB}
\equiv\sum_{k}^{N}\psi_{k}|b\hspace{0.5mm};k\rangle_{A}|\overline{b}\hspace{0.5mm};k\rangle_{B}\ ,
}
and a Hermitian operator $\mathcal{O}_{AB}$ that can be written as
\ba{\label{ew2.eq3}
\mathcal{O}_{AB}
&=\sum_{a\hspace{0.2mm},\hspace{0.2mm}b}^{N_{A}}\sum_{\overline{b}}^{N_{B}}
\mathcal{O}^{a,b}
| a\hspace{0.3mm},\hspace{0.3mm}\overline{b}\rangle_{AB\hspace{0.7mm}AB}\langle b\hspace{0.3mm},\hspace{0.3mm}\overline{b}|\nonumber\\
&=\sum_{a\hspace{0.2mm},\hspace{0.2mm}b}^{N_{A}}\sum_{\overline{b}}^{N_{B}}\sum_{i,k}^{N}
\psi_{i}
\mathcal{O}^{a,b}\psi_{k}
|a\hspace{0.5mm};i\rangle_{A\hspace{0.4mm}} |\overline{b}\hspace{0.5mm};i\rangle_{B\hspace{0.7mm}A}\langle b\hspace{0.5mm};k|_{\hspace{0.6mm}B}\langle \overline{b}\hspace{0.5mm};k|\ ,
}
where the matrix elements satisfy $\mathcal{O}^{a,b\ast}=\mathcal{O}^{b,a}$. 
The shrunk operator $\mathcal{O}_{AB\rightarrow A}$ that satisfies the conditions \eqref{Hermiticitycond1}-\eqref{Hermiticitycond3} with $\mathcal{O}_{AB}$ and $| b\hspace{0.3mm},\hspace{0.3mm}\overline{b}\rangle_{AB}$ for all possible $b$'s and $\overline{b}$'s is given by 
\ba{\label{ew2.eq4}
\mathcal{O}_{AB\rightarrow A}
=\sum_{a\hspace{0.2mm},\hspace{0.2mm}b}^{N_{A}}\sum_{a\hspace{0.2mm}}^{N_{A}}
\mathcal{O}^{a,b}|a\hspace{0.5mm};i\rangle_{A\hspace{0.4mm}A}\langle b\hspace{0.5mm};i|\ .
}
This $\mathcal{O}_{AB\rightarrow A}$ is obviously a Hermitian operator on $A$. 
We can easily see  that  $\mathcal{O}_{AB\rightarrow A}$ is the shrunk operator for $\mathcal{O}_{AB}$ on $| b\hspace{0.3mm},\hspace{0.3mm}\overline{b}\rangle_{AB}$ for all possible $b$'s and $\overline{b}$'s in the following manner,
\ba{
\mathcal{O}_{AB}| b\hspace{0.3mm},\hspace{0.3mm}\overline{b}\rangle_{AB}
&=\sum_{a\hspace{0.2mm}}^{N_{A}}\mathcal{O}^{a,b}| a\hspace{0.3mm},\hspace{0.3mm}\overline{b}\rangle_{AB}
=\sum_{a\hspace{0.2mm}}^{N_{A}}\sum_{i}^{N}
\psi_{i}\mathcal{O}^{a,b}|a\hspace{0.5mm};i\rangle_{A\hspace{0.4mm}} |\overline{b}\hspace{0.5mm};i\rangle_{B}\nonumber\\
&=\mathcal{O}_{AB\rightarrow A}\otimes I_{B}\hspace{0.5mm}| b\hspace{0.3mm},\hspace{0.3mm}\overline{b}\rangle_{AB}\ 
}
where we have just used the definitions \eqref{ew2.eq2}-\eqref{ew2.eq4}.
Thus $\mathcal{O}_{AB\rightarrow A}$ is a Hermitian shrunk operator for $\mathcal{O}_{AB}$ on the subspace in  $\mathcal{H}_{A}\otimes \mathcal{H}_{B}$ spanned by $\{| b\hspace{0.3mm},\hspace{0.3mm}\overline{b}\rangle_{AB}\}^{b=1,\cdots, N_{A}}_{\overline{b}=1,\cdots,N_{B}}$.
\vspace{-5.3mm}

\rightline{$\square$}
\end{example}
%%%%%%%%   Example  %%%%%%%%%%%%%%%%%%%%

This is an example of the shrinking that preserves the Hermiticity which holds not only on a state in $\mathcal{H}_{A}\otimes \mathcal{H}_{B}$ but also on the nontrivial subspace in $\mathcal{H}_{A}\otimes \mathcal{H}_{B}$.
This example captures how the shrinking that preserves the Hermiticity is realized on the code subspace of the AdS/CFT-like subalgebra code in section \ref{adscft}.

%We end this section by showing 
The conditions for $\mathcal{O}_{AB}$ to shrink the supports to $A$ and also on $B$ while preserving the Hermiticity follow immediately from Corollary \ref{shrinktoAandB} and Lemma \ref{preservingHermiticity}:

%\newpage
%%%%%%%%%  corollary %%%%%%%%%%%%%%
\noindent\rule{\columnwidth}{1pt}\vspace{-2mm}
%\begin{screen}
\begin{corollary} \label{shrinkonA&onBpreservingHermiticity}%\ \\
\ For $\mathcal{O}_{AB}\in \mathcal{S}_{\mathcal{L}}(A\hspace{0.5mm}; |\psi\rangle_{AB})\hspace{0.5mm}\cap\hspace{0.5mm}\mathcal{S}_{\mathcal{L}}(B\hspace{0.5mm}; |\psi\rangle_{AB})$,  $^{\exists}\mathcal{O}_{AB\rightarrow A}\in  \mathcal{A}(\mathcal{H}_{A})$ and 

\noindent $^{\exists}\mathcal{O}_{AB\rightarrow B}\in  \mathcal{A}(\mathcal{H}_{B})$ s.t.
\ba{
\mathcal{O}_{AB} |\psi\rangle_{AB}=\mathcal{O}_{AB\rightarrow A}\otimes I_{B} |\psi\rangle_{AB} = I_{A}\otimes\mathcal{O}_{AB\rightarrow B} |\psi\rangle_{AB}\ ,
}
if $\mathcal{O}_{AB}$, $\mathcal{O}_{AB\rightarrow A}$ and $\mathcal{O}_{AB\rightarrow B}$ defined in Corollary \ref{shrinktoAandB} satisfy the following conditions,
%\begin{enumerate}
%\item 
%\normalfont
\ba{
&\text{{\rm (i)}}\ \ \sum_{j}^{N}\psi_{i}^{-1}\mathcal{O}^{ki,jj\ast}\psi_{j}
=\sum_{j}^{N}\psi_{k}^{-1}\mathcal{O}^{ik,jj}\psi_{j}
\ ,\hspace{3mm} \text{for\ \  $1\le i \le N$\ , $1\le k \le N$}. \label{Hermiticitycond1onAB}\\[5pt]
&\text{{\rm (ii)}}\ \ \overline{\mathcal{O}}^{i,k}_{A}=0 ,\hspace{3mm} \text{for\ \  $1\le i \le N$\ , $N< k \le |A|$}\ ,\ \\[3pt]
&\hspace{8.2mm} \overline{\mathcal{O}}^{i,k}_{B}=0 ,\hspace{3mm} \text{for\ \  $1\le i \le N$\ , $N< k \le |B|$}\ .\\[9pt]
&\text{{\rm (iii)}}\ \ \overline{\mathcal{O}}^{k,i\ast}=\overline{\mathcal{O}}^{i,k}\ ,
\hspace{3mm} \text{for\ \  $N< i \le |A|$\ , $N< k \le |A|$}\ .\label{Hermiticitycond3onAB}
}
%\end{enumerate}
\end{corollary}
\vspace{-5mm}
\noindent\rule{\columnwidth}{1pt}
%%%%%%%%%%%%%%%%%%%%%%%%%%

So far we have studied the conditions for the shrinking of operators for general operators $\mathcal{O}_{AB}$ acting on $\mathcal{H}_{A}\otimes\mathcal{H}_{B}$.
We can also study the conditions for ``teleporting" of the support from $B$ to $A$ just by restricting $\mathcal{O}_{AB}$ to the local operator on $B$, $i.e.$, the conditions for the existence of $\mathcal{O}_{B\rightarrow A}$ for $\mathcal{O}_{B}$ such that
\ba{
I_{A}\otimes\mathcal{O}_{B}|\psi\rangle_{AB}=\mathcal{O}_{B\rightarrow A}\otimes I_{B} |\psi\rangle_{AB}\ .
}
The results are summarized in Appendix \ref{sec:Teleporting}.
As an application, the teleporting on the thermofield double state is calculated in Appendix \ref{sec:TFD}.

%%%%%%%%%%%%%%%%%%%%%%%%%%%%%%%%%%%%%%%%%%%%%%%%%%%%%%%%%%%%%%%%%%%%%%%%%%%%%%%%%%%%%%%%%%%%%%%%%%%%%%%%%
%%%%%%%%%%%%%%%%%%%%%%%%%%%%%%%%%%%%%%%%%%%%%%%%%%%%%%%%%%%%%%%%%%%%%%%%%%%%%%%%%%%%%%%%%%%%%%%%%%%%%%%%%
%%%%%%%%%%%%%%%%%%%%%%%%%%%%%%%%%%%%%%%%%%%%%%%%%%%%%%%
\section{Systematic construction of decoders in quantum error-correcting codes against erasure errors}\label{Systematicconstruction}
In this section we start with reviewing briefly the ``quantum secret sharing schemes" as a typical example of  quantum error-correcting codes that are able to correct erasure errors, especially by focusing on the encoding and the decoding procedures.
Secondly, we show a useful expression of a logical basis in the quantum error-correcting codes against erasure errors in subsection \ref{The Schmidt decomposition of logical bases}. 
Finally, we show how to systematically construct a decoder in subsection \ref{A formula for the construction of decoders}.

%%%%%%%%%%%%%%%%%%%%%%%%%%%%%%%%%%%%%%%%%%%%%%%%%%%%%%%%%%%%%%%%%%%%%%%%%%%%%%%%%%%%%%%%%%%%%%%%%%%%%%%%%%%%%%
\subsection{Review of encoding and decoding procedures in quantum secret sharing schemes}\label{introQTS}
%$((K, 2K-1))$ threshold scheme as most efficient QSS 
Suppose we have a \emph{secret} quantum state $|\psi\rangle_{S}$ which we want to protect against general errors caused by its environment or robbers. 
Following the idea of quantum error-correcting codes, we encode the secret quantum information into larger degrees of freedom, which are divided into $n$ shares. 
Then a $((k, n))$ \emph{threshold} scheme, with $k\leq n$, is defined as a quantum error-correcting code that has the following property: any $k$ shares are sufficient to reconstruct the secret quantum information perfectly, but from any $k-1$ shares, no information  about the secret can be extracted. 
We will focus on the $((k, 2k-1))$ threshold scheme because of the following two reasons shown in \cite{Cleve:1999qg}.
First, the $((k, n))$ threshold scheme with $n\geq 2k$ does not exist due to the ``non-cloning theorem'' \cite{Wootters:1982zz,Dieks:1982dj} because one can copy an unknown quantum state by using the scheme. Second, the $((k, n))$ threshold schemes with $n<2k-1$ can be obtained from the $((k, 2k-1))$ threshold schemes just by discarding $2k-n-1$ shares.

In order to explain the encoding and decoding processes in the quantum secret sharing schemes briefly and to introduce some notations, let us consider a simple $((k, 2k-1))$ threshold scheme in which the secret quantum state and all shares are \emph{qudits}.
This code encodes single secret qudit (a Hilbert space of dimension $d$) into $2k-1$ qudits (a Hilbert space of dimension $d^{2k-1}$) to protect the secret state against erasures of any $k-1$ or less qudits.  The encoder of this code is given by a unitary operator that maps the basis of single qudit, $\{ \hspace{0.5mm}|i\rangle\}_{i=1,\cdots ,d}$, into the \textit{logical} basis, $\{ \hspace{0.5mm}|\bar{i}\rangle\}_{i=1,\cdots ,d}$
\footnote{Precisely speaking, when we encode, we need to prepare an irrelevant additional state to make the encoder unitary transformation. For example, we can use $|1\rangle^{\otimes 2k-2}$ as such an additional state, then the encoder is defined as 
$U_{enc} |i\rangle_{S}|1\rangle^{\otimes 2k-2}= |\bar{i}\rangle.$
We write this encoding as eq.\eqref{logicalencode}.
},%footnote ends
\ba{\label{logicalencode}
 U_{enc} : |i\rangle_{S}\ \rightarrow\ |\bar{i}\rangle\ .
}
Each $ |\bar{i}\rangle$ is composed of $2k-1$ qudits. The Hilbert space spanned by the logical basis is called \textit{code subspace} $\mathcal{H}_{code}$. Then an arbitrary single secret qudit $|\psi\rangle_{S}= \sum_{i=1}^{d}\psi_{i}|i\rangle_{S}$ is mapped as
\ba{
 U_{enc} : \ |\psi\rangle_{S}\ \rightarrow\ |\overline{\psi}\rangle=\sum_{i=1}^{d}\psi_{i}|\bar{i}\rangle\ .
}
Suppose any $k-1$ qudits in $|\overline{\psi}\rangle$ are erased, and write the Hilbert space of those erased qudits as $\mathcal{H}_{E}$ and that of remaining unerased qudits as $\mathcal{H}_{\overline{E}}$. We pick a single qudit from $k$ unerased qudits, and decompose the Hilbert space of $\overline{E}$ as $\mathcal{H}_{\overline{E}}=\mathcal{H}_{\overline{E}_{R}}\otimes\mathcal{H}_{\overline{E}_{S}}$ where  $\mathcal{H}_{\overline{E}_{S}}$ denotes the Hilbert space of the picked single qudit and $\mathcal{H}_{\overline{E}_{R}}$ is the Hilbert space of remaining $k-1$ qudits in $\overline{E}$. Then the ability to protect the erasure error is ensured by the existence of a unitary operator $U_{\overline{E}}$ on $\mathcal{H}_{\overline{E}}$ such that 
\ba{
I_{E}\otimes U_{\overline{E}}|\bar{i}\rangle=|\chi\rangle_{E\overline{E}_{R}}|i\rangle_{\overline{E}_{S}}\ \ \text{for}\  i=1,\cdots ,d\label{defdecoder}
}
where $|\chi\rangle_{E\overline{E}_{R}}$ is a state on $\mathcal{H}_{E}\otimes\mathcal{H}_{\overline{E}_{R}}$, which does not depend on index $i$. If such a unitary operator exists, we can recover the encoded state $|\psi\rangle_{S}$ on $\overline{E}_{S}$ without touching the erased qudits by implementing the unitary transformation as
\ba{
I_{E}\otimes U_{\overline{E}}|\overline{\psi}\rangle=|\chi\rangle_{E\overline{E}_{R}}|\psi\rangle_{\overline{E}_{S}}\ .
}
One can easily see how these procedures actually work in one of the simplest examples, the three-qutrit code of \cite{Cleve:1999qg}.

Thus a central problem of the theoretical construction of quantum secret sharing schemes is to find the pairs of the logical basis $\{ \hspace{0.5mm}|\bar{i}\rangle\}_{i=1,\cdots ,d}$ and the \textit{decoder} $U_{\overline{E}}$ which satisfies the condition \eqref{defdecoder}.

\subsection{The Schmidt decomposition of logical bases}\label{The Schmidt decomposition of logical bases}
%%%%%%%%%%%%%  lemma   %%%%%%%%%%%%%
%\begin{screen}
\rule{\columnwidth}{1pt}\vspace{-2mm}
\begin{lemma}\label{logicalbasis}%\ \\
%\noindent\rule{\columnwidth}{1pt}\vspace{-2mm}
\ Let $\mathcal{H}_{S}$ and $\mathcal{H}$ be  finite-dimensional Hilbert spaces, with a tensor product structure $\mathcal{H}=\mathcal{H}_{E}\otimes\mathcal{H}_{\overline{E}}$ with $|\overline{E}|=|S|\times |E|$. Moreover let $|\phi\rangle$ be a state defined as
\ba{
|\phi\rangle\equiv\frac{1}{\sqrt{|S|}}\sum_{i=1}^{|S|}|i\rangle_{S}|\overline{i}\rangle}
where $\{ \hspace{0.5mm}|i\rangle_{S}\}_{i=1,\cdots, |S|}$ is an orthonormal basis on $\mathcal{H}_{S}$ ,and $\{|\overline{i}\rangle\}_{i=1,\cdots, |S|}$ is a set of orthonormal states on $\mathcal{H}$. If the state $|\phi\rangle$ satisfies
\ba{
\mathop{\mathrm{Tr}}_{\overline{E}}|\phi\rangle\langle\phi|=\frac{1}{|S|}I_{S}\otimes\frac{1}{|E|}I_{E},\label{noinf}
}
then the state $|\overline{i}\rangle$ can be expanded as
\ba{\label{expansionoflogical}
|\overline{i}\rangle=\frac{1}{\sqrt{|E|}}\sum_{k=1}^{|E|}|k\rangle_{E}|k;\overline{i}\rangle_{\overline{E}},\ \  \  i=1,\cdots, |S| 
}
where $\{ \hspace{0.5mm}|k\rangle_{E}\}_{k=1,\cdots, |E|}$ is an arbitrary orthonormal basis on $\mathcal{H}_{E}$, and $\{ \hspace{0.5mm}|k;\overline{i}\rangle_{\overline{E}}\}_{k=1,\cdots, |E|}^{i=1,\cdots, |S|}$ is  an orthonormal basis on $\mathcal{H}_{\overline{E}}$ such that
\ba{
 _{E}\langle k|l\rangle_{E}=\delta_{k,l}\ ,\  _{\overline{E}}\langle k;\overline{i}|l;\overline{j}\rangle_{\overline{E}}=\delta_{k,l}\delta_{i,j}.\label{orthogonality}
}
\end{lemma}\vspace{-3mm}
%\noindent\rule{\columnwidth}{1pt}
%\end{screen}
\vspace{-2mm}
\noindent\rule{\columnwidth}{1pt}
%%%%%%%%%%%%%%%%%%%%%%%%%%%%%
%\vspace{3mm}

This Lemma itself is a little abstract but it has the following physical meanings from the viewpoint of the quantum error-correcting codes. 
As in the example in subsection \ref{introQTS}, $S$ denotes the system of the secret state. $\{|\overline{i}\rangle\}_{i=1,\cdots, |S|}$ is a logical basis on the code subspace that is a subspace in $\mathcal{H}$, $i.e.$, the encoder is defined as
\ba{
U_{enc} : |i\rangle_{S}\ \rightarrow\ |\bar{i}\rangle.
}
$E$  and $\overline{E}$ are the degrees of freedom(d.o.f) which are supposed to be erased and not to be erased respectively when the maximum error occurs within the range that can be decoded. Then, imposing the condition $|\overline{E}|=|S|\times |E|$ corresponds to considering a $((k, 2k-1))$ threshold scheme because the condition says that the number of d.o.f (for instance qudits) of $\overline{E}$ is larger by just the number of the d.o.f of $S$ than that of $E$. 
In other words, when the maximum error occurs, we need to collect just one more d.o.f that has the same number of d.o.f of the secret state than the erased d.o.f in order to decode successfully.  For example, if we  set $|S|=d$ and $|E|=d^{K-1}$, then the condition says  $|\overline{E}|=d^{K}$, which is the same situation as the example in subsection \ref{introQTS}. This means we need just one more qudit than erased $k-1$ qudits in order to decode. 
On the other hand, the condition \eqref{noinf} says that the erased d.o.f on $E$ carry no information about the secret state, in other words the remaining d.o.f on $\overline{E}$ contain all information about the secret
\footnote{It would be possible to generalize the condition \ref{noinf} to $\mathop{\mathrm{Tr}}_{\overline{E}}|\phi\rangle\langle\phi|=\rho_{S}\otimes\rho_{E}$ as in \cite{Harlow:2016vwg} if we add additional d.o.f that are irrelevant to encoding and decoding procedures. But we do not consider such cases for simplicity. } .
The condition is needed to ensure the existence of a decoder \cite{Schumacher:1996dy,Grassl:1996eh,Harlow:2016vwg}.

\begin{proof}\ \\
 Each $|\overline{i}\rangle$ can be expressed in the Schmidt basis as
 \ba{\label{logicalschmidt}
|\bar{i}\rangle=\sum_{k=1}^{|E|}\sqrt{p_{k}^{(i)}}|k';\overline{i}\rangle_{E}|k';\overline{i}\rangle_{\overline{E}}
}
where $\{ \hspace{0.5mm}|k';\overline{i}\rangle_{E}\}_{k=1,\cdots, |E|}$  is an orthonormal basis on $\mathcal{H}_{E}$ and $\{ \hspace{0.5mm}|k';\overline{i}\rangle_{\overline{E}}\}_{k=1,\cdots, |E|}$ is a set of orthonormal  states on $\mathcal{H}_{\overline{E}}$, such that
\ba{
_{E} \langle k';\overline{i}|l';\overline{i}\rangle_{E}=\  _{\overline{E}}\langle k';\overline{i}|l';\overline{i}\rangle_{\overline{E}}=\delta_{k',l'}\ ,
}
and $p_{k}^{(i)}$'s are real numbers satisfying $\sum_{k=1}^{|E|}p_{k}^{(i)}=1$ for each $i=1,\cdots, d$. 
Substituting eq.\eqref{logicalschmidt} into the condition \eqref{noinf}, one can easily find
\ba{\label{erasedoneoninf}
%&\mathop{\mathrm{Tr}}_{\overline{E}}|\phi\rangle\langle\phi|=
\frac{1}{|S|}\sum_{i,j}^{|S|}|i\rangle_{S\hspace{0.5mm}S}\langle j|\otimes \sum_{k,l}^{|E|}\sqrt{p_{k}^{(i)}p_{l}^{(j)}}\ _{\overline{E}}\langle l;\overline{j}|k;\overline{i}\rangle_{\overline{E}}|k;\overline{i}\rangle_{E\hspace{0.5mm}E}\langle l;\overline{j}|
=\frac{1}{|S|}I_{S}\otimes\frac{1}{|E|}I_{E}\ .
}
By multiplying the both sides of the above equation by $ _{S}\langle i|$ from the lefthand side  and $|i\rangle_{S}$ from the righthand side, the Schmidt coefficients are determined as $p_{k}^{(i)}=\frac{1}{|E|}$.
Then the Schmidt basis for $|\overline{i}\rangle$ is not uniquely determined since all the Schmidt coefficients are degenerate. In other words, we are free to apply the following simultaneous unitary transformations in $\mathcal{H}_{E}$ and $\mathcal{H}_{\overline{E}}$,
\ba{
|k';\overline{i}\rangle_{E}=\sum_{l}^{|E|}|l;\overline{i}\rangle_{E} U_{lk}^{(\overline{i})}\ ,\ 
|k';\overline{i}\rangle_{\overline{E}}=\sum_{l}^{|E|}|l;\overline{i}\rangle_{\overline{E}} U_{lk}^{(\overline{i})\ast}
}
where $U_{lk}^{(\overline{i})}$ is an arbitrary $|E|\times|E|$ unitary matrix. Using the above transformation, we can use the Schmidt basis such that the $\{ \hspace{0.5mm}|l;\overline{i}\rangle_{E}\}_{l,\cdots, |E|}$ is independent of the index $i$, such  as
\ba{\label{chosenbasisonE}
|l;\overline{i}\rangle_{E}=|l\rangle_{E}
}
where $\{ \hspace{0.5mm}|l\rangle_{E}\}_{k=1,\cdots, |E|}$ is an arbitrary complete orthogonal basis on $\mathcal{H}_{E}$. Then the logical basis can be expressed as
\ba{
|\bar{i}\rangle=\frac{1}{\sqrt{|E|}}\sum_{k=1}^{|E|}|k\rangle_{E}|k;\overline{i}\rangle_{\overline{E}}\label{isch}
}
Finally, substituting eq.\eqref{chosenbasisonE} into the condition \eqref{erasedoneoninf}, we get
\ba{
\sum_{i,j}^{|S|} \sum_{k,l}^{|E|}\ _{\overline{E}}\langle l;\overline{j}|k;\overline{i}\rangle_{\overline{E}}\frac{1}{|S|}|i\rangle_{R\hspace{0.5mm}R}\langle j|\otimes\frac{1}{|E|}|k\rangle_{E\hspace{0.5mm}E}\langle l| 
=\frac{1}{|S|}I_{R}\otimes\frac{1}{|E|}I_{E} \label{orthobarE}\ .
}
By multiplying the both sides of \eqref{orthobarE} by $ _{R}\langle i|_{E}\langle k|$ from the lefthand side, and $|j\rangle_{R}|l\rangle_{E}$ from the righthand side, the condition, $ _{\overline{E}}\langle l;\overline{j}|k;\overline{i}\rangle_{\overline{E}}=\delta_{k,l}\delta_{i,j}$, is obtained. 
Then $\{ \hspace{0.5mm}|k;\overline{i}\rangle_{\overline{E}}\}_{k=1,\cdots, |E|}^{i=1,\cdots, |S|}$ is an orthonormal basis on $\mathcal{H}_{\overline{E}}$ since it  contains $|\overline{E}|(=|S|\times |E|)$ orthogonal states.
\end{proof}

\subsection{A formula for the construction of decoders}\label{A formula for the construction of decoders}
%%%%%%%%theorem%%%%%%%%%
\rule{\columnwidth}{1pt}\vspace{-2mm}
%\begin{screen}
\begin{theorem} \label{makedecoder}%\ \\
\ Let $U_{E\overline{E}}$ be a unitary operator on $\mathcal{H}$, with tensor product structures $\mathcal{H}=\mathcal{H}_{E}\otimes\mathcal{H}_{\overline{E}}$ and $\mathcal{H}_{\overline{E}}=\mathcal{H}_{\overline{E}_{R}}\otimes\mathcal{H}_{\overline{E}_{S}}$, such that for $i=1,\cdots, |S|$,
\ba{\label{trivialdecoder}
 U_{E\overline{E}}|\bar{i}\rangle=|\chi\rangle_{E\overline{E}_{R}}|i\rangle_{\overline{E}_{S}}
}
where $|\bar{i}\rangle$ is the state defined in \eqref{expansionoflogical} and $|\chi\rangle_{E\overline{E}_{R}}$ is a state on $\mathcal{H}_{E\overline{E}_{R}}$ such that
\ba{
\mathop{\mathrm{Tr}}_{\overline{E}_{R}}|\chi\rangle_{E\overline{E}_{R}\hspace{0.5mm}E\overline{E}_{R}}\langle \chi|=\frac{1}{|E|}I_{E}\label{noinfR}\ .
}
Then a unitary operator (decoder) on $\mathcal{H}_{\overline{E}}$ that satisfies
\ba{
I_{E}\otimes U^{dec}_{\overline{E}}|\bar{i}\rangle=|\chi\rangle_{E\overline{E}_{R}}|i\rangle_{\overline{E}_{S}}
}
for $i=1,\cdots, |S|$ can be written  as
\ba{
%U^{dec}_{\overline{E}}=\sum_{i,j=1}^{|S|}\sum_{k,l=1}^{|E|}|k;i\rangle_{\overline{E}\hspace{0.5mm} E}\langle l|_{\bar{E}}\langle k;i|U_{E\overline{E}}|\bar{j}\rangle_{\hspace{0.8mm}\overline{E}}\langle l;j| \label{constructionofdec}
U^{dec}_{\overline{E}}=\sum_{j=1}^{|S|}\sum_{l=1}^{|E|}\ _{E}\langle l|U_{E\overline{E}}|\bar{j}\rangle_{\hspace{0.8mm}\overline{E}}\langle l;j| \label{constructionofdec}
}
where $\{ \hspace{0.5mm}|k\rangle_{E}\}_{k=1,\cdots, |E|}$ and $\{ \hspace{0.5mm}|k;\overline{i}\rangle_{\overline{E}}\}_{k=1,\cdots, |E|}^{i=1,\cdots, |S|}$ are the bases on $\mathcal{H}_{E}$ and $\mathcal{H}_{\overline{E}}$ respectively, which are the Schmidt bases of $|\bar{i}\rangle$ defined in Lemma \ref{logicalbasis}.
\end{theorem}
%\end{screen}
\vspace{-5mm}
\noindent\rule{\columnwidth}{1pt}
%\ \\
%%%%%%%%%%%%%%

The operator $U_{E\overline{E}}$ is a \emph{trivial} decoder in the sense that it decodes the code by acting on all the physical degrees of freedom on the logical basis as \eqref{trivialdecoder}. We cannot implement the trivial decoder after a single error has occurred. On the other hand, $U^{dec}_{\overline{E}}$  can recover the encoded state on $\overline{E}_{S}$ by acting only on $\overline{E}$.  In other words, $U^{dec}_{\overline{E}}$ is a decoder available even after  any error has occurred on $E$. The formula  \eqref{constructionofdec} provides a method of constructing such decoders from trivial decoders
\footnote{How to shrink supports of general linear operators that act within a code subspace was given in \cite{Almheiri:2014lwa}. Thr result \eqref{constructionofdec} is not the same as the formula in \cite{Almheiri:2014lwa} since the trivial decoder maps states in a code subspace into states outside the code subspace.
}. 
The condition \eqref{noinfR} is the statement that the state on $E\overline{E}_{R}$ after decoding on $\overline{E}_{S}$ has no information about the encoded state.  This condition is necessary to guarantee the unitarity of the decoder, otherwise the decoding violates the non-cloning theorem.

 \begin{proof}\ \\
 Using the bases $\{ \hspace{0.5mm}|k\rangle_{E}\}_{k=1,\cdots, |E|}$ on $\mathcal{H}_{E}$ and $\{ \hspace{0.5mm}|k;\overline{i}\rangle_{\overline{E}}\}_{k=1,\cdots, |E|}^{i=1,\cdots, |S|}$ on $\mathcal{H}_{\overline{E}}$, we can expand a trivial decoder $U_{E\overline{E}}$ as
 \ba{\label{expansionoftrivialdec}
 U_{E\overline{E}}=\sum_{i,j}^{|S|}\sum_{k,l,m,n}^{|E|} U_{k,(l:i),m,(n;j)}|k\rangle_{E}|l;i\rangle_{\overline{E}\hspace{0.5mm}E}\langle m|_{\overline{E}}\langle n;j|
 }
 where $U_{k,(l:i),m,(n;j)}$ is defined by the above equation as the matrix elements of a given trivial decoder.
It follows from Lemma \ref{generalgatetel} that the operator $U_{E\overline{E}\rightarrow \overline{E}}(\bar{j})$ in $\mathcal{L}(\mathcal{H}_{\overline{E}})$ that satisfies
\ba{\label{actiondec}
U_{E\overline{E}}|\bar{j}\rangle_{E\overline{E}}=I_{E}\otimes U_{E\overline{E}\rightarrow \overline{E}}(\bar{j})|\bar{j}\rangle_{E\overline{E}}\ ,
}
can be written as
\ba{
U_{E\overline{E}\rightarrow \overline{E}}(\bar{j})&\equiv \sum_{i}^{|S|}\sum_{k,l,m}^{|E|}U_{k,(l;i),m,(m;j)}|l;i\rangle_{\overline{E}\hspace{0.5mm}\overline{E}}\langle k;j|\\
%&=\sum_{i}^{|S|}\sum_{k,l}^{|E|}|l;i\rangle_{\overline{E}\hspace{0.5mm} E}\langle k|_{\bar{E}}\langle l;i|U_{E\overline{E}}|\bar{j}\rangle_{\hspace{0.5mm}\overline{E}}\langle k;j|\ ,
&=\sum_{k}^{|E|}\ _{E}\langle k|U_{E\overline{E}}|\bar{j}\rangle_{\hspace{0.5mm}\overline{E}}\langle k;j|\ ,
}
where we have dropped the irrelevant operator that vanishes when it acts on $|\bar{j}\rangle_{E\overline{E}}$.
Now we define
\ba{\label{expansiondec}
U^{dec}_{\overline{E}}\equiv\sum_{j}^{|E|} U_{E\overline{E}\rightarrow \overline{E}}(\bar{j})
=\sum_{i,j}^{|S|}\sum_{k,l,m}^{|E|}U_{k,(l;i),m,(m;j)}|l;i\rangle_{\overline{E}\hspace{0.5mm}\overline{E}}\langle k;j|\ .
}
The orthogonality condition \eqref{orthogonality} ensures that $U^{dec}_{\overline{E}}$ has the same action as $U_{E\overline{E}}$ on code space, $i.e.$
\ba{\label{dec}
I_{E}\otimes U^{dec}_{\overline{E}}\sum_{i}^{|S|}\psi_{i}|\bar{i}\rangle=\sum_{i}^{|S|}\psi_{i}\ U_{E\overline{E}\rightarrow \overline{E}}(\bar{i})|\bar{i}\rangle
=U_{E\overline{E}}\sum_{i}^{|S|}\psi_{i}|\bar{i}\rangle\ .
}
Using the properties \eqref{trivialdecoder}, \eqref{actiondec} and \eqref{dec}, we can see that $U^{dec}_{\overline{E}}$ is actually a decoder:
\ba{\label{decoder}
I_{E}\otimes U^{dec}_{\overline{E}}\sum_{i}^{|S|}\psi_{i}|\bar{i}\rangle=\sum_{i}^{|S|}\psi_{i}|\chi\rangle_{E\overline{E}_{R}}|i\rangle_{\overline{E}_{S}}=|\chi\rangle_{E\overline{E}_{R}}|\psi\rangle_{\overline{E}_{S}}\ .
}
In order to implement this decoding procedure in the real world, $U_{E\overline{E}\rightarrow \overline{E}}$ needs to be a unitary operator acting on $\mathcal{H}_{\overline{E}}$. 
To see the unitarity, we examine the constraint on $U_{E\overline{E}}$ from the condition \eqref{noinfR} as follows. First, we multiply $\delta_{j,j'}$ to the both sides of eq.\eqref{noinfR} and rewrite it with $U_{E\overline{E}}$ in the following manner,
\ba{\label{contrainttotrivdec}
\frac{\delta_{j,j'}}{|E|}I_{E}&=\delta_{j,j'}\mathop{\mathrm{Tr}}_{\overline{E}_{R}}|\chi\rangle_{E\overline{E}_{R}\hspace{0.5mm}E\overline{E}_{R}}\langle \chi|
=\mathop{\mathrm{Tr}}_{\overline{E}_{R}\overline{E}_{S}}|\chi\rangle_{E\overline{E}_{R}}|j\rangle_{\overline{E}_{S}\hspace{0.7mm}E\overline{E}_{R}}\langle \chi|_{\hspace{0.5mm}\overline{E}_{S}}\hspace{-0.5mm}\langle j'|\nonumber\\
&=\mathop{\mathrm{Tr}}_{\overline{E}}U_{E\overline{E}}|\bar{j}\rangle\langle \bar{j'}|U_{E\overline{E}}^{\dagger}
}
where we have used $\mathop{\mathrm{Tr}}_{\overline{E}_{S}}|j\rangle_{S\hspace{0.5mm}S}\langle j'|=\delta_{j,j'}$ in the second equality and eq.\eqref{trivialdecoder} in the third equality. Secondly, we substitute the expression \eqref{expansionoftrivialdec} into the righthand side of eq.\eqref{contrainttotrivdec} and we get
\ba{
\frac{\delta_{j,j'}}{|E|}I_{E}&=\mathop{\mathrm{Tr}}_{\overline{E}}\sum_{i,i'}^{|S|}\ \sum_{k,k',l,l',m,m'}^{|E|}U_{k,(l;i),m,(m;j)}|k\rangle_{E}|l;i\rangle_{\overline{E}\hspace{0.5mm}E}\langle k'|_{\overline{E}}\langle l';i'|U^{\ast}_{k',(l';i'),m',(m';j')}\nonumber\\
&=\sum_{i}^{|S|}\sum_{k,k',l,m,m'}^{|E|}U_{k,(l;i),m,(m;j)}U^{\ast}_{k',(l;i),m',(m';j')}|k\rangle_{E\hspace{0.5mm}E}\langle k'|\label{constraintonUEbarE}
}
where we have just performed the trace over $\overline{E}$ in the second equality. Since $\{ \hspace{0.5mm}|k\rangle_{E}\}_{k=1,\cdots, |E|}$ is the basis on $\mathcal{H}_{E}$, it follows from \eqref{constraintonUEbarE} that 
\ba{
\sum_{i}^{|S|}\sum_{l,m,m'}^{|E|}U_{k,(l;i),m,(m;j)}U^{\ast}_{k',(l;i),m',(m';j')}=\frac{1}{|E|}\delta_{k,k'}\delta_{j,j'}.\label{costraintonU}
}
Finally, we are able to check the unitarity of $U^{dec}_{\overline{E}}$ in the following manner,
\ba{
 U^{dec\dagger}_{\overline{E}} U^{dec}_{\overline{E}}\nonumber
%=&\sum_{i',j'=1}^{|S|}\sum_{k',l',m'=1}^{|E|}U^{\ast}_{l',(k';i'),m',(m';j')}|l';j'\rangle_{\overline{E}\hspace{0.5mm}\overline{E}}\langle k';i'|\sum_{i,j=1}^{|S|}\sum_{k,l,m=1}^{|E|}U_{l,(k;i),m,(m;j)}|k;i\rangle_{\overline{E}\hspace{0.5mm}\overline{E}}\langle l;j|\nonumber\\
=&\sum_{i,j,j'}^{|S|}\ \sum_{k,l,l',m,m'}^{|E|} U^{\ast}_{l',(k;i),m',(m';j')}U_{l,(k;i),m,(m;j)}|l';j'\rangle_{\overline{E}\hspace{0.5mm}\overline{E}}\langle l;j|\nonumber\\
=&\sum_{j,j'}^{|S|}\sum_{l,l'}^{|E|}\frac{1}{|E|}\delta_{l',l}\delta_{j',j}|l';j'\rangle_{\overline{E}\hspace{0.5mm}\overline{E}}\langle l;j|=I_{E}
}
where we have used eq.\eqref{expansiondec} in the first equality and the condition \eqref{costraintonU} in the third equality.
\end{proof}

This theorem gives a reinterpretation of code subspaces against erasure errors as follows.
As it was already noted that in the equation,
\ba{
\mathcal{O}_{AB}|\psi\rangle_{AB}=\mathcal{O}_{AB\rightarrow A}(\psi) \otimes I_{B} |\psi\rangle_{AB}\ ,
}
the shrunk operator $\mathcal{O}_{AB\rightarrow A}(\psi)$  generally depends on the state $|\psi\rangle_{AB}$. 
In other words, for a general state $|\phi\rangle_{AB}$ in $\mathcal{H}_{AB}$,
\ba{
\mathcal{O}_{AB}|\phi\rangle_{AB}\neq\mathcal{O}_{AB\rightarrow A}(\psi) \otimes I_{B} |\phi\rangle_{AB}\ .
}
However there is still a possibility of the existence of the non-trivial subspaces in $\mathcal{H}_{AB}$  on which $\mathcal{O}_{AB\rightarrow A}(\psi)$ acts  in the same way as $\mathcal{O}_{AB}$ does.
Moreover it is possible to consider the subspaces on which both the action of $\mathcal{O}_{AB}$  and the unitarity of $\mathcal{O}_{AB}$ are preserved through the shrinking of $\mathcal{O}_{AB}$ to $\mathcal{O}_{AB\rightarrow A}(\psi)$.
Such a subspace associated with $U_{E\overline{E}}$ and $U_{\overline{E}}^{dec}$ becomes a code subspace against erasure errors.

%%%%%%%%%%%%%%%%%%%%%%%%%%%%%%%%%%%%%%%%%%%%%%%%%%%%%%
%%%%%%%%%%%%%%%%%%%%%%%%%%%%%%%%%%%%%%%%%%%%%%%%%%%%%%

\section{Implications for the operator dictionary in AdS/CFT}\label{adscft}

In  \cite{Harlow:2016vwg}, a type of quantum error correcting-codes called the \textit{subalgebra code with complementary recovery} was introduced.
This code embeds the ``bulk Hilbert space" $\mathcal{H}_{a}\otimes \mathcal{H}_{\overline{a}}$ into the ``boundary Hilbert space" $\mathcal{H}_{A}\otimes \mathcal{H}_{\overline{A}}$ such that the operators on $a$ can be reconstructed on $A$ and the operators on $\overline{a}$ can be reconstructed on $\overline{A}$.
This property is called \textit{complementary recovery}. 
It was shown that the Ryu-Takayanagi-like formula and also the equivalence of the bulk and boundary relative entropies exactly hold for the subalgebra code with complementary recovery.
It was also shown that in general there exist the logical operators that are reconstructable on $A$ and also on $\overline{A}$. 
Such operators must be the central operators of the reconstructed algebras on $A$ and on $\overline{A}$.
These results imply the validity of the entanglement wedge reconstruction and  the existence of the bulk operators which are reconstructable on a boundary subregion and also on the complementary of the subregion in AdS/CFT correspondence.
An concrete example of the subalgebra code with complementary recovery with nontrivial centers was proposed in \cite{Donnelly:2016qqt}.

In this section, we study the subalgebra code with complementary recovery under the restrictions of the bulk (logical) Hilbert space into a subspace.
We first explain our model in subsection \ref{sec:ourmodel}.
Secondly, we review the basic operator dictionary and also how the Ryu-Takayanagi-like formula holds in a concrete way in subsection \ref{sec:review}.
Then we will discuss the emergence of the central operators from the restriction of   bulk Hilbert space in subsection \ref{sec:emergenceofcenter}.
Finally, we prove a theorem for the code, which implies the validity of the entanglement wedge reconstruction and also its converse statement with nontrivial centers in subsection \ref{sec:EWRwithcenters}.

%%%%%%%%%%%%%%%%  subsection  %%%%%%%%%%%%%%%%%%%%%%%%%%
%%%%%%%%%%%%%%%%%%%%%%%%%%%%%%%%%%%%%%%%%%%%%%%%%%%%%%%%%%%%%%%%%%%%%%%%%%%%%%%%%%%%%%%%%%%%%%%%%%%%%%%%%%%%%%%%%%
\subsection{Subalgebra code with complementary recovery under restriction of logical Hilbert space to subspace}\label{sec:ourmodel}

First, let us define the logical space as
\ba{
\mathcal{H}_{bulk}^{tot}\equiv \mathcal{H}_{bulk}^{}\otimes \mathcal{H}_{bulk}^{ (gr)}\ .
}
where $\mathcal{H}_{bulk}^{(gr)}$ and $\mathcal{H}_{bulk}^{}$ are finite dimensional Hilbert spaces that  imitate the Hilbert space of gravity and the Hilbert space of all other d.o.f than gravity, respectively.
We will assume that the bulk Hilbert space has the following tensor factorization,
\ba{
&\mathcal{H}_{bulk}^{}\equiv \mathcal{H}_{\mathcal{E}_{A}}^{}\otimes \mathcal{H}_{\mathcal{E}_{\overline{A}}}^{}\ ,\\[7pt]
&\mathcal{H}_{bulk}^{(gr)}\equiv \mathcal{H}_{\mathcal{E}_{A}}^{(gr)}\otimes \mathcal{H}_{\mathcal{E}_{\overline{A}}}^{(gr)}\ .
}
where $\mathcal{E}_{A}$ (or $\mathcal{E}_{\overline{A}}$) corresponds to  the entanglement wedge of boundary subregion $A$ (or its complementary subregion $\overline{A}$) on a bulk time-slice.
$\mathcal{H}_{\mathcal{E}_{A/\overline{A}}}^{(gr)}$ and  $\mathcal{H}_{\mathcal{E}_{A/\overline{A}}}^{}$ correspond to the Hilbert space of gravity on $\mathcal{E}_{A/\overline{A}}$ and the Hilbert space of all the other d.o.f on $\mathcal{E}_{A/\overline{A}}$, respectively.
Defining 
\ba{
&\mathcal{H}_{\mathcal{E}_{A}}^{tot}\equiv \mathcal{H}_{\mathcal{E}_{A}}^{}\otimes \mathcal{H}_{\mathcal{E}_{A}}^{(gr)}\ , \\[7pt]
&\mathcal{H}_{\mathcal{E}_{\overline{A}}}^{tot}\equiv \mathcal{H}_{\mathcal{E}_{\overline{A}}}^{}\otimes \mathcal{H}_{\mathcal{E}_{\overline{A}}}^{(gr)}\ ,
}
we can write $\mathcal{H}_{bulk}^{}$  as 
\ba{
\mathcal{H}_{bulk}^{tot}\equiv\mathcal{H}_{\mathcal{E}_{A}}^{tot}\otimes\mathcal{H}_{\mathcal{E}_{\overline{A}}}^{tot}\ .
}
The embedding space is a finite dimensional Hilbert space with the following tensor decomposition,
\ba{
\mathcal{H}_{bdry}^{}\equiv\mathcal{H}_{A}^{}\otimes\mathcal{H}_{\overline{A}}^{}.
}
$\mathcal{H}_{A}^{}$ and  $\mathcal{H}_{\overline{A}}^{}$ play the role of the Hilbert space associated with a boundary subregion $A$ and its complement region $\overline{A}$ in AdS/CFT, respectively.

Our encoder $U_{code}: \mathcal{H}_{bulk}^{tot} \mapsto \mathcal{H}_{bdry}^{}$ takes the following form,
\ba{\label{Ucode}
U_{code}\equiv U_{A}U_{\overline{A}}
}
with the two isometry maps
\footnote{
The ``isometry maps" means the following conditions,
\ba{\label{isometrycondcode}
%U_{\gamma_{A}_{A\overline{A}}}^{\dagger}U_{\gamma_{A}_{A\overline{A}}}=I_{\gamma_{A}}\ \  ,\ \ 
 U_{A}^{\dagger}U_{A}=I_{\mathcal{E}_{A}}\ \ ,\ \ U_{\overline{A}}^{\dagger}U_{\overline{A}}=I_{\mathcal{E}_{\overline{A}}}
}
}:
\ba{
%&U_{\gamma_{A}_{A\overline{A}}}: \mathcal{H}_{\gamma_{A}} \mapsto \mathcal{H}_{\gamma_{A}}\otimes \mathcal{H}_{\gamma_{A}_{\overline{A}}}\ , \\[5pt]
&U_{A}\ : \mathcal{H}_{\mathcal{E}_{A}}^{tot} \mapsto \mathcal{H}_{A}^{}\ , \\[5pt]
&U_{\overline{A}}\ :  \mathcal{H}_{\mathcal{E}_{\overline{A}}}^{tot}\mapsto \mathcal{H}_{\overline{A}}^{}\ .
}
The encoder $U_{code}$ plays the role of the duality map in AdS/CFT.
When we input a state $\ket{\Psi}_{bulk}^{(gr)} \in \mathcal{H}_{bulk}^{(gr)}$ into $U_{code}$, we obtain
\ba{\label{UcodeonPsi}
U_{code}^{(\Psi)}\equiv U_{code}\ket{\Psi}_{bulk}^{(gr)}=U_{A}U_{\overline{A}}\ket{\Psi}_{bulk}^{(gr)}.
}
This map, $U_{code}^{(\Psi)}:\mathcal{H}_{bulk} \mapsto \mathcal{H}_{code}$, is an isometry map and it takes the form of the encoder of the subalgebra code with complementary recovery in \cite{Harlow:2016vwg}.

If we assume that the AdS/CFT correspondence is a exact duality in which the bulk and the boundary operators have an exact one-to-one correspondence, the representation of dual boundary operator on full boundary Hilbert space should be uniquely fixed.
%The non-uniqueness of the representation (and the supports) of the boundary operators should breakdown full Hilbert space.
In other words, the non-uniqueness of the representation (and the supports) of the boundary operators is valid only on the (code) subspace that is dual to a subspace of the full bulk Hilbert space.
The bulk Hilbert space is also constrained by the gauge constraints in AdS/CFT.
In order to imitate these restrictions on the bulk Hilbert space, \vspace{1mm} we restrict the logical space to a subspace in $\mathcal{H}_{bulk}$, which we will call $\widetilde{\mathcal{H}}_{bulk} (\subset \mathcal{H}_{bulk})$.
Note that $\widetilde{\mathcal{H}}_{bulk}$ generally does not \vspace{1mm}factorize into the Hilbert spaces on $\mathcal{E}_{A}$ and on  $\mathcal{E}_{\overline{A}}$, $i.e.$,
for any $ \widetilde{\mathcal{H}}_{\mathcal{E}_{A}}\subset \mathcal{H}_{\mathcal{E}_{A}}$ and $\widetilde{\mathcal{H}}_{\mathcal{E}_{\overline{A}}}\subset \mathcal{H}_{\mathcal{E}_{\overline{A}}}$, 
\ba{\label{notfactrize}
\widetilde{\mathcal{H}}_{bulk}^{} \neq \widetilde{\mathcal{H}}_{\mathcal{E}_{A}}^{}\otimes\widetilde{\mathcal{H}}_{\mathcal{E}_{\overline{A}}}^{}\ ,
} 
which reflects the universal entanglement in low energy states.
Let us define the code subspace as
\ba{
\widetilde{\mathcal{H}}_{code}^{(\Psi)}\equiv U_{code}^{(\Psi)}\widetilde{\mathcal{H}}_{bulk}^{}\ ,
}
which plays the role of the boundary Hilbert space dual to the bulk semiclassical Hilbert space with the fixed background geometry.
We will see that the central operators of the reconstructed algebra on $A$ and on $\overline{A}$ can emerge from the restriction from $\mathcal{H}_{bulk}$ to $\widetilde{\mathcal{H}}_{bulk}$.

It was shown in \cite{Harlow:2016vwg} that the bulk reconstruction with complementary recovery, the Ryu-Takayanagi-like formula and also the equivalence of the bulk and boundary relative entropies are exactly hold for this type of codes.
Before proceeding, let us review the basic operator dictionary  and also how the Ryu-Takayanagi-like formula holds in this code in a  concrete way.

%%%%%%%%%%%%%%%%%%%%%%%%%%%%%%%%%%%%%%%%%%%%%%%%%%%%%%%%%%%%%%%%%%%
\subsection{Review of the operator dictionary and  Ryu-Takayanagi formula in subalgebra code with complementary recovery}\label{sec:review}
%the complementary recovery code
In this subsection,  based on \cite{Pastawski:2015qua, Harlow:2016vwg, Donnelly:2016qqt}, we review the basic rule about the operator dictionary and also how the Ryu-Takayanagi-like formula holds in this code in a concrete way.

First,\vspace{1mm} we prepare orthonormal bases, $\{|b\hspace{0.5mm}\rangle_{\mathcal{E}_{A}}\}_{b=1,\cdots,|\mathcal{E}_{A}^{}|}$ on $\mathcal{H}_{\mathcal{E}_{A}}^{}$ 
and $\{|\overline{b}\hspace{0.5mm}\rangle_{\mathcal{E}_{\overline{A}}}\}_{\overline{b}=1,\cdots,|\mathcal{E}_{\overline{A}}^{}|}$ on $\mathcal{H}_{\mathcal{E}_{\overline{A}}}^{}$.
Then $\{| b\hspace{0.3mm},\hspace{0.3mm}\overline{b}\rangle_{bulk}\equiv |b\hspace{0.5mm}\rangle_{\mathcal{E}_{A}}|\overline{b}\hspace{0.5mm}\rangle_{\mathcal{E}_{\overline{A}}}\}^{b=1,\cdots,|\mathcal{E}_{A}|}_{\overline{b}=1,\cdots,|\mathcal{E}_{\overline{A}}|}$  is the  orthonormal logical basis on $\mathcal{H}_{bulk}$.\vspace{1mm}
We also prepare orthonormal bases, $\{|k\hspace{0.5mm}\rangle_{\mathcal{E}_{A}}^{(gr)}\}_{k=1,\cdots,|\mathcal{E}_{A}^{(gr)}|}$ on $\mathcal{H}_{\mathcal{E}_{A}}^{(gr)}$ and $\{|k\hspace{0.5mm}\rangle_{\mathcal{E}_{\overline{A}}}^{(gr)}\}_{k=1,\cdots,|\mathcal{E}_{\overline{A}}^{(gr)}|}$ on $\mathcal{H}_{\mathcal{E}_{\overline{A}}}^{(gr)}$
\footnote{
$|\mathcal{E}_{A/\overline{A}}^{(gr)}|$ is the dimensionality of the Hilbert space $\mathcal{H}_{\mathcal{E}_{A/\overline{A}}}^{(gr)}$.
}.

Secondly, we write $\ket{\Psi}_{bulk}^{(gr)}$ in the Schmidt decomposition as
\ba{
\ket{\Psi}_{bulk}^{(gr)}
=\sum_{k}^{|\gamma_{A}|}\Psi^{k}\ket{k}_{\mathcal{E}_{A}}^{(gr)}\ket{k}_{\mathcal{E}_{\overline{A}}}^{(gr)}\ .
}
Defining 
\ba{\label{U(k)}
U_{A}(k)\equiv U_{A}\ket{k}_{\mathcal{E}_{A}}^{(gr)}\ \ ,\ \ 
U_{\overline{A}}(k)\equiv U_{\overline{A}}\ket{k}_{\mathcal{E}_{\overline{A}}}^{(gr)}\ ,
}
we can write $U_{code}^{(\Psi)}$ defined in eq.\eqref{UcodeonPsi} as
\ba{
U_{code}^{(\Psi)}= U_{code}\ket{\Psi}_{bulk}^{(gr)}=\sum_{k}^{|\gamma_{A}|}\Psi^{k}U_{A}(k)U_{\overline{A}}(k)\ .
}
Since $U_{A}$ and $U_{\overline{A}}$ are isometry maps, the following properties hold,
\ba{\label{ortU(k)}
U_{A}^{\dagger}(i)U_{A}(k)=\delta_{i,k}I_{\mathcal{E}_{A}}\ \ ,\ \ U_{\overline{A}}^{\dagger}(i)U_{\overline{A}}(k)=\delta_{i,k}I_{\mathcal{E}_{\overline{A}}}\ .
}
Then each bulk basis is mapped by $U_{code}$ into the boundary state  as
\ba{\label{logicalbasisonhcode}
| b\hspace{0.3mm},\hspace{0.3mm}\overline{b}\rangle_{bdry}^{(\Psi)}
&\equiv U_{code}^{(\Psi)}| b\hspace{0.3mm},\hspace{0.3mm}\overline{b}\rangle_{bulk}
=\sum_{k}^{|\gamma_{A}|}\Psi^{k}U_{A}(k)|b\hspace{0.5mm}\rangle_{\mathcal{E}_{A}}U_{\overline{A}}(k) |\overline{b}\hspace{0.5mm}\rangle_{\mathcal{E}_{\overline{A}}}\nonumber\\
&\equiv \sum_{k}^{|\gamma_{A}|}\Psi^{k}|b\hspace{0.5mm};k\rangle_{A}|\overline{b}\hspace{0.5mm};k\rangle_{\overline{A}}\ ,
}
where we have defined 
\ba{\label{defoflogicalonA&Abar}
|b\hspace{0.5mm};k\rangle_{A}\equiv U_{A}(k)|b\hspace{0.5mm}\rangle_{\mathcal{E}_{A}}\ \ ,\ \ 
 |\overline{b}\hspace{0.5mm};k\rangle_{\overline{A}}\equiv U_{\overline{A}}(k)|\overline{b}\hspace{0.5mm}\rangle_{\mathcal{E}_{\overline{A}}}\ .
}
These are orthogonal states in the sense that
\ba{\label{logicalorthonormality}
\ ^{\ (\Psi)}_{bdry}\langle a\hspace{0.3mm},\hspace{0.3mm}\overline{a}| b\hspace{0.3mm},\hspace{0.3mm}\overline{b}\rangle_{bdry}^{(\Psi)}
=\delta_{a,b}\delta_{\overline{a},\overline{b}}
\ \ ,\ \ 
\ _{A}\langle a\hspace{0.5mm};i|b\hspace{0.5mm};k\rangle_{A}
=\delta_{a,b}\delta_{i,k}
\ \ ,\ \ 
 _{\overline{A}}\langle \overline{a}\hspace{0.5mm};i|\overline{b}\hspace{0.5mm};k\rangle_{\overline{A}}
=\delta_{\overline{a},\overline{b}}\delta_{i,k}
}
because $U_{code}$, $U_{A}$ and $U_{\overline{A}}$ are isometry maps.
$\{| b\hspace{0.3mm},\hspace{0.3mm}\overline{b}\rangle_{bdry}^{(\Psi)}\}^{b=1,\cdots,|\mathcal{E}_{A}|}_{\overline{b}=1,\cdots,|\mathcal{E}_{\overline{A}}|}$  is the  orthonormal logical basis on $\mathcal{H}_{code}^{(\Psi)}$.

%%%%%%%%%%%%%%%%%%%%%%%%%%%%%%%%%%%%%%%%%%%%%%%%%%%%%%%%%%%%%%%%%%%%
\subsubsection{The operator dictionary}
Let us review the basic rule about the operator dictionary between bulk operators and boundary operators.
A bulk state $|\psi\rangle_{bulk} \in \mathcal{H}_{bulk}$ can be mapped by $U_{code}^{(\Psi)}$ into the boundary state $|\psi\rangle_{bdry}$ in $\widetilde{\mathcal{H}}_{code}^{(\Psi)}$ as
\ba{\label{defdualState}
|\psi\rangle_{bdry}\equiv U_{code}^{(\Psi)}\hspace{0.5mm}|\psi\rangle_{bulk}\ .
}
We say that ``$|\psi\rangle_{bdry}$ and $|\psi\rangle_{bulk}$ are dual to each other" by the above relation.
For operators, we say that ``$\phi_{bdry}$ is dual to $\phi_{bulk}$" or  ``$\phi_{bulk}$ can be reconstructed on the boundary as $\phi_{bdry}$" if $\phi_{bdry}$  acts on boundary states  in the same way as $\phi_{bulk}$ does on bulk states under the identification \eqref{defdualState}, $i.e.$, if
\ba{\label{defdualOp}
 _{bdry}\hspace{-1mm}\bra{\psi^{'}}\phi_{bdry}\ket{\psi}_{bdry}= _{bulk}\hspace{-1mm}\bra{\psi^{'}}\phi_{bulk}\ket{\psi}_{bulk}
}
holds for any state, $\ket{\psi}_{bulk}, \ket{\psi}_{bulk}^{'} \in \mathcal{H}_{bulk}^{}$.
Putting the dual relation \eqref{defdualState} into eq.\eqref{defdualOp}, we obtain
\ba{
\phi_{bdry}=U_{code}^{(\Psi)}\hspace{0.5mm}\phi_{bulk} \hspace{0.5mm}U_{code}^{(\Psi)\dagger}
}
In other words, the boundary operator $\phi_{bdry}$ dual to a  bulk operator $\phi_{bulk}$ can be defined as the operators that satisfy
\ba{
U_{code}^{(\Psi)}\hspace{0.5mm}\phi_{bulk}|\psi\rangle_{bulk}=\phi_{bdry}|\psi\rangle_{bdry}\ .
}
for any state $|\psi\rangle_{bulk}$ in the bulk Hilbert space.
Moreover, our code has a property called the ``complementary recovery". 
This property basically says that
for $\phi_{\mathcal{E}_{A}}\otimes I_{\mathcal{E}_{\overline{A}}}\in  \mathcal{A}(\mathcal{H}_{bulk})$ and $I_{\mathcal{E}_{A}}\otimes \phi_{\mathcal{E}_{\overline{A}}}\in  \mathcal{A}(\mathcal{H}_{bulk})$, there exist $\phi_{A}\otimes I_{\overline{A}}\in  \mathcal{A}(\mathcal{H}_{code})$ and $I_{A}\otimes \phi_{\overline{A}}\in  \mathcal{A}(\mathcal{H}_{code})$ such that
\ba{
&U_{code}^{(\Psi)}(\hspace{0.5mm}\phi_{\mathcal{E}_{A}}\otimes I_{\mathcal{E}_{\overline{A}}})\ket{\psi}_{bulk}=\phi_{A}\otimes I_{\overline{A}}\ket{\psi}_{bdry}\ ,\\[7pt]
&U_{code}^{(\Psi)}(\hspace{0.5mm}I_{\mathcal{E}_{A}}\otimes \phi_{\mathcal{E}_{\overline{A}}})\ket{\psi}_{bulk}=I_{A}\otimes \phi_{\overline{A}}\ket{\psi}_{bdry}\ ,
}
for any $\ket{\psi}_{bulk}\in \mathcal{H}_{bulk}$.
This property corresponds to the entanglement wedge reconstruction in AdS/CFT. 
We will see how this complementary recovery works with more details in  subsection \ref{sec:EWRwithcenters}.

\subsubsection{Ryu-Takayanagi formula}

Next let us review  how the the Ryu-Takayanagi-like formula emerges in this code.
Any bulk state on $\mathcal{H}_{bulk}^{}$ can be rewritten in the Schmidt basis as\vspace{-2mm}
\ba{
|\psi\rangle_{bulk}^{}=\sum_{s}^{N}\sqrt{p_{s}}|s\rangle_{\mathcal{E}_{A}}|s\rangle_{\mathcal{E}_{\overline{A}}}
}
where the Schmidt basis $|s\rangle_{\mathcal{E}_{A/\overline{A}}}$ can be written in terms of the previous basis  as
\ba{
|s\rangle_{\mathcal{E}_{A}}\equiv \sum_{b}^{|\mathcal{E}_{A}|}
U^{b\hspace{0.5mm},\hspace{0.5mm} s}
|b\hspace{0.5mm}\rangle_{\mathcal{E}_{A}}\ ,\ 
|s\rangle_{\mathcal{E}_{\overline{A}}}\equiv \sum_{\overline{b}}^{|\mathcal{E}_{\overline{A}}|}
\overline{U}^{\overline{b}\hspace{0.5mm},\hspace{0.5mm} s}|\overline{b}\hspace{0.5mm}\rangle_{\mathcal{E}_{\overline{A}}}.
}
with some proper isometry matrixes $U,\ \overline{U}$.
%$\sum_{b}^{|\mathcal{E}_{A}|} \sum_{\overline{b}}^{|\mathcal{E}_{\overline{A}}|}U^{s, b}\phi^{ b\hspace{0.2mm},\hspace{0.2mm} \overline{b}}\overline{U}^{\overline{s}, \overline{b}}=\sqrt{p_{s}}\delta_{s, \overline{s}}$.
The entanglement entropy of $\mathcal{E}_{A}$ for the bulk state $|\psi\rangle_{bulk}$ is given by $S^{(bulk)}_{\mathcal{E}_{A}}(\psi)=-\sum_{s=1}^{N}p_{s}\log p_{s}$.
The dual boundary state is given by\vspace{-2mm}
\ba{
|\psi\rangle_{bdry}\label{dualofbulkentangled}
&=U_{code}^{(\Psi)}|\psi\rangle_{bulk}^{}
%=\sum_{b}^{|\mathcal{E}_{A}|}\sum_{ \overline{b}}^{|\mathcal{E}_{\overline{A}}|}\sum_{k}^{|\gamma_{A}|}\Psi^{b\hspace{0.5mm}, \overline{b}\hspace{0.5mm}}
%|b\hspace{0.5mm};k\rangle_{A}| \overline{b}\hspace{0.5mm};k\rangle_{\overline{A}}\nonumber\\
=\sum_{k}^{|\gamma_{A}|}\sum_{s}^{N}\Psi^{k}\sqrt{p_{s}}
|s, k\rangle_{A}| s, k\rangle_{\overline{A}}
}
where we defined 
\ba{
|s, k\rangle_{A}\equiv  \sum_{b}^{|\mathcal{E}_{A}|}U^{b\hspace{0.5mm},\hspace{0.5mm}s}|b\hspace{0.5mm};k\rangle_{A}\ ,\ 
|s, k\rangle_{\overline{A}}\equiv  \sum_{\overline{b}}^{|\mathcal{E}_{\overline{A}}|}\overline{U}^{\overline{b}\hspace{0.5mm},\hspace{0.5mm} s}|\overline{b}\hspace{0.5mm};k\rangle_{\overline{A}}
}
These states are orthogonal in the sense that
\ba{
\ _{A}\langle r, i|s, k\rangle_{A}\ = \ _{\overline{A}}\langle r, i|s, k\rangle_{\overline{A}}=\ \delta_{r,s}\delta_{i,k}\   ,
}
which follows from the isometry condition of $U$ and $\overline{U}$.
The density matrix for $A$ can be written as
\ba{
\rho_{A}
=\mathop{\mathrm{Tr}}_{\overline{A}}|\psi\rangle_{bdry\hspace{0.5mm}bdry}\langle \psi|
=\sum_{k}^{|\gamma_{A}|}\sum_{s=1}^{N}|\Psi^{k}|^{2}p_{s}\hspace{0.5mm}|s, k\rangle_{A\hspace{0.5mm}A}\langle s, k|\  .
}
Then one can easily find that the entanglement entropy of the subsystem $A$ for $|\Psi\rangle_{bdry}$ is given by\vspace{-2mm}
\ba{\label{RT+bulk}
S_{A}=-\sum_{k}^{|\gamma_{A}|}|\Psi^{k}|^{2}\log |\Psi^{k}|^{2}
-\sum_{s=1}^{N}p_{s}\log p_{s}.%-\sum_{s}^{N}p_{s}\log p_{s}\ .
}
Let us write the entanglement entropy of $\ket{\Psi}_{bulk}^{(gr)}$ between $\mathcal{E}_{A}$ and $\mathcal{E}_{\overline{A}}$ as $S_{\mathcal{E}_{A}}(\Psi)$ and define the ``Area operator" as
\ba{
\mathcal{L}_{A}(\Psi)\equiv S_{\mathcal{E}_{A}}(\Psi)I_{\mathcal{E}_{A}}\ .
}
Then the entanglement entropy in \eqref{RT+bulk} can  be rewritten as
\ba{
S_{A}={\mathop{\mathrm{Tr}}_{\mathcal{E}_{A}}}(\rho_{\mathcal{E}_{A}}\mathcal{L}_{A}(\Psi))+S^{(bulk)}_{\mathcal{E}_{A}}(\psi)\ .
}
The first term is universal in the sense that  it does not depend on the information of the bulk state $|\psi\rangle_{bulk}^{}$ at all.
The second term in eq.\eqref{RT+bulk} is exactly the bulk entanglement entropy of $|\psi\rangle_{bulk}^{}$ between $\mathcal{E}_{A}$ and $\mathcal{E}_{\overline{A}}$.
Thus the  result analogous to the quantum corrections to the holographic entanglement entropy in \cite{Faulkner:2013ana} is also reproduced.

Based on \cite{Pastawski:2015qua}, let us also review   how the universal term is proportional to the ``Area" of the entangling surface.
Based on the tensor network picture, the entanglement of $\ket{\Psi}_{bulk}^{GR}$ between $\mathcal{E}_{A}$ and $\mathcal{E}_{\overline{A}}$ are created by the Bell pairs straddling the entangling surface of $A$ and connecting the two boundary point on $A$ and on $\overline{A}$.
Then it would natural to assume $|\gamma_{A}|\le v^{\text{Area}(\gamma_{A})/L_{AdS}}$ with some ``coarse graining scale, $L_{AdS}$" and $v$ is the dimension of the Hilbert space of  local d.o.f on the boundary.
If we assume  that $\ket{\Psi}_{bulk}^{(gr)}$ is the maximally entanglement state within the bound,  it can be written as
\ba{
\ket{\Psi}_{bulk}^{(gr)}
=\frac{1}{|\gamma_{A}|}\sum_{k}^{|\gamma_{A}|}\ket{k}_{\mathcal{E}_{A}}^{(gr)}\ket{k}_{\mathcal{E}_{\overline{A}}}^{(gr)}\ ,
}
with $|\gamma_{A}|= v^{\text{Area}(\gamma_{A})/L_{AdS}}$.
Then the universal term can be written as
\footnote{
In the holographic code model in \cite{Pastawski:2015qua}, $v$ corresponds to the dimension of the Hilbert space associated to the each tensor leg of the perfect tensors, and $L_{AdS}$ corresponds to the length of one side of the ``tiles" which are uniformly tiling the two-dimensional hyperbolic space. 
The entangling surface $\gamma_{A}$ also corresponds to the \textit{greedy} geodesic.
Under this identification, the result \eqref{universalterm(area)} exactly holds in the model.
}
\ba{\label{universalterm(area)}
{\mathop{\mathrm{Tr}}_{\mathcal{E}_{A}}}(\rho_{\mathcal{E}_{A}}\mathcal{L}_{A}(\Psi))
=\log |\gamma_{A}|=\frac{\log v}{L_{AdS}} \ \text{Area}(\gamma_{A})\ .
}
Since $v$ is the dimension of local Hilbert spaces, $c\equiv \log v$ is roughly the number of d.o.f on each local spacetime point on the boundary.
Defining $\frac{c}{L_{AdS}}\equiv\frac{1}{4G_{N}}$
\footnote{
 This relation is analogous  to the formula between the radius of curvature of AdS and the central charge, $c=\frac{3L_{AdS}}{2G_{N}}$ in \cite{Brown:1986nw}.
}, we then obtain the ``Ryu-Takayanagi formula",
\ba{
{\mathop{\mathrm{Tr}}_{\mathcal{E}_{A}}}(\rho_{\mathcal{E}_{A}}\mathcal{L}_{A}(\Psi))
=\frac{\text{Area}(\gamma_{A})}{4G_{N}}\ .
}

%%%%%%%%%%%%%%%%%%%%%%%%%%%%%%%%%%%%%%%%%%%%%%%%%%%%%%%%%%%%%%%%%%%%%%%%%%%%%%%%%%%%%%%%%%%%%%%%%%%%%%%%%%%%%%%%%%
\subsection{Emergence of central operators }\label{sec:emergenceofcenter}
In this subsection, we point out that the central operators of the algebras associated with subregions can emerge as a consequence of the restriction of the Hilbert space.

First, let us define a set of operators, $\mathcal{S}(A\hspace{0.5mm}; \mathcal{H})$, as follows:

\vspace{-1mm}
%\newpage
%%%%%%%%% Definition%%%%%%%%%%%%%%
\noindent\rule{\columnwidth}{1pt}\vspace{-2mm}
%\begin{screen}
\begin{definition} %\ \\
\ Given a subspace  $\mathcal{H}$ in  $\mathcal{H}_{A\overline{A}}=\mathcal{H}_{A}\otimes\mathcal{H}_{\overline{A}}$,  $\mathcal{S}_{A\overline{A}}(A\hspace{0.5mm}; \mathcal{H})$ is  defined as
\ba{
\bullet\  \mathcal{S}&_{A\overline{A}}(A;\mathcal{H}) \\
\equiv &\{\ \mathcal{O}_{_{A\overline{A}}}\in \mathcal{A}(\mathcal{H})\ |\  ^{\exists}\mathcal{O}_{A\overline{A}\rightarrow A}\  s.t.\ 
^{\forall}|\psi\rangle_{A\overline{A}}\in \mathcal{H}\ , \mathcal{O}_{A\overline{A}}|\psi\rangle_{A\overline{A}}=\mathcal{O}_{A\overline{A}\rightarrow A}\otimes I_{\overline{A}}\hspace{1mm}|\psi\rangle_{A\overline{A}}\}\ .\nonumber%\\[5mm]
%\bullet\ \mathcal{V}&_{A\overline{A}}(\mathcal{H})
%\equiv \{\ \mathcal{O}_{A\overline{A}}\in \mathcal{A}(\mathcal{H}_{A\overline{A}})\ |\
 %^{\forall}|\psi\rangle_{A\overline{A}}\in \mathcal{H}\ ,\mathcal{O}_{A\overline{A}}|\psi\rangle_{A\overline{A}}=0\}
 }
\end{definition}
\vspace{-5mm}
\noindent\rule{\columnwidth}{1pt}
%%%%%%%%%%%%%%%%%%%%%%%%%%

$\mathcal{S}_{A\overline{A}}(A\hspace{0.5mm};\mathcal{H})$ is the set of the Hermitian operators that act within $\mathcal{H}$ and are able to shrink the supports to $A$ not only on a state in $\mathcal{H}$ but also on any state in $\mathcal{H}$.
Then  Lemma  \ref{CenterforGaugeTheory} holds:

%%%%%%%%%  Lemma %%%%%%%%%%%%%%
\noindent\rule{\columnwidth}{1pt}\vspace{-2mm}
%\begin{screen}a
\begin{lemma} \label{CenterforGaugeTheory}%\ \\
\ For $\mathcal{H}_{bulk}^{}=\mathcal{H}_{\mathcal{E}_{A}}^{}\otimes \mathcal{H}_{\mathcal{E}_{\overline{A}}}^{}$\ , 
\ba{
\mathcal{S}_{bulk}(\mathcal{E}_{A};\mathcal{H}_{bulk})
=\{\  \mathcal{O}_{bulk} \in  \mathcal{A}(\mathcal{H}_{bulk})\ |\ ^{\exists}\mathcal{O}_{\mathcal{E}_{A}}\ s.t. \ \mathcal{O}_{bulk}=\mathcal{O}_{\mathcal{E}_{A}}\otimes I_{\mathcal{E}_{\overline{A}}}\ \}
}
\end{lemma}
\vspace{-5mm}
\noindent\rule{\columnwidth}{1pt}
%%%%%%%%%%%%%%%%%%%%%%%%%%

This lemma describes the trivial fact that for $\mathcal{H}_{bulk}^{}\equiv \mathcal{H}_{\mathcal{E}_{A}}^{}\otimes \mathcal{H}_{\mathcal{E}_{\overline{A}}}^{}$, the operators in $ \mathcal{A}(\mathcal{H}_{bulk})$ that can shrink their support to $\mathcal{E}_{A}$ on any state in $\mathcal{H}_{bulk}$ are just the operators that are supported on $\mathcal{E}_{A}$.

\vspace{0.5mm}Once we restrict the Hilbert space from $\mathcal{H}_{bulk}$ to the subspace $\widetilde{\mathcal{H}}_{bulk}(\subset \mathcal{H}_{bulk})$, then $ \mathcal{S}_{bulk}(\mathcal{E}_{A};\widetilde{\mathcal{H}}_{bulk})$ can be nontrivial 
%for the subspace $\widetilde{\mathcal{H}}_{bulk}\subset \mathcal{H}_{bulk}$ 
because of the property \eqref{notfactrize}.
In fact, in the case of the shrinking that preserves the unitarity, we have already seen that the trivial decoder can shrink its support\vspace{1mm} to only $\overline{E}$ not only  on a state but also on any state in $\mathcal{H}_{code}(\subset\mathcal{H}_{E}\otimes\mathcal{H}_{\overline{E}}$)\ .
\vspace{1mm}Note that if there are elements in $ \mathcal{S}_{bulk}(\mathcal{E}_{A};\widetilde{\mathcal{H}}_{bulk})\cap  \mathcal{S}_{bulk}(\mathcal{E}_{\overline{A}};\widetilde{\mathcal{H}}_{bulk})$, \vspace{1mm}then the elements commute with all the operators in $ \mathcal{A}(\widetilde{\mathcal{H}}_{bulk})$ supported only on $\mathcal{E}_{A}$ \vspace{1mm}and also all the operators in $ \mathcal{A}(\widetilde{\mathcal{H}}_{bulk})$ supported only on $\mathcal{E}_{\overline{A}}$.
\vspace{1mm}In this sense, $ \mathcal{S}_{bulk}(\mathcal{E}_{A};\widetilde{\mathcal{H}}_{bulk})\cap  \mathcal{S}_{bulk}(\mathcal{E}_{\overline{A}};\widetilde{\mathcal{H}}_{bulk})$ is the center of the algebra on $\mathcal{E}_{A}$ and also on $\mathcal{E}_{\overline{A}}$.

 Here let us see how the \vspace{1mm}restriction of   the Hilbert space by the ``gauge constraints" gives the nontrivial elements in  $ \mathcal{S}_{bulk}(\mathcal{E}_{A};\widetilde{\mathcal{H}}_{bulk})\cap  \mathcal{S}_{bulk}(\mathcal{E}_{\overline{A}};\widetilde{\mathcal{H}}_{bulk})$ \cite{Casini:2013rba, Radicevic:2014kqa, Soni:2015yga, Ma:2015xes}.
For example, let us define $\widetilde{\mathcal{H}}_{bulk}$ as
\ba{
\widetilde{\mathcal{H}}_{bulk}\equiv \{ \ket{\psi}_{bulk}\in \mathcal{H}_{bulk}\ |\ G \ket{\psi}_{bulk}= \ket{\psi}_{bulk}\ \}\ ,
}
with a unitary operator $G\equiv G_{\mathcal{E}_{A}}\otimes G_{\mathcal{E}_{\overline{A}}}$, supported on both $\mathcal{H}_{\mathcal{E}_{A}}$ and $\mathcal{H}_{\mathcal{E}_{\overline{A}}}$, which we  will call the ``gauge transformation operator".
In this case, $\widetilde{\mathcal{H}}_{bulk}$ and $\mathcal{H}_{bulk}$ play the role of the physical Hilbert space and the extended Hilbert space, respectively.
Then for any state $\ket{\psi}_{bulk}\in \widetilde{\mathcal{H}}_{bulk}$, we have
\ba{\label{Gaugeconstraint}
G_{\mathcal{E}_{A}}\otimes G_{\mathcal{E}_{\overline{A}}} \ket{\psi}_{bulk}= \ket{\psi}_{bulk}\ .
}
By acting on eq.\eqref{Gaugeconstraint} with $I_{\mathcal{E}_{A}}\otimes G_{\mathcal{E}_{\overline{A}}}^{\dagger}$ from the lefthand side, we obtain
\ba{
G_{\mathcal{E}_{A}}\otimes I_{\mathcal{E}_{\overline{A}}} \ket{\psi}_{bulk}= I_{\mathcal{E}_{A}}\otimes G_{\mathcal{E}_{\overline{A}}}^{\dagger}\ket{\psi}_{bulk}\ .
}
This means $G_{\mathcal{E}_{A}}\otimes I_{\mathcal{E}_{\overline{A}}}\ ,\  I_{\mathcal{E}_{A}}\otimes G_{\mathcal{E}_{\overline{A}}}^{\dagger}\in  \mathcal{S}_{bulk}(\mathcal{E}_{A};\widetilde{\mathcal{H}}_{bulk})\cap  \mathcal{S}_{bulk}(\mathcal{E}_{\overline{A}};\widetilde{\mathcal{H}}_{bulk})$
\footnote{
These central operators $G_{\mathcal{E}_{A}}\otimes I_{\mathcal{E}_{\overline{A}}}$ ,  $I_{\mathcal{E}_{A}}\otimes G_{\mathcal{E}_{\overline{A}}}^{\dagger}$ are the same operators in the sense that they act on all the physical states in exactly the  same way.
If we consider a  general subspace $\widetilde{\mathcal{H}}_{bulk}\subset\mathcal{H}_{bulk}$ ($e.g.$  low-energy subspace), there can exist the central operators that act on physical states differently.
}.
For the concrete examples of these central operators in gauge theories, please see \cite{Casini:2013rba, Radicevic:2014kqa, Soni:2015yga, Ma:2015xes, Harlow:2015lma}.
We will see in the next subsection that the central boundary operators in the reconstructed algebras on $A$ and on $\overline{A}$ are dual to these central operators in $\mathcal{S}_{bulk}(\mathcal{E}_{A};\widetilde{\mathcal{H}}_{bulk})\cap  \mathcal{S}_{bulk}(\mathcal{E}_{\overline{A}};\widetilde{\mathcal{H}}_{bulk})$ in the bulk theories.

%%%%%%%%%%%%%%%%%%%%%%%%%%%%%%%%%%%%%%%%%%%%%%%%%%%%%%%%%%%%%%%%%%%%%%%%%%%%%%%%%%%%%%%%%%%%%%%%%%%%%%%%%%%%%%%%%%
\subsection{The entanglement wedge reconstruction and its converse statement with centers}\label{sec:EWRwithcenters}

In this subsection, we prove a theorem for our code which implies the validity of the entanglement wedge reconstruction and also its converse statement with nontrivial centers:
%Next, we will prove the following main theorem about the operator dictionary in this code.

%%%%%%%%%  Theorem %%%%%%%%%%%%%%
\noindent\rule{\columnwidth}{1pt}\vspace{-2mm}
%\begin{screen}a
\begin{theorem} \label{EWRwithCenter}%\ \\
\ For $\phi_{bdry}\in\mathcal{A}(\widetilde{\mathcal{H}}_{code}^{(\Psi)})$, the following two statements are equivalent:
\ba{
(A)\  \ & ^{\exists}\phi_{bulk}\in 
 \mathcal{S}_{bulk}(\hspace{0.5mm} \mathcal{E}_{A};\widetilde{\mathcal{H}}_{bulk}\hspace{0.5mm} )\ s.t. \ \ %\nonumber\\[5pt]
%\label{linktoA}\\[5pt]
 \phi_{bdry}%P_{code}
 =\hspace{0.5mm}U_{code}^{(\Psi)}\hspace{0.5mm}\phi_{bulk}\hspace{0.5mm}U_{code}^{(\Psi)\dagger}\  \label{dualrelation}
 %=\hspace{0.5mm}U_{code}(\hspace{0.5mm}\phi_{\mathcal{E}_{A}}\otimes I_{\mathcal{E}_{\overline{A}}}\hspace{0.5mm})U_{code}^{\dagger}\ . \label{dualrelation}
\\[0.3cm]
(B)\ \ & \phi_{bdry}\in 
 \mathcal{S}_{bdry}(\hspace{0.5mm} A;\widetilde{\mathcal{H}}_{code}^{(\Psi)}\hspace{0.5mm} ) \label{Statement(B)inEWR}
}
\end{theorem}
\vspace{-5mm}
\noindent\rule{\columnwidth}{1pt}
%%%%%%%%%%%%%%%%%%%%%%%%%%

The  $(A) \Rightarrow (B)$ direction of the theorem says that if a bulk operator in $ \mathcal{A}(\widetilde{\mathcal{H}}_{bulk})$ can shrink its support to only $\mathcal{E}_{A}$, then it can be reconstructed as a boundary operator that can shrink its support to only $A$ in the code subspace $\widetilde{\mathcal{H}}_{code}^{(\Psi)}$.
This corresponds to the statement of the entanglement wedge reconstruction hypothesis in AdS/CFT. 
This direction has been  proven  in \cite{Harlow:2016vwg}, and here let us show again in a concrete way by following the idea of \cite{Pastawski:2015qua}. 
On the other hand, the converse direction $(B)\Rightarrow(A)$ says that if a boundary operator in $\mathcal{A}(\widetilde{\mathcal{H}}_{code}^{(\Psi)})$ can shrink its support to $A$, then the dual bulk operator must be able to shrink its support to only $A$ on $\widetilde{\mathcal{H}}_{bulk}$.
Note that we can replace ($\mathcal{E}_{A}$, $A$) with ($\mathcal{E}_{\overline{A}}$, $\overline{A}$) in Theorem \ref{EWRwithCenter} since $U_{code}$ equally treats ($\mathcal{E}_{A}$, $A$) and ($\mathcal{E}_{\overline{A}}$, $\overline{A}$) in the definition \ref{Ucode}.
\begin{proof}\ \\
\noindent$(A)\Rightarrow(B)$: 
It follows from the assumption, $\phi_{bulk}\in \mathcal{S}_{bulk}(\hspace{0.5mm} \mathcal{E}_{A};\widetilde{\mathcal{H}}_{bulk}\hspace{0.5mm} )$, that there exists an operator $\phi_{\mathcal{E}_{A}}\otimes I_{\mathcal{E}_{\overline{A}}}\in  \mathcal{A}(\widetilde{\mathcal{H}}_{bulk})$ such that
\ba{\label{EWRwithCentereq1}
\phi_{bulk}\ket{\psi}_{bulk}=\phi_{\mathcal{E}_{A}}\otimes I_{\mathcal{E}_{\overline{A}}} \ket{\psi}_{bulk}\ ,
}
for any state $\ket{\psi}_{bulk}\in \widetilde{\mathcal{H}}_{bulk}$.
Acting on  eq.\eqref{EWRwithCentereq1} with $U_{code}^{(\Psi)}$ from the lefthand side, we obtain
\ba{\label{EWReq1}
(l.h.s)&=U_{code}^{(\Psi)}\phi_{bulk}\ket{\psi}_{bulk}
=\phi_{bdry}\ket{\psi}_{bdry}
\ ,
}
where we have inserted, $I_{bulk}=U_{code}^{(\Psi)\dagger}U_{code}^{(\Psi)}$ in the first equality and used $\ket{\psi}_{bdry}=U_{code}^{(\Psi)}\ket{\psi}_{bulk}$ and $\phi_{bdry}=U_{code}^{(\Psi)}\phi_{bulk}U_{code}^{(\Psi)\dagger}$ in the second equality.
%%%%%%%%%%%%%%%%%%%%%%%%%%%%%%%%%%%
%%%%%%%%%%%    figure    %%%%%%%%%%%%
\begin{figure}[tbp]
\begin{center}
\resizebox{155mm}{!}{\includegraphics{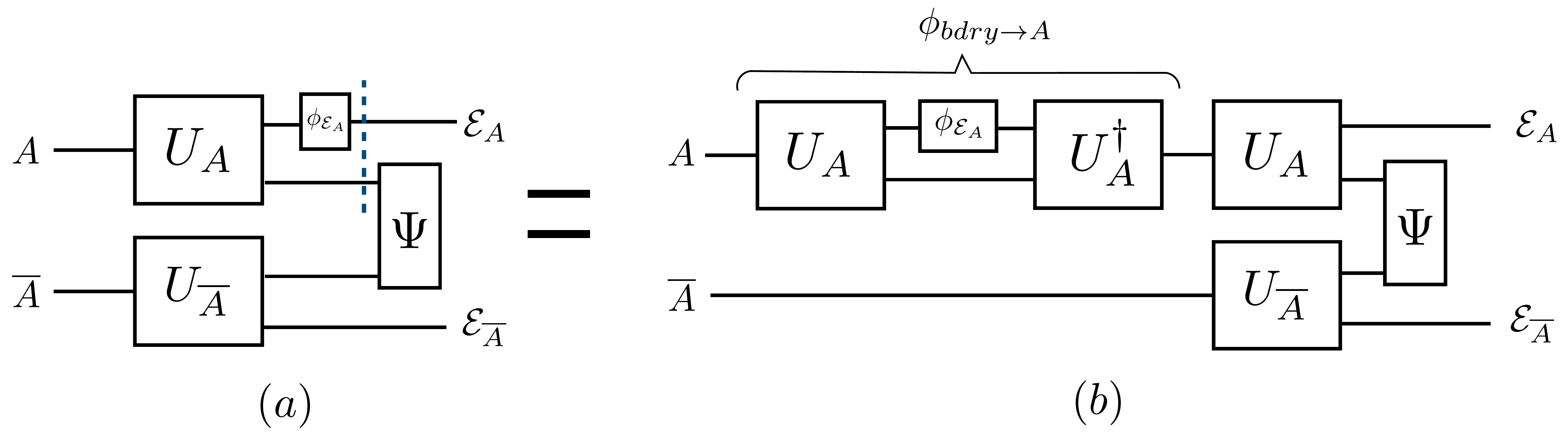}}
%\resizebox{155mm}{!}{\includegraphics{FigEWC5.eps}}
%\resizebox{105mm}{!}{\includegraphics{Fig_1b_v4.pdf}}
\caption{\small 
Fig.$(a)$ represents the tensor networks expression of $U_{code}^{(\Psi)}(\hspace{0.5mm}\phi_{\mathcal{E}_{A}}\otimes I_{\mathcal{E}_{\overline{A}}}\hspace{0.5mm})$, which corresponds to eq.\eqref{eq:figa} without $U_{code}^{(\Psi)\dagger}\ket{\psi}_{bdry}$.
Inserting $I_{\mathcal{E}_{A}}\otimes I_{\mathcal{E}_{A}^{(gr)}}=U_{A}^{\dagger}U_{A}$ on the dashed line in Fig.(a), we obtain Fig.$(b)$, which corresponds to eq.\eqref{eq:figb} (or \eqref{eq:EWC4}) without $U_{code}^{(\Psi)\dagger}\ket{\psi}_{bdry}$.
}
%Figure\ref{fig:Ew}
\label{fig:EWC}
\end{center}
\end{figure} 
%%%%%%%%%%%%%%%%%%%%%%%%%%%%%% 
The righthand side  is given by
\ba{
(r.h.s)&=U_{code}^{(\Psi)}(\hspace{0.5mm}\phi_{\mathcal{E}_{A}}\otimes I_{\mathcal{E}_{\overline{A}}}\hspace{0.5mm}) U_{code}^{(\Psi)\dagger}\ket{\psi}_{bdry}\label{eq:figa}\\
&=\sum_{j,k}^{|\gamma_{A}|}\Psi^{j}\Psi^{k}\hspace{0.5mm}
U_{A}(j)\hspace{0.5mm}\phi_{\mathcal{E}_{A}}^{}U_{A}^{\dagger}(k)
\otimes U_{\overline{A}}(j)\hspace{0.5mm}I_{\mathcal{E}_{\overline{A}}}\hspace{0.5mm}U_{\overline{A}}^{\dagger}(k)\ket{\psi}_{bdry}\nonumber\\
&=\sum_{i}^{|\mathcal{E}_{A}^{(gr)}|}\sum_{j,k}^{|\gamma_{A}|}\Psi^{j}\Psi^{k}\hspace{0.5mm}
U_{A}(i)\hspace{0.5mm}\phi_{\mathcal{E}_{A}}^{}\hspace{0.5mm}U_{A}^{\dagger}(i)\hspace{0.5mm}U_{A}(j)\hspace{0.5mm}U_{A}^{\dagger}(k)\otimes U_{\overline{A}}(j)\hspace{0.5mm}I_{\mathcal{E}_{\overline{A}}}\hspace{0.5mm}U_{\overline{A}}^{\dagger}(k)\ket{\psi}_{bdry}\label{eq:figb}\\[5pt]
&=(\hspace{0.5mm}\phi_{bdry\rightarrow A}\otimes I_{\overline{A}}\hspace{0.5mm})U_{code}^{(\Psi)}U_{code}^{(\Psi)\dagger}\ket{\psi}_{bdry}\label{eq:EWC4}\\[5pt]
&=\hspace{0.5mm}\phi_{bdry\rightarrow A}\otimes I_{\overline{A}}\hspace{0.5mm}\ket{\psi}_{bdry}\label{eq:EWC}
}
where we have used %the definition  in the second equality and 
eq.\eqref{ortU(k)} in the third equality and have defined 
\ba{\label{reconstructedonA}
\phi_{bdry\rightarrow A}\equiv U_{A}(\hspace{0.5mm}\phi_{\mathcal{E}_{A}}\otimes I_{\mathcal{E}_{A}^{(gr)}}\hspace{0.5mm})U_{A}^{\dagger}
=\sum_{i}^{|\mathcal{E}_{A}^{(gr)}|}U_{A}(i)\hspace{0.5mm}\phi_{\mathcal{E}_{A}}^{}\hspace{0.5mm}U_{A}^{\dagger}(i)
}
in the fourth equality
 \footnote{
 $I_{\mathcal{E}_{A}^{(gr)}}$ is the identity operator on $\mathcal{H}_{\mathcal{E}_{A}}^{(gr)}$.
 }.
We have also used the property that $U_{code}^{(\Psi)}U_{code}^{(\Psi)\dagger}$ acts on any state in $\widetilde{\mathcal{H}}_{code}^{(\Psi)}$ as the identity operator (because of $U_{code}^{(\Psi)}U_{code}^{(\Psi)\dagger}U_{code}^{(\Psi)}=U_{code}^{(\Psi)}$) in the final equality.
These equation from \eqref{eq:figa} to \eqref{eq:EWC} can be represented in the tensor networks as Figure.\ref{fig:EWC}.
Thus, we obtain for any state $\ket{\psi}_{bdry}$ in $\widetilde{\mathcal{H}}_{code}^{(\Psi)}$,
\ba{
\phi_{bdry}\ket{\psi}_{bdry}=\phi_{bdry\rightarrow A}\otimes I_{\overline{A}}\hspace{0.5mm}\ket{\psi}_{bdry}\ .
}
Thus we conclude $\phi_{bdry}\in \mathcal{A}(\hspace{0.5mm}\widetilde{\mathcal{H}}_{code}^{(\Psi)}\hspace{0.5mm} ) \cap \mathcal{S}_{bdry}(\hspace{0.5mm} A;\widetilde{\mathcal{H}}_{code}^{(\Psi)}\hspace{0.5mm} )$.

\noindent$(B)\Rightarrow(A)$:
From the assumption \eqref{Statement(B)inEWR}, there exists an operator $\phi _{A}\otimes I_{\overline{A}}\in \mathcal{A}(\widetilde{\mathcal{H}}_{code}^{(\Psi)})$  such that
\ba{\label{EWReq5}
\phi_{bdry}\hspace{0.5mm}\ket{\psi}_{bdry}=\phi _{A}\otimes I_{\overline{A}}\hspace{0.5mm}\ket{\psi}_{bdry}\ ,
}
for any state $\ket{\psi}_{bdry}$ in $\widetilde{\mathcal{H}}_{code}^{(\Psi)}$.
Furthermore, from the assumption of $\phi_{bdry}\in \mathcal{A}(\hspace{0.5mm}\mathcal{H}_{code}\hspace{0.5mm} )$, there exists a bulk operator $\phi_{bulk}$  such that
\ba{\label{EWReq6}
U_{code}^{(\Psi)}\hspace{0.5mm}\phi_{bulk}\hspace{0.5mm}U_{code}^{(\Psi)\dagger}=\phi_{bdry}\ .
}
When we act on  eq.\eqref{EWReq5} with $U_{code}^{(\Psi)\dagger}$ from the lefthand side, the lefthand side of the equation is given by
\ba{
(l.h.s)&=U_{code}^{(\Psi)\dagger}\hspace{0.5mm}\phi_{bdry}\hspace{0.5mm}U_{code}^{(\Psi)}\ket{\psi}_{bulk}\\
&=\phi_{bulk}\ket{\psi}_{bulk}
}
where we have used eq.\eqref{EWReq6} in the second equality.
The righthand side is give by
\ba{
(r.h.s)&=U_{code}^{(\Psi)\dagger}(\hspace{0.5mm} \phi _{A}\otimes I_{\overline{A}}\hspace{0.5mm})U_{code}^{(\Psi)}\ket{\psi}_{bulk}\\
&=\sum_{i,k}^{|\gamma_{A}|}\Psi^{i}\Psi^{k}U_{A}^{\dagger}(i)U_{\overline{A}}^{\dagger}(i)(\hspace{0.5mm} \phi _{A}\otimes I_{\overline{A}}\hspace{0.5mm})U_{A}(k)U_{\overline{A}}(k)\ket{\psi}_{bulk}\\
&=\sum_{i}^{|\gamma_{A}|} (\Psi^{i})^{2}U_{A}^{\dagger}(i)\phi _{A}U_{A}(i)\otimes I_{\mathcal{E}_{\overline{A}}}\ket{\psi}_{bulk}
%&=\phi _{\mathcal{E}_{A}}\otimes I_{\mathcal{E}_{\overline{A}}}\ket{\psi}_{bulk}
}
where we have used eq.\eqref{ortU(k)} in the third equality.
Defining $\phi _{\mathcal{E}_{A}}\equiv\sum_{i}^{|\gamma_{A}|} (\Psi^{i})^{2}U_{A}^{\dagger}(i)\phi _{A}U_{A}(i)$, we thus obtain 
\ba{
\phi_{bulk}\ket{\psi}_{bulk}=\phi _{\mathcal{E}_{A}}\otimes I_{\mathcal{E}_{\overline{A}}}\ket{\psi}_{bulk}\ .
}
for any state $\ket{\psi}_{bulk}\in \widetilde{\mathcal{H}}_{bulk}$.
This means  $ \phi_{bulk}\in  \mathcal{S}_{bulk}(\hspace{0.5mm} \mathcal{E}_{A};\widetilde{\mathcal{H}}_{bulk}\hspace{0.5mm} )$.
\end{proof}

Since the bulk operators supported only on $\mathcal{E}_{A}$ are trivially included in $ \mathcal{S}_{bulk}(\hspace{0.5mm} \mathcal{E}_{A};\widetilde{\mathcal{H}}_{bulk}\hspace{0.5mm} )$, the result below  immediately follows from Theorem \ref{EWRwithCenter}\ :

%%%%%%%%  theorem EWC without centers  %%%%%%%%%%%%%%%%%%%%%%%%%%%
\noindent\rule{\columnwidth}{1pt}\vspace{-1.5mm}
\begin{corollary} \label{EWRtheoremwithoutcenters}%\ \\
%\ For $\phi_{bdry}\in\mathcal{A}(\widetilde{\mathcal{H}}_{code}^{(\Psi)})$,
%\ba{
\hspace{2mm}  \text{If}\ \ $ ^{\exists\ } \phi_{\mathcal{E}_{A}}\otimes I_{\mathcal{E}_{\overline{A}}} \in \mathcal{A}(\widetilde{\mathcal{H}}_{bulk})
 \ 
 s.t.\   U_{code}^{(\Psi)}(\phi_{\mathcal{E}_{A}}\otimes I_{\mathcal{E}_{\overline{A}}})U_{code}^{(\Psi)\dagger}=\hspace{0.5mm}\phi_{bdry}\hspace{0.5mm}$%\nonumber\\[3mm]
\vspace{3mm}

\hspace{26mm}   \text{then} \  \ 
$ \ \phi_{bdry}\in  \mathcal{S}_{bdry}(A;\widetilde{\mathcal{H}}_{code}^{(\Psi)})\cap \mathcal{A}(\widetilde{\mathcal{H}}_{code}^{(\Psi)})$.
% \phi_{bdry}\  \text{is dual to a bulk operator with the support only on}  \ \mathcal{E}_{A}.
%}
\end{corollary}\vspace{-5mm}
\noindent\rule{\columnwidth}{1pt}\vspace{2mm}
%%%%%%%%%%%%%%%%%%%%%%%%%%%%%%%%%%%%%%%%%%%%
This says that the bulk operators supported only on $\mathcal{E}_{A}$ can be reconstructed on $A$.
This corresponds to more conventional entanglement wedge reconstruction hypothesis in AdS/CFT.
However the converse statement does not generally hold because there possibly exist the bulk operators that have a part of their supports  on $\mathcal{E}_{\overline{A}}$ but can shrink their supports to $\mathcal{E}_{A}$ in $\widetilde{\mathcal{H}}_{bulk}$.

Here, let us see how this bulk reconstruction with complementary recovery works in more concrete way from the viewpoint of the shrinking in Lemma \ref{generalgatetel}.
Any bulk operator supported on $\mathcal{E}_{A}$, $\phi_{\mathcal{E}_{A}}\otimes I_{\mathcal{E}_{\overline{A}}} \in \mathcal{A}(\widetilde{\mathcal{H}}_{bulk}^{})$, can be written in the bulk basis as
\ba{
\phi_{\mathcal{E}_{A}}^{}
=\sum_{a\hspace{0.2mm},\hspace{0.2mm}b}^{|\mathcal{E}_{A}|} \phi^{a,b}\hspace{0.5mm}
| a\rangle_{\mathcal{E}_{A}\hspace{0.7mm}\mathcal{E}_{A}}\langle b|
}
with  $ \phi^{b,a\ast}=\phi^{a,b}$. 
Then the boundary operators on $\widetilde{\mathcal{H}}_{code}^{(\Psi)}$ dual to $\phi_{\mathcal{E}_{A}}$ are given by
\ba{\label{codetocode}
\phi_{bdry}
&=U_{code}^{(\Psi)}(\hspace{0.5mm}\phi_{\mathcal{E}_{A}}^{}\otimes I_{\mathcal{E}_{\overline{A}}}^{}\hspace{0.5mm})U^{(\Psi)\dagger}_{code}
=\sum_{a\hspace{0.2mm},\hspace{0.2mm}b}^{|\mathcal{E}_{A}|}\sum_{ \overline{b}}^{|\mathcal{E}_{\overline{A}}|}
\phi^{a,b}\hspace{0.5mm}
| a\hspace{0.3mm},\hspace{0.3mm}\overline{b}\rangle_{bdry\hspace{0.7mm}bdry}^{(\Psi)\hspace{3.3mm}(\Psi)}\langle b\hspace{0.3mm},\hspace{0.3mm}\overline{b}|\ 
}
where we have used  $I_{\mathcal{E}_{\overline{A}}}^{}=\sum_{\overline{b}}^{|\mathcal{E}_{\overline{A}}|}\ket{\overline{b}}_{\mathcal{E}_{\overline{A}}\hspace{0.7mm}\mathcal{E}_{\overline{A}}}\hspace{-1mm}\bra{\overline{b}}$ and the definition \eqref{logicalbasisonhcode}.

On the other hand, $\phi_{\mathcal{E}_{A}}^{}$ has been reconstructed on $A$ as \eqref{reconstructedonA}, which can be written explicitly as
\ba{\label{dualbulkoponE_A}
\phi_{bdry\rightarrow A}\equiv \sum_{i}^{|\gamma_{A}|}U_{A}(i)\phi_{\mathcal{E}_{A}}^{}U_{A}^{\dagger}(i)
=\sum_{a\hspace{0.2mm},\hspace{0.2mm}b}^{|\mathcal{E}_{A}|}\sum_{i}^{|\gamma_{A}|}
\phi^{a, b\hspace{0.5mm}}
|a\hspace{0.5mm};i\rangle_{A\hspace{0.7mm}A}\langle b\hspace{0.5mm};i|\ .
} 
We already proved in Theorem \ref{EWRwithCenter} that for $\phi_{bdry}$ defined in \eqref{codetocode}, $\phi_{bdry\rightarrow A}$ defined in eq.\eqref{dualbulkoponE_A} satisfies
\ba{\label{shrinkingoncode}
\phi_{bdry}\ket{\psi}_{bdry}=\phi_{bdry\rightarrow A}\otimes I_{\overline{A}}\hspace{0.5mm}\ket{\psi}_{bdry}\ ,
}
for any state $\ket{\psi}_{bdry} \in \widetilde{\mathcal{H}}_{code}^{(\Psi)}$\ .
Note that since $\phi_{bdry\rightarrow A}$ is the Hermitain operator if $\phi_{bdry}$ is the one, the shrinking in \eqref{shrinkingoncode} preserves the Hermiticity.
We can check that eq.\eqref{shrinkingoncode} actually holds easily in the same way in Example \ref{essenceofShrinkHermitian} as follows.
When $\phi_{bdry}$ defined in \eqref{codetocode} acts on a general state, $\sum_{b}^{|\mathcal{E}_{A}|}\sum_{ \overline{b}}^{|\mathcal{E}_{\overline{A}}|}\psi^{(b, \overline{b})}| b\hspace{0.3mm},\hspace{0.3mm}\overline{b}\rangle_{bdry}^{(\Psi)}$, in $\widetilde{\mathcal{H}}_{code}^{(\Psi)}$, we obtain
%%%
\ba{
\phi_{bdry}\sum_{b}^{|\mathcal{E}_{A}|}\sum_{ \overline{b}}^{|\mathcal{E}_{\overline{A}}|}\psi^{(b, \overline{b})}| b\hspace{0.3mm},\hspace{0.3mm}\overline{b}\rangle_{bdry}^{(\Psi)}
=&\sum_{b}^{|\mathcal{E}_{A}|}\sum_{ \overline{b}}^{|\mathcal{E}_{\overline{A}}|}\psi^{(b, \overline{b})}
\sum_{a}^{|\mathcal{E}_{A}|}\phi^{a, b}\hspace{0.5mm}
| a\hspace{0.3mm},\hspace{0.3mm}\overline{b}\rangle_{bdry}^{(\Psi)} \nonumber\\
=&\sum_{b}^{|\mathcal{E}_{A}|}\sum_{ \overline{b}}^{|\mathcal{E}_{\overline{A}}|}\psi^{(b, \overline{b})}\ 
\phi_{bdry\rightarrow A}\otimes I_{\overline{A}}\ 
| b\hspace{0.3mm},\hspace{0.3mm}\overline{b}\rangle_{bdry}^{(\Psi)} \nonumber\\
=&\  \phi_{bdry\rightarrow A}\otimes I_{\overline{A}}\hspace{0.5mm}\sum_{b}^{|\mathcal{E}_{A}|}\sum_{ \overline{b}}^{|\mathcal{E}_{\overline{A}}|}\psi^{(b, \overline{b})}| b\hspace{0.3mm},\hspace{0.3mm}\overline{b}\rangle_{bdry}^{(\Psi)}
}
where we have used the definition \eqref{dualbulkoponE_A} in the second equality and also the final equality.
%Th is same  is Example \ref{essenceofShrinkHermitian}

It is worth emphasizing again that the single operator $\phi_{bdry\rightarrow A}$ for $\phi_{bdry}$ satisfies the condition \eqref{shrinkingoncode} not only just on a single state in $\widetilde{\mathcal{H}}_{code}^{(\Psi)}$ but also on any state in $\widetilde{\mathcal{H}}_{code}^{(\Psi)}$, just as the decoder defined in \eqref{expansiondec} does.
In this sense, the condition \eqref{shrinkingoncode} for  $\phi_{bdry\rightarrow A}(\mathcal{H}_{\text{code}})$ can hold without acting on a particular state only if they are supposed to act only on the code subspace. 
In other words, the following holds
\ba{
\phi_{bdry}\widetilde{P}_{code}^{(\Psi)}=\phi_{bdry\rightarrow A}\otimes I_{\overline{A}}\hspace{0.5mm}\widetilde{P}_{code}^{(\Psi)}\ .
}
where $\widetilde{P}_{code}^{(\Psi)}$ is the projection operator onto $\widetilde{\mathcal{H}}_{code}^{(\Psi)}$.
This gives an explanation for why 
the AdS Rindler-wedge reconstruction can be derived without acting on a particular state, as was done in \cite{Hamilton:2006az}, even though the shrunk operators are generally state-dependent.

The consequence  below also immediately follows from Theorem \ref{EWRwithCenter}\ :

%%%%%%%%%  Theorem %%%%%%%%%%%%%%
\noindent\rule{\columnwidth}{1pt}\vspace{-2mm}
%\begin{screen}
\begin{corollary} \label{}%\ \\
\ For $\phi_{bdry}\in\mathcal{A}(\widetilde{\mathcal{H}}_{code}^{(\Psi)})$, the following two statements are equivalent:
\ba{
(A)\  \ & ^{\exists}\phi_{bulk}\in 
 \mathcal{S}_{bulk}(\hspace{0.5mm} \mathcal{E}_{A};\widetilde{\mathcal{H}}_{bulk}\hspace{0.5mm} )\cap \mathcal{S}_{bulk}(\hspace{0.5mm} \mathcal{E}_{\overline{A}};\widetilde{\mathcal{H}}_{bulk}\hspace{0.5mm} )\ s.t. \ \ %\nonumber\\[5pt]
%\label{linktoA}\\[5pt]
 \phi_{bdry}%P_{code}
 =\hspace{0.5mm}U_{code}^{(\Psi)}\hspace{0.5mm}\phi_{bulk}\hspace{0.5mm}U_{code}^{(\Psi)\dagger}\  \label{dualrelationcenter}
\\[0.3cm]
(B)\ \ & \phi_{bdry}\in \mathcal{S}_{bdry}(\hspace{0.5mm} A;\widetilde{\mathcal{H}}_{code}^{(\Psi)}\hspace{0.5mm} ) \cap \mathcal{S}_{bdry}(\hspace{0.5mm} \overline{A};\widetilde{\mathcal{H}}_{code}^{(\Psi)}\hspace{0.5mm} )\ \label{Statement(B)inEWRcenter}
}
\end{corollary}
\vspace{-5mm}
\noindent\rule{\columnwidth}{1pt}
%%%%%%%%%%%%%%%%%%%%%%%%%%

This theorem says that bulk operators in $\mathcal{A}( \widetilde{\mathcal{H}}_{bulk})$ are reconstructable on $A$ and also on $\overline{A}$ if and only if they can shrink their support to $\mathcal{E}_{A}$ and also on $\mathcal{E}_{\overline{A}}$ in $\widetilde{\mathcal{H}}_{bulk}$.
Thus, the central operators of the reconstructed algebras on $\mathcal{E}_{A}$ and $\mathcal{E}_{\overline{A}}$ can emerge.

We end this section with a comment about the generalization of $\mathcal{A}(\widetilde{\mathcal{H}}_{code}^{(\Psi)})$.
Let us define $\widehat{\mathcal{A}}(\widetilde{\mathcal{H}}_{code}^{(\Psi)}\subset \mathcal{H}_{bdry})$ as
\ba{
\widehat{\mathcal{A}}(\widetilde{\mathcal{H}}_{code}^{(\Psi)}\subset \mathcal{H}_{bdry})
\equiv\{\ \mathcal{O}_{bdry}\in \mathcal{A}(\mathcal{H}_{bdry})\ |\ 
\mathcal{O}_{bdry}\widetilde{\mathcal{H}}_{code}^{(\Psi)}\subseteq\widetilde{\mathcal{H}}_{code}^{(\Psi)}\ ,\ 
\mathcal{O}_{bdry}\widetilde{\mathcal{H}}_{code}^{(\Psi)\bot}\subseteq\widetilde{\mathcal{H}}_{code}^{(\Psi)\bot}\ \}
}
where $\widetilde{\mathcal{H}}_{code}^{(\Psi)\bot}$ is the subspace in $\mathcal{H}_{bdry}$ orthogonal to $\widetilde{\mathcal{H}}_{code}^{(\Psi)}$, $i.e.$,
\ba{
\widetilde{\mathcal{H}}_{code}^{(\Psi)\bot}
=\{\ |\psi\rangle_{bdry} \in \mathcal{H}_{bdry}\ |\ _{bdry}\langle \psi|\phi\rangle_{bdry}=0\ ,\  ^{\forall}|\phi\rangle_{bdry} \in \widetilde{\mathcal{H}}_{code}^{(\Psi)}\ \}\ .
}
$\widehat{\mathcal{A}}(\widetilde{\mathcal{H}}_{code}^{(\Psi)}\subset \mathcal{H}_{bdry})$ is the more  general  set of Hermitian operators on $\mathcal{H}_{bdry}$ 
such that the algebra on $\widetilde{\mathcal{H}}_{code}^{(\Psi)}$ is closed in itself.
 $\mathcal{A}(\widetilde{\mathcal{H}}_{code}^{(\Psi)})$ can be obtained by projecting the support of  $\widehat{\mathcal{A}}(\widetilde{\mathcal{H}}_{code}^{(\Psi)}\subset \mathcal{H}_{bdry})$ onto $\widetilde{\mathcal{H}}_{code}^{(\Psi)}$,
\ba{
\mathcal{A}(\widetilde{\mathcal{H}}_{code}^{(\Psi)})=\{\ \mathcal{O}_{bdry}\widetilde{P}_{code}^{(\Psi)}\ |\ \mathcal{O}_{bdry} \in \widehat{\mathcal{A}}(\widetilde{\mathcal{H}}_{code}^{(\Psi)}\subset \mathcal{H}_{bdry})\ \}\ .
}
where $\widetilde{P}_{code}^{(\Psi)}$ is the projection operator onto $\mathcal{A}(\widetilde{\mathcal{H}}_{code}^{(\Psi)})$ .
Then Theorem \ref{EWRwithCenter} can be restated as
\ba{
%& ^{\exists}\phi_{\mathcal{E}_{A}}\in \mathcal{A}(\mathcal{H}_{\mathcal{E}_{A}})\ 
 %s.t.\   U_{hcode}(\phi_{\mathcal{E}_{A}}\otimes I_{\mathcal{E}_{\overline{A}}})U_{hcode}^{\dagger}=\phi_{bdry}\hspace{0.5mm}P_{code} \nonumber
  & ^{\exists}\phi_{bulk}\in 
 \mathcal{S}_{bulk}(\hspace{0.5mm} \mathcal{E}_{A};\widetilde{\mathcal{H}}_{bulk}\hspace{0.5mm} )\ s.t. \ \ %\nonumber\\[5pt]
%\label{linktoA}\\[5pt]
 \phi_{bdry}\widetilde{P}_{code}^{(\Psi)}
 =\hspace{0.5mm}U_{code}^{(\Psi)}\hspace{0.5mm}\phi_{bulk}\hspace{0.5mm}U_{code}^{(\Psi)\dagger}\\[3mm]
   \Leftrightarrow \  \ 
 &\ \phi_{bdry}\in \widehat{\mathcal{A}}(\widetilde{\mathcal{H}}_{code}^{(\Psi)}\subset \mathcal{H}_{bdry}) \cap \mathcal{S}_{bdry}(A;\widetilde{\mathcal{H}}_{code}^{(\Psi)}).
% \phi_{bdry}\  \text{is dual to a bulk operator with the support only on}  \ \mathcal{E}_{A}.
}
This would be a more natural expression in the context of AdS/CFT in the following sense.
The light local operators in CFT will be the natural candidates for acting within the low-energy code subspace. 
However they will not vanish in general when they act on some high energy states outside the code subspace. 
Therefore, the light local operators may be not the elements of $\mathcal{A}(\widetilde{\mathcal{H}}_{code}^{(\Psi)})$ but the elements of  $\widehat{\mathcal{A}}(\widetilde{\mathcal{H}}_{code}^{(\Psi)}\subset \mathcal{H}_{bdry})$.

%%%%%%%%%%%%%%%%%%%%%%%%%%%%%%%%%%%%%%%%%%%%%%%%%%%%

%%%%%%%%%%%%%%%%%%%%%%%%%%%%%%%%%%%%%%%%%%%%%%%%%%%%
\section{Summary and Discussion}\label{SandD}%Summary and Discussion

We have shown how to shrink the supports of operators  on a bipartite pure state.
 Based on this results, we have shown a formula for constructing the decoder against erasure errors.
 We have also pointed out that the nontrivial central operators of the algebras on subsystems can emerge when the Hilbert space is restricted to a subspace that has no tensor factorization into the subsystems.
 Finally we have proven a theorem which implies the validity of not only the entanglement wedge reconstruction but also its converse statement with the central operators. 

In section \ref{Shrinkingteleporting}, we have specified the class of operators that can shrink their supports from $AB$ to only  $A$ (or $B$) on an arbitrary bipartite pure state $|\psi\rangle_{AB}$, in Lemma \ref{generalgatetel}.
In other words, for an operator $\mathcal{O}_{AB}$ in the class, there exists an operator $\mathcal{O}_{AB\rightarrow B}(\psi)$ that satisfies
\ba{
\mathcal{O}_{AB}|\psi\rangle_{AB}=I_{A} \otimes \mathcal{O}_{AB\rightarrow B}(\psi)|\psi\rangle_{AB}.}
As a consequence of Lemma \ref{generalgatetel}, we have shown that any nonlocal operator on $AB$ with $|A|\ge |B|$ can shrink its support to $A$ on any bipartite state with the non-vanishing Schmidt coefficients in Corollary \ref{maximalgatetel}.
We have also specified the conditions for preserving  unitarity and Hermiticity of $\mathcal{O}_{AB\rightarrow B}(\psi)$ through the shrinking of the support.

In section \ref{Systematicconstruction}, we have shown a general expression of logical states against erasure errors in Lemma \ref{logicalbasis}. 
Then we have shown a formula to make decoders from the given logical basis in Theorem \ref{makedecoder}. 

In section \ref{adscft}, we have studied the subalgebra code with complementary recovery under the restrictions on the bulk (logical) Hilbert space.
We have pointed out that the central operators of the algebras associated with subregions can emerge when we  restrict the Hilbert space in subsection \ref{sec:emergenceofcenter}.
Finally, we have shown a theorem in the code, which implies the validity of the entanglement wedge reconstruction and also its converse statement with nontrivial centers in Theorem \ref{EWRwithCenter}.

\vspace{7mm}
We have concretely specified the set of operators that can shrink their supports on a state in section \ref{Shrinkingteleporting}.
However, the set of operators that can shrink not only on a state but also on a nontrivial subspace (namely $\mathcal{S}_{A\overline{A}}(A\hspace{0.5mm};\mathcal{H})$ for $\mathcal{H}\subset\mathcal{H}_{A}\otimes\mathcal{H}_{\overline{A}}$ in the case for preserving Hermiticity,) was not specified in general.
We have just observed that such shrinking is realized on the code subspace of the quantum error-correcting code in section \ref{Systematicconstruction} and section \ref{adscft}.
I leave the specification of $\mathcal{S}_{A\overline{A}}(A\hspace{0.5mm};\mathcal{H})$ to  future work.

The shrinking of operators might be also useful for implementing nonlocal quantum operations.
When we want to perform a unitary operation $\mathcal{U}_{AB}$ to  states in $\mathcal{H}(\subset\mathcal{H}_{A}\otimes\mathcal{H}_{B})$, the experimental construction of the unitary shrunk operation  $\mathcal{U}_{AB\rightarrow A}$ makes it possible to perform $\mathcal{U}_{AB}$ locally on the subsystem $A$.
One known example is implementing the decoder of the quantum error-correcting codes for the erasure errors as we have seen in section \ref{Systematicconstruction}.
The experimental implementations of the measurement of the Hermitian shrunk operator $\mathcal{O}_{AB\rightarrow A}$ for $\mathcal{O}_{AB}$ on $\mathcal{H}$ would also make it possible to measure the observable $\mathcal{O}_{AB}$ for states in $\mathcal{H}$ with the local measurement on the subsystem $A$.
In this sense, the complete specification of $\mathcal{S}_{A\overline{A}}(A\hspace{0.5mm};\mathcal{H})$ may be important and possibly gives powerful tool in some cases.

We have analyzed the properties of shrinking supports of operators in finite dimensional system.
The generalization to infinite dimensional system, in particular quantum field theories (QFT) with UV cut-off or local QFT, may be interesting and useful for further  understanding of the non-uniqueness of boundary operators in the bulk reconstruction in AdS/CFT. 
In local QFT case the generalization would give the concrete formula of constructing the shrunk operators in the Reeh-Schlieder theorem.

Regarding the quantum error-correcting code model for AdS/CFT in section \ref{adscft}, we fixed the state $\ket{\Psi}_{bulk}^{(gr)}$ in $\mathcal{H}_{bulk}^{(gr)}$ and also fixed the factorization $\mathcal{H}_{bulk}=\mathcal{H}_{\mathcal{E}_{A}}\otimes\mathcal{H}_{\mathcal{E}_{\overline{A}}}$ independently.
This state plays the role of the background geometry as we have seen that the area term of the Ryu-Takayanagi formula originates from the entanglement of the state between $\mathcal{H}_{\mathcal{E}_{A}}$ and $\mathcal{H}_{\mathcal{E}_{\overline{A}}}$.
In this sense, how to factorize $\mathcal{H}_{bulk}$ into $\mathcal{H}_{\mathcal{E}_{A}}$ and  $\mathcal{H}_{\mathcal{E}_{\overline{A}}}$ should be determined by $\ket{\Psi}_{bulk}^{(gr)}$ in some way since the division of the bulk spacetime into the entanglement wedges, $\mathcal{E}_{A}$ and $\mathcal{E}_{\overline{A}}$, is determined geometrically by the Hubeny-Rangamani-Takayanagi surface $\gamma_{A}$ \cite{Hubeny:2007xt}. 
Extremizing the entanglement entropy 
between $\mathcal{H}_{\mathcal{E}_{A}}$ and $\mathcal{H}_{\mathcal{E}_{\overline{A}}}$ may be a possible candidate for determining the factorization \cite{Engelhardt:2014gca}.
Incorporating the effect of the small gravitational perturbative corrections into the model is also beyond the scope of this paper.
I hope to come back to these questions in the future.

%%%%%%%%%%%%%%%%%%%%%%%%%%%%%%%%%%%%%
%\subsection{Reeh-Schlieder theorem}

%%%%%%%%%%%%%%%%%%%%%%%%%%%%%%%%%%%%%%%%%%%%%%%%%%%%%%%%%%%%%%
\section*{Acknowledgements}
%Thanks

I would like to thank Kin-ya Oda, Tomonori Ugajin and Satoshi Yamaguchi for useful conversations which led to this work. 
I am particularly grateful to  Satoshi Yamaguchi for many valuable discussions.
I would also like to express my gratitude to Keisuke Fujii, Masato Koashi, Sotaro Sugishita, Tadashi Takayanagi and Satoshi Yamaguchi for useful comments on a draft of this manuscript.
I acknowledge financial support from the JSPS fellowship.

%%%%%%%%%%%%%%%%%%%%%%%%%%%%%%%%%%%%%%%%%%%%%%%%%%%%%%%%%%%%%%

%%%%%%%%%%%%%%%%  Appendix  %%%%%%%%%%%%%%%%%%%%%%%%%%%%
%%%%%%%%%%%%%%%%%%%%%%%%%%%%%%%%%%%%%%%%%%%%%%%%%%%
\appendix

%%%%%%%%%%%%%%%%%%%%%%%%%%%%%%%%%%%%%%%%%%%%
%%%%%%%%%%%%%%%%%%%%%%%%%%%%%%%%%%%%%%%%%%%%
\section{Teleporting the supports of local operators }\label{sec:Teleporting}
We study the consequences of Lemma \ref{generalgatetel} about local operators. 
To do this, we just consider the case where
\ba{\label{localcase}
\mathcal{O}^{ij,kl}=\delta^{ik}\mathcal{O}^{jl}\ .
}
Then $\mathcal{O}_{AB}$ defined in eq.\eqref{operatorABpsiA} reduces to
\ba{
\mathcal{O}_{AB}=I_{A}\otimes \mathcal{O}_{B}
}
where we defined 
\ba{
\mathcal{O}_{B}\equiv\sum_{j=1}^{N}\sum_{l=1}^{N}\mathcal{O}^{jl}\hspace{0.5mm}|j\rangle_{B\hspace{0.5mm}B}\langle l|
+\sum_{j=1}^{|B|}\sum_{l>N}^{|B|}\mathcal{O}^{jl}\hspace{0.5mm}|j\rangle_{B\hspace{0.5mm}B}\langle l|.
}
Substituting eq.\eqref{localcase} into eq.\eqref{shrunkopABtoA}, we obtain corollary \ref{teleportatioinofop}:
%\newpage
%\vspace{5mm}

%%%%%%%  corollary  %%%%%%%%%%%%
\noindent\rule{\columnwidth}{1pt}\vspace{-2mm}
%\begin{screen}
\begin{corollary} \label{teleportatioinofop}
\ For a given arbitrary state, $|\psi\rangle_{AB}$, in $\mathcal{H}_{AB}$ with the Schmidt rank $N$ that is written in the Schmidt basis as \eqref{state}, the following two statements are equivalent,
\ba{
(A)\  \ &\text{For}\  \mathcal{O}_{B}\in\mathcal{L}(\mathcal{H}_{B}), \  ^{\exists}\mathcal{O}_{B\rightarrow A}(\psi)\in\mathcal{L}(\mathcal{H}_{A}) \ s.t. \nonumber\\[5pt]
\label{teleportationlocalop}
& I_{A}\otimes \mathcal{O}_{B}\hspace{0.5mm}|\psi\rangle_{AB}=\mathcal{O}_{B\rightarrow A}(\psi) \otimes I_{B}\hspace{0.5mm}|\psi\rangle_{AB}\ . \\[0.3cm]
(B)\ \ &\text{$\mathcal{O}_{B}$ and  $\mathcal{O}_{B\rightarrow A}(\psi)$ can be written as follows with $\mathcal{O}^{jl}, \overline{\mathcal{O}}^{ij}\in\mathbb{C}$},\\
&\mathcal{O}_{B}\equiv\sum_{j=1}^{N}\sum_{l=1}^{N}\mathcal{O}^{jl}\hspace{0.5mm}|j\rangle_{B\hspace{0.5mm}B}\langle l|
+\overline{\mathcal{O}}_{B}(\psi)
\ ,\ \\%[-0.7cm]
&\label{teleportedop}
\mathcal{O}_{B\rightarrow A}(\psi)\equiv\sum_{i,j=1}^{N}\psi^{-1}_{j}\mathcal{O}^{ji}\psi_{i}\hspace{0.5mm}|i\rangle_{A\hspace{0.5mm}A}\langle j|
+\sum_{i}^{|A|}\sum_{j>N}^{|A|}\overline{\mathcal{O}}^{ij}
\hspace{0.5mm}|i\rangle_{A\hspace{0.5mm}A}\langle j|\\
%+\overline{\mathcal{O}}_{A}(\psi)\ , \hspace{4.5cm} \\
&\text{where}\ \ I_{A}\otimes\overline{\mathcal{O}}_{B}(\psi)\in \mathcal{V}(\hspace{0.5mm}|\psi\rangle_{AB}\hspace{0.5mm}) .\nonumber
%\ \text{and} \ \  \overline{\mathcal{O}}_{A}(\psi)\otimes I_{B} \in \mathcal{V}(\hspace{0.5mm}|\psi\rangle_{AB}\hspace{0.5mm}).\nonumber
}
\end{corollary}
%\end{screen}
\vspace{-5mm}
\noindent\rule{\columnwidth}{1pt}

\noindent In this sense, Lemma \ref{generalgatetel} provides a method of ``teleporting" the support of local operators from $A$ to $B$. 
As an application of this formula, we calculate the teleportation of the supports of local operators on the thermofield double state in Appendix \ref{sec:TFD}.

By substituting eq.\eqref{localcase} into Lemma \ref{preservingUnitarity} and Lemma \ref{preservingHermiticity},  we obtain the conditions for preserving the unitarity and the Hermiticity through this teleporting as follows.

%%%%%%%%%  corollary %%%%%%%%%%%%%%
\noindent\rule{\columnwidth}{1pt}\vspace{-2mm}
%\begin{screen}
\begin{corollary} \label{preservingUnitarityinTele}%\ \\
\ In Corollary \ref{teleportatioinofop}, $\mathcal{O}_{B\rightarrow A}$ becomes a unitary operator on $\mathcal{H}_A$ if  $\mathcal{O}_{B}$ and $ \overline{\mathcal{O}}_{A}(\psi)$  satisfy the following conditions,
%\begin{enumerate}
%\item 
%\normalfont
\ba{
&\text{{\rm (i)}}\ \ \sum_{j}^{N}\psi_{i}^{-1}\psi_{k}^{-1}\mathcal{O}^{i,j\ast}\mathcal{O}^{k,j}\psi_{n}^{2}=\delta_{i,k}\ ,\hspace{3mm} 
%\text{for\ \ $i , k=1,\cdots,N$}\\
\text{for\ \  $1\le i \le N$\ , $1\le k \le N$}, \\
&\text{{\rm (ii)}}\ \ \sum_{j}^{N}\psi_{i}^{-1}\mathcal{O}^{i,j\ast}\overline{\mathcal{O}}^{j,k}\psi_{j}=0\ ,\hspace{3mm}  
%\text{for\ \ $i=1,\cdots,N$\ ,\ $k=N+1,\cdots,|A|$}\\
\text{for\ \  $1\le i \le N$\ , $N< k \le |A|$},\\
&\text{{\rm (iii)}}\ \ \sum_{j}^{|A|}\overline{\mathcal{O}}^{j,i\ast}\overline{\mathcal{O}}^{j,k}=\delta_{i,k}\ ,
\hspace{3mm} \text{for\ \  $N< i \le |A|$\ , $N< k \le |A|$}.
}
%\end{enumerate}
\end{corollary}
\vspace{-5mm}
\noindent\rule{\columnwidth}{1pt}
%%%%%%%%%%%%%%%%%%%%%%%%%%

%%%%%%%%%  corollary %%%%%%%%%%%%%%
\noindent\rule{\columnwidth}{1pt}\vspace{-2mm}
%\begin{screen}
\begin{corollary} \label{preservingHermiticityinTele}%\ \\
\ In Corollary \ref{teleportatioinofop}, $\mathcal{O}_{B\rightarrow A}$ becomes a Hermitian operator on $\mathcal{H}_A$ if  $\mathcal{O}_{B}$ and $ \overline{\mathcal{O}}_{A}(\psi)$  satisfy the following conditions,
%\begin{enumerate}
%\item 
%\normalfont
\ba{
&\text{{\rm (i)}}\ \ \psi_{i}^{-1}\mathcal{O}^{i,k\ast}\psi_{k}
=\psi_{k}^{-1}\mathcal{O}^{k,i}\psi_{i}
\ ,\hspace{3mm} \text{for\ \  $1\le i \le N$\ , $1\le k \le N$}, \label{HermiticitycondTele1}\\
&\text{{\rm (ii)}}\ \ \overline{\mathcal{O}}^{i,k}=0 ,\hspace{3mm} \text{for\ \  $1\le i \le N$\ , $N< k \le |A|$}, \\
&\text{{\rm (iii)}}\ \ \overline{\mathcal{O}}^{k,i\ast}=\overline{\mathcal{O}}^{i,k}\ ,
\hspace{3mm} \text{for\ \  $N< i \le |A|$\ , $N< k \le |A|$}.\label{HermiticitycondTele3}
}
%\end{enumerate}
\end{corollary}
\vspace{-5mm}
\noindent\rule{\columnwidth}{1pt}
%%%%%%%%%%%%%%%%%%%%%%%%%%

%%%%%%%%%%%%%%%%%%%%%%%%%%%%%%
%%%%%%%%%%%%%%%%%%%%%%%%%%%%%%%%%%%%%%%%%%%%%%%%%%%

%%%%%%%%%%%%%%%%%%%%%%%%%%%%%%
%%%%%%%%%%%%%%%%%%%%%%%%%%%%%%%%%%%%%%%%%%%%%%%%%%%
\section{Teleporting the supports of local operators on the thermofield double state}\label{sec:TFD}
In this section, we calculate the teleported operator on the thermofield doubled (TFD) state \cite{Maldacena:2001kr} based on the formula in Corollary \ref{teleportatioinofop} and see that the answer reproduces the known result.

The thermofield double(TFD) state is defined as an entangled state on two independent copies of a quantum system such that the density matrix for either system behaves as a thermal state at inverse temperature $\beta$, $i.e.,$
\ba{
|TFD\rangle=\frac{1}{\sqrt{Z(\beta)}}\sum_{n}e^{-\frac{\beta E_{n}}{2}}|E_{n}\rangle_{L}|E_{n}\rangle_{R}
}
where $|E_{n}\rangle_{L/R}$ is the energy eigenstate of the left/right quantum system and $Z(\beta)$ is the canonical partition function at inverse temperature $\beta$. This is written in the Schmidt decomposition and we define the Schmidt coefficients as
$
\psi_{n}\equiv\frac{1}{\sqrt{Z(\beta)}}e^{-\frac{\beta E_{n}}{2}}.
$
%\ba{
%(\mathcal{O}_{L\rightarrow R})_{mn}=\psi_{n}^{-1}\mathcal{O}_{nm}\psi_{m}=\mathcal{O}_{nm}e^{-\frac{\beta}{2}(E_{m}-E_{n})}
%}
Then if we consider an operator on the left quantum system,
\ba{
\mathcal{O}_{L}\equiv\sum_{m,n}\mathcal{O}_{mn}|E_{m}\rangle_{L\hspace{0.5mm}L}\langle E_{n}|,
}
it follows from \eqref{teleportedop} and \eqref{teleportationlocalop} that the operator which satisfies
\ba{
\mathcal{O}_{L}\otimes I_{R}|TFD\rangle=I_{L}\otimes\mathcal{O}_{L\rightarrow R}(TFD)|TFD\rangle
}
can be written as
\ba{
&\mathcal{O}_{L\rightarrow R}(TFD)
=\sum_{m,n}\psi_{n}^{-1}\mathcal{O}_{nm}\psi_{m}|E_{m}\rangle_{R\hspace{0.5mm}R}\langle E_{n}|
%=\sum_{m,n}\mathcal{O}_{nm}e^{-\frac{\beta}{2}(E_{m}-E_{n})}|E_{m}\rangle_{R\hspace{0.5mm}R}\langle E_{n}|\\
=e^{-\frac{\beta}{2}H_{R}}(\mathcal{O}_{R})^{T}e^{\frac{\beta}{2}H_{R}}
}
where we defined
$
\mathcal{O}_{R}\equiv\sum_{m,n}\mathcal{O}_{mn}|E_{m}\rangle_{R\hspace{0.5mm}R}\langle E_{n}|,
$
which is the identical copy of $\mathcal{O}_{L}$ to the right quantum system. This result has already been known and used in the context of AdS/CFT correspondence (see $e.g.$ \cite{Hartman:2013qma, Caputa:2015waa, Papadodimas:2013jku}).

%%%%%%%%%%%%%%%%%%%%%%%%%%%%%%%%%%%%%%%%%%%%%%%%%%%%%%%%%%%%%%
%%%%%%%%%%%%%%%%%%%%%%%%%%%%%%%%%%%%%%%%%%%%%%%%%%%%%%%%%%%%%%

\bibliographystyle{utphys}
\bibliography{Ref}

\providecommand{\href}[2]{#2}\begingroup\raggedright\begin{thebibliography}{10}

\bibitem{PhysRevA.52.R2493}
P.~W. Shor, ``Scheme for reducing decoherence in quantum computer memory,''
  \href{http://dx.doi.org/10.1103/PhysRevA.52.R2493}{{\em Phys. Rev. A}
  {\bfseries 52} (Oct, 1995) R2493--R2496}.
  \url{https://link.aps.org/doi/10.1103/PhysRevA.52.R2493}.

\bibitem{Grassl:1996eh}
M.~Grassl, T.~Beth, and T.~Pellizzari, ``{Codes for the quantum erasure
  channel},'' \href{http://dx.doi.org/10.1103/PhysRevA.56.33}{{\em Phys. Rev.}
  {\bfseries A56} (1997) 33},
\href{http://arxiv.org/abs/quant-ph/9610042}{{\ttfamily arXiv:quant-ph/9610042
  [quant-ph]}}.
%%CITATION = QUANT-PH/9610042;%%.

\bibitem{Hillery:1998yq}
M.~Hillery, V.~Buzek, and A.~Berthiaume, ``{Quantum secret sharing},''
  \href{http://dx.doi.org/10.1103/PhysRevA.59.1829}{{\em Phys. Rev.} {\bfseries
  A59} (1999) 1829},
\href{http://arxiv.org/abs/quant-ph/9806063}{{\ttfamily arXiv:quant-ph/9806063
  [quant-ph]}}.
%%CITATION = QUANT-PH/9806063;%%.

\bibitem{Cleve:1999qg}
R.~Cleve, D.~Gottesman, and H.-K. Lo, ``{How to share a quantum secret},''
  \href{http://dx.doi.org/10.1103/PhysRevLett.83.648}{{\em Phys. Rev. Lett.}
  {\bfseries 83} (1999) 648--651},
\href{http://arxiv.org/abs/quant-ph/9901025}{{\ttfamily arXiv:quant-ph/9901025
  [quant-ph]}}.
%%CITATION = QUANT-PH/9901025;%%.

\bibitem{Maldacena:1997re}
J.~M. Maldacena, ``{The Large N limit of superconformal field theories and
  supergravity},'' \href{http://dx.doi.org/10.1023/A:1026654312961,
  10.4310/ATMP.1998.v2.n2.a1}{{\em Int. J. Theor. Phys.} {\bfseries 38} (1999)
  1113--1133}, \href{http://arxiv.org/abs/hep-th/9711200}{{\ttfamily
  arXiv:hep-th/9711200 [hep-th]}}.
[Adv. Theor. Math. Phys.2,231(1998)].
%%CITATION = HEP-TH/9711200;%%.

\bibitem{Hamilton:2006az}
A.~Hamilton, D.~N. Kabat, G.~Lifschytz, and D.~A. Lowe, ``{Holographic
  representation of local bulk operators},''
  \href{http://dx.doi.org/10.1103/PhysRevD.74.066009}{{\em Phys. Rev.}
  {\bfseries D74} (2006) 066009},
\href{http://arxiv.org/abs/hep-th/0606141}{{\ttfamily arXiv:hep-th/0606141
  [hep-th]}}.
%%CITATION = HEP-TH/0606141;%%.

\bibitem{Morrison:2014jha}
I.~A. Morrison, ``{Boundary-to-bulk maps for AdS causal wedges and the
  Reeh-Schlieder property in holography},''
  \href{http://dx.doi.org/10.1007/JHEP05(2014)053}{{\em JHEP} {\bfseries 05}
  (2014) 053},
\href{http://arxiv.org/abs/1403.3426}{{\ttfamily arXiv:1403.3426 [hep-th]}}.
%%CITATION = ARXIV:1403.3426;%%.

\bibitem{Hubeny:2012wa}
V.~E. Hubeny and M.~Rangamani, ``{Causal Holographic Information},''
  \href{http://dx.doi.org/10.1007/JHEP06(2012)114}{{\em JHEP} {\bfseries 06}
  (2012) 114},
\href{http://arxiv.org/abs/1204.1698}{{\ttfamily arXiv:1204.1698 [hep-th]}}.
%%CITATION = ARXIV:1204.1698;%%.

\bibitem{Czech:2012bh}
B.~Czech, J.~L. Karczmarek, F.~Nogueira, and M.~Van~Raamsdonk, ``{The Gravity
  Dual of a Density Matrix},''
  \href{http://dx.doi.org/10.1088/0264-9381/29/15/155009}{{\em Class. Quant.
  Grav.} {\bfseries 29} (2012) 155009},
\href{http://arxiv.org/abs/1204.1330}{{\ttfamily arXiv:1204.1330 [hep-th]}}.
%%CITATION = ARXIV:1204.1330;%%.

\bibitem{Wall:2012uf}
A.~C. Wall, ``{Maximin Surfaces, and the Strong Subadditivity of the Covariant
  Holographic Entanglement Entropy},''
  \href{http://dx.doi.org/10.1088/0264-9381/31/22/225007}{{\em Class. Quant.
  Grav.} {\bfseries 31} no.~22, (2014) 225007},
\href{http://arxiv.org/abs/1211.3494}{{\ttfamily arXiv:1211.3494 [hep-th]}}.
%%CITATION = ARXIV:1211.3494;%%.

\bibitem{Headrick:2014cta}
M.~Headrick, V.~E. Hubeny, A.~Lawrence, and M.~Rangamani, ``{Causality \&
  holographic entanglement entropy},''
  \href{http://dx.doi.org/10.1007/JHEP12(2014)162}{{\em JHEP} {\bfseries 12}
  (2014) 162},
\href{http://arxiv.org/abs/1408.6300}{{\ttfamily arXiv:1408.6300 [hep-th]}}.
%%CITATION = ARXIV:1408.6300;%%.

\bibitem{Jafferis:2015del}
D.~L. Jafferis, A.~Lewkowycz, J.~Maldacena, and S.~J. Suh, ``{Relative entropy
  equals bulk relative entropy},''
  \href{http://dx.doi.org/10.1007/JHEP06(2016)004}{{\em JHEP} {\bfseries 06}
  (2016) 004},
\href{http://arxiv.org/abs/1512.06431}{{\ttfamily arXiv:1512.06431 [hep-th]}}.
%%CITATION = ARXIV:1512.06431;%%.

\bibitem{Almheiri:2014lwa}
A.~Almheiri, X.~Dong, and D.~Harlow, ``{Bulk Locality and Quantum Error
  Correction in AdS/CFT},''
  \href{http://dx.doi.org/10.1007/JHEP04(2015)163}{{\em JHEP} {\bfseries 04}
  (2015) 163},
\href{http://arxiv.org/abs/1411.7041}{{\ttfamily arXiv:1411.7041 [hep-th]}}.
%%CITATION = ARXIV:1411.7041;%%.

\bibitem{Mintun:2015qda}
E.~Mintun, J.~Polchinski, and V.~Rosenhaus, ``{Bulk-Boundary Duality, Gauge
  Invariance, and Quantum Error Corrections},''
  \href{http://dx.doi.org/10.1103/PhysRevLett.115.151601}{{\em Phys. Rev.
  Lett.} {\bfseries 115} no.~15, (2015) 151601},
\href{http://arxiv.org/abs/1501.06577}{{\ttfamily arXiv:1501.06577 [hep-th]}}.
%%CITATION = ARXIV:1501.06577;%%.

\bibitem{Pastawski:2015qua}
F.~Pastawski, B.~Yoshida, D.~Harlow, and J.~Preskill, ``{Holographic quantum
  error-correcting codes: Toy models for the bulk/boundary correspondence},''
  \href{http://dx.doi.org/10.1007/JHEP06(2015)149}{{\em JHEP} {\bfseries 06}
  (2015) 149},
\href{http://arxiv.org/abs/1503.06237}{{\ttfamily arXiv:1503.06237 [hep-th]}}.
%%CITATION = ARXIV:1503.06237;%%.

\bibitem{Hayden:2016cfa}
P.~Hayden, S.~Nezami, X.-L. Qi, N.~Thomas, M.~Walter, and Z.~Yang,
  ``{Holographic duality from random tensor networks},''
  \href{http://dx.doi.org/10.1007/JHEP11(2016)009}{{\em JHEP} {\bfseries 11}
  (2016) 009},
\href{http://arxiv.org/abs/1601.01694}{{\ttfamily arXiv:1601.01694 [hep-th]}}.
%%CITATION = ARXIV:1601.01694;%%.

\bibitem{Freivogel:2016zsb}
B.~Freivogel, R.~Jefferson, and L.~Kabir, ``{Precursors, Gauge Invariance, and
  Quantum Error Correction in AdS/CFT},''
  \href{http://dx.doi.org/10.1007/JHEP04(2016)119}{{\em JHEP} {\bfseries 04}
  (2016) 119},
\href{http://arxiv.org/abs/1602.04811}{{\ttfamily arXiv:1602.04811 [hep-th]}}.
%%CITATION = ARXIV:1602.04811;%%.

\bibitem{Harlow:2016vwg}
D.~Harlow, ``{The Ryu-Takayanagi Formula from Quantum Error Correction},''
  \href{http://dx.doi.org/10.1007/s00220-017-2904-z}{{\em Commun. Math. Phys.}
  {\bfseries 354} no.~3, (2017) 865--912},
\href{http://arxiv.org/abs/1607.03901}{{\ttfamily arXiv:1607.03901 [hep-th]}}.
%%CITATION = ARXIV:1607.03901;%%.

\bibitem{Donnelly:2016qqt}
W.~Donnelly, B.~Michel, D.~Marolf, and J.~Wien, ``{Living on the Edge: A Toy
  Model for Holographic Reconstruction of Algebras with Centers},''
  \href{http://dx.doi.org/10.1007/JHEP04(2017)093}{{\em JHEP} {\bfseries 04}
  (2017) 093},
\href{http://arxiv.org/abs/1611.05841}{{\ttfamily arXiv:1611.05841 [hep-th]}}.
%%CITATION = ARXIV:1611.05841;%%.

\bibitem{Pastawski:2016qrs}
F.~Pastawski and J.~Preskill, ``{Code properties from holographic
  geometries},'' \href{http://dx.doi.org/10.1103/PhysRevX.7.021022}{{\em Phys.
  Rev.} {\bfseries X7} no.~2, (2017) 021022},
\href{http://arxiv.org/abs/1612.00017}{{\ttfamily arXiv:1612.00017
  [quant-ph]}}.
%%CITATION = ARXIV:1612.00017;%%.

\bibitem{Ryu:2006bv}
S.~Ryu and T.~Takayanagi, ``{Holographic derivation of entanglement entropy
  from AdS/CFT},'' \href{http://dx.doi.org/10.1103/PhysRevLett.96.181602}{{\em
  Phys. Rev. Lett.} {\bfseries 96} (2006) 181602},
\href{http://arxiv.org/abs/hep-th/0603001}{{\ttfamily arXiv:hep-th/0603001
  [hep-th]}}.
%%CITATION = HEP-TH/0603001;%%.

\bibitem{Takayanagi:2017knl}
T.~Takayanagi and K.~Umemoto, ``{Entanglement of purification through
  holographic duality},''
  \href{http://dx.doi.org/10.1038/s41567-018-0075-2}{{\em Nature Phys.}
  {\bfseries 14} no.~6, (2018) 573--577},
\href{http://arxiv.org/abs/1708.09393}{{\ttfamily arXiv:1708.09393 [hep-th]}}.
%%CITATION = ARXIV:1708.09393;%%.

\bibitem{Nguyen:2017yqw}
P.~Nguyen, T.~Devakul, M.~G. Halbasch, M.~P. Zaletel, and B.~Swingle,
  ``{Entanglement of purification: from spin chains to holography},''
  \href{http://dx.doi.org/10.1007/JHEP01(2018)098}{{\em JHEP} {\bfseries 01}
  (2018) 098},
\href{http://arxiv.org/abs/1709.07424}{{\ttfamily arXiv:1709.07424 [hep-th]}}.
%%CITATION = ARXIV:1709.07424;%%.

\bibitem{Hirai:2018jwy}
H.~Hirai, K.~Tamaoka, and T.~Yokoya, ``{Towards Entanglement of Purification
  for Conformal Field Theories},''
  \href{http://dx.doi.org/10.1093/ptep/pty063}{{\em PTEP} {\bfseries 2018}
  no.~6, (2018) 063B03},
\href{http://arxiv.org/abs/1803.10539}{{\ttfamily arXiv:1803.10539 [hep-th]}}.
%%CITATION = ARXIV:1803.10539;%%.

\bibitem{Dong:2016eik}
X.~Dong, D.~Harlow, and A.~C. Wall, ``{Reconstruction of Bulk Operators within
  the Entanglement Wedge in Gauge-Gravity Duality},''
  \href{http://dx.doi.org/10.1103/PhysRevLett.117.021601}{{\em Phys. Rev.
  Lett.} {\bfseries 117} no.~2, (2016) 021601},
\href{http://arxiv.org/abs/1601.05416}{{\ttfamily arXiv:1601.05416 [hep-th]}}.
%%CITATION = ARXIV:1601.05416;%%.

\bibitem{PhysRevLett.98.100502}
C.~B\'eny, A.~Kempf, and D.~W. Kribs, ``Generalization of quantum error
  correction via the heisenberg picture,''
  \href{http://dx.doi.org/10.1103/PhysRevLett.98.100502}{{\em Phys. Rev. Lett.}
  {\bfseries 98} (Mar, 2007) 100502}.
  \url{https://link.aps.org/doi/10.1103/PhysRevLett.98.100502}.

\bibitem{PhysRevA.76.042303}
C.~B\'eny, A.~Kempf, and D.~W. Kribs, ``Quantum error correction of
  observables,'' \href{http://dx.doi.org/10.1103/PhysRevA.76.042303}{{\em Phys.
  Rev. A} {\bfseries 76} (Oct, 2007) 042303}.
  \url{https://link.aps.org/doi/10.1103/PhysRevA.76.042303}.

\bibitem{article}
H.~Reeh and S.~Schlieder, ``Bemerkungen zur
  unita$\ddot{\text{a}}$r$\ddot{\text{a}}$quivalenz von lorentzinvarianten
  feldern. nuovo cim. x 22, 1051,''
  \href{http://dx.doi.org/10.1007/BF02787889}{{\em Il Nuovo Cimento} {\bfseries
  22} (12, 1961) 1051--1068}.

\bibitem{Witten:2018lha}
E.~Witten, ``{APS Medal for Exceptional Achievement in Research: Invited
  article on entanglement properties of quantum field theory},''
  \href{http://dx.doi.org/10.1103/RevModPhys.90.045003}{{\em Rev. Mod. Phys.}
  {\bfseries 90} no.~4, (2018) 045003},
\href{http://arxiv.org/abs/1803.04993}{{\ttfamily arXiv:1803.04993 [hep-th]}}.
%%CITATION = ARXIV:1803.04993;%%.

\bibitem{Wootters:1982zz}
W.~K. Wootters and W.~H. Zurek, ``{A single quantum cannot be cloned},''
\href{http://dx.doi.org/10.1038/299802a0}{{\em Nature} {\bfseries 299} (1982)
  802--803}.
%%CITATION = NATUA,299,802;%%.

\bibitem{Dieks:1982dj}
D.~Dieks, ``{COMMUNICATION BY EPR DEVICES},''
\href{http://dx.doi.org/10.1016/0375-9601(82)90084-6}{{\em Phys. Lett.}
  {\bfseries A92} (1982) 271--272}.
%%CITATION = PHLTA,A92,271;%%.

\bibitem{Schumacher:1996dy}
B.~Schumacher and M.~A. Nielsen, ``{Quantum data processing and error
  correction},'' \href{http://dx.doi.org/10.1103/PhysRevA.54.2629}{{\em Phys.
  Rev.} {\bfseries A54} (1996) 2629},
\href{http://arxiv.org/abs/quant-ph/9604022}{{\ttfamily arXiv:quant-ph/9604022
  [quant-ph]}}.
%%CITATION = QUANT-PH/9604022;%%.

\bibitem{Faulkner:2013ana}
T.~Faulkner, A.~Lewkowycz, and J.~Maldacena, ``{Quantum corrections to
  holographic entanglement entropy},''
  \href{http://dx.doi.org/10.1007/JHEP11(2013)074}{{\em JHEP} {\bfseries 11}
  (2013) 074},
\href{http://arxiv.org/abs/1307.2892}{{\ttfamily arXiv:1307.2892 [hep-th]}}.
%%CITATION = ARXIV:1307.2892;%%.

\bibitem{Brown:1986nw}
J.~D. Brown and M.~Henneaux, ``{Central Charges in the Canonical Realization of
  Asymptotic Symmetries: An Example from Three-Dimensional Gravity},''
\href{http://dx.doi.org/10.1007/BF01211590}{{\em Commun. Math. Phys.}
  {\bfseries 104} (1986) 207--226}.
%%CITATION = CMPHA,104,207;%%.

\bibitem{Casini:2013rba}
H.~Casini, M.~Huerta, and J.~A. Rosabal, ``{Remarks on entanglement entropy for
  gauge fields},'' \href{http://dx.doi.org/10.1103/PhysRevD.89.085012}{{\em
  Phys. Rev.} {\bfseries D89} no.~8, (2014) 085012},
\href{http://arxiv.org/abs/1312.1183}{{\ttfamily arXiv:1312.1183 [hep-th]}}.
%%CITATION = ARXIV:1312.1183;%%.

\bibitem{Radicevic:2014kqa}
D.~Radicevic, ``{Notes on Entanglement in Abelian Gauge Theories},''
\href{http://arxiv.org/abs/1404.1391}{{\ttfamily arXiv:1404.1391 [hep-th]}}.
%%CITATION = ARXIV:1404.1391;%%.

\bibitem{Soni:2015yga}
R.~M. Soni and S.~P. Trivedi, ``{Aspects of Entanglement Entropy for Gauge
  Theories},'' \href{http://dx.doi.org/10.1007/JHEP01(2016)136}{{\em JHEP}
  {\bfseries 01} (2016) 136},
\href{http://arxiv.org/abs/1510.07455}{{\ttfamily arXiv:1510.07455 [hep-th]}}.
%%CITATION = ARXIV:1510.07455;%%.

\bibitem{Ma:2015xes}
C.-T. Ma, ``{Entanglement with Centers},''
  \href{http://dx.doi.org/10.1007/JHEP01(2016)070}{{\em JHEP} {\bfseries 01}
  (2016) 070},
\href{http://arxiv.org/abs/1511.02671}{{\ttfamily arXiv:1511.02671 [hep-th]}}.
%%CITATION = ARXIV:1511.02671;%%.

\bibitem{Harlow:2015lma}
D.~Harlow, ``{Wormholes, Emergent Gauge Fields, and the Weak Gravity
  Conjecture},'' \href{http://dx.doi.org/10.1007/JHEP01(2016)122}{{\em JHEP}
  {\bfseries 01} (2016) 122},
\href{http://arxiv.org/abs/1510.07911}{{\ttfamily arXiv:1510.07911 [hep-th]}}.
%%CITATION = ARXIV:1510.07911;%%.

\bibitem{Hubeny:2007xt}
V.~E. Hubeny, M.~Rangamani, and T.~Takayanagi, ``{A Covariant holographic
  entanglement entropy proposal},''
  \href{http://dx.doi.org/10.1088/1126-6708/2007/07/062}{{\em JHEP} {\bfseries
  07} (2007) 062},
\href{http://arxiv.org/abs/0705.0016}{{\ttfamily arXiv:0705.0016 [hep-th]}}.
%%CITATION = ARXIV:0705.0016;%%.

\bibitem{Engelhardt:2014gca}
N.~Engelhardt and A.~C. Wall, ``{Quantum Extremal Surfaces: Holographic
  Entanglement Entropy beyond the Classical Regime},''
  \href{http://dx.doi.org/10.1007/JHEP01(2015)073}{{\em JHEP} {\bfseries 01}
  (2015) 073},
\href{http://arxiv.org/abs/1408.3203}{{\ttfamily arXiv:1408.3203 [hep-th]}}.
%%CITATION = ARXIV:1408.3203;%%.

\bibitem{Maldacena:2001kr}
J.~M. Maldacena, ``{Eternal black holes in anti-de Sitter},''
  \href{http://dx.doi.org/10.1088/1126-6708/2003/04/021}{{\em JHEP} {\bfseries
  04} (2003) 021},
\href{http://arxiv.org/abs/hep-th/0106112}{{\ttfamily arXiv:hep-th/0106112
  [hep-th]}}.
%%CITATION = HEP-TH/0106112;%%.

\bibitem{Hartman:2013qma}
T.~Hartman and J.~Maldacena, ``{Time Evolution of Entanglement Entropy from
  Black Hole Interiors},''
  \href{http://dx.doi.org/10.1007/JHEP05(2013)014}{{\em JHEP} {\bfseries 05}
  (2013) 014},
\href{http://arxiv.org/abs/1303.1080}{{\ttfamily arXiv:1303.1080 [hep-th]}}.
%%CITATION = ARXIV:1303.1080;%%.

\bibitem{Caputa:2015waa}
P.~Caputa, J.~Sim$\Acute{\text{o}}$n, A.~$\Check{\text{S}}$tikonas,
  T.~Takayanagi, and K.~Watanabe, ``{Scrambling time from local perturbations
  of the eternal BTZ black hole},''
  \href{http://dx.doi.org/10.1007/JHEP08(2015)011}{{\em JHEP} {\bfseries 08}
  (2015) 011},
\href{http://arxiv.org/abs/1503.08161}{{\ttfamily arXiv:1503.08161 [hep-th]}}.
%%CITATION = ARXIV:1503.08161;%%.

\bibitem{Papadodimas:2013jku}
K.~Papadodimas and S.~Raju, ``{State-Dependent Bulk-Boundary Maps and Black
  Hole Complementarity},''
  \href{http://dx.doi.org/10.1103/PhysRevD.89.086010}{{\em Phys. Rev.}
  {\bfseries D89} no.~8, (2014) 086010},
\href{http://arxiv.org/abs/1310.6335}{{\ttfamily arXiv:1310.6335 [hep-th]}}.
%%CITATION = ARXIV:1310.6335;%%.

\end{thebibliography}\endgroup
\end{document}